\documentclass[a4paper,11pt]{article}
\pdfoutput=1 

\usepackage{jheppub} 

\usepackage[T1]{fontenc} 
\usepackage{multirow}
\usepackage{setspace}

\title{First-order electroweak phase transition in a complex singlet model with $\mathbb{Z}_3$ symmetry}

\author[a,b,1]{Cheng-Wei Chiang}
\author[a,2]{Bo-Qiang Lu}

\affiliation[a]{Department of Physics, National Taiwan University, Taipei, Taiwan 10617, Republic of China}
\affiliation[b]{Institute of Physics, Academia Sinica, Taipei, Taiwan 11529, Republic of China}

\emailAdd{chengwei@phys.ntu.edu.tw}
\emailAdd{bqlu@phys.ntu.edu.tw}

\abstract{We consider an extension of the Standard Model with a complex singlet scalar, where a global $U(1)$ symmetry is explicitly broken to $\mathbb{Z}_3$ symmetry.
We study the two-step electroweak phase transition in the model and find that it can be of first-order 
if the heavy scalar mass falls in the range of $1-2$~TeV and the mixing angle $\left | \theta \right |\gtrsim 0.2$ ($11.5^{\circ}$).
The Higgs signal strength measurements at the LHC, on the other hand, restrict the mixing angle 
$\left | \theta \right |\lesssim 0.4$ ($23^{\circ}$).  Future colliders including high-luminosity LHC can probe the 
remaining parameter space of first-order phase transition in this scenario. After the $U(1)$ symmetry breaking,
the pseudo-Goldstone boson becomes a dark matter candidate due to a hidden $\mathbb{Z}_2$ symmetry of the model. 
We find that the pseudo-Goldstone boson can make up a small fraction of the observed dark matter and escape from the constraints of current direct detection.
We also show that the stochastic gravitational wave signals from the phase transition are potentially discoverable with 
future space-based interferometers.
}

\begin{document} 
\maketitle
\flushbottom

\section{Introduction}
\label{sec:intro}

The Standard Model (SM) of particle physics has been completed since the discovery of the Higgs boson at the LHC in 2012~\cite{ATLAS2012PLB, CMS2012PLB}. 
However, it is widely believed that new physics is required to explain various phenomena beyond SM, such as the existence of dark matter (DM) and the origin of neutrino mass and mixing.
We concern ourselves in this work two prominent particle physics puzzles.
One is the origin of baryon number asymmetry in the Universe~\cite{Planck2015AA}. 
In the content of electroweak (EW) baryogenesis~\cite{Kuzmin1985PLB, Cohen1990PLB, Cline2000, Morrissey2012NJP},
the baryon number asymmetry is generated outside the EW broken phase and then captured by the bubbles' expansion in the
progress of electroweak phase transition (EWPT).  Furthermore, a sufficiently strong first-order EWPT is required 
to suppress the washout of baryon number asymmetry through sphalerons~\cite{Morrissey2012NJP, Cline2006}. 
However, the EWPT in the SM with the observed Higgs boson mass is found to be a crossover~\cite{Onofrio2014PRL}.
The other is the existence of DM.
There is overwhelming evidence from cosmological and astrophysical observations that show about 85\% of the matter 
in the Universe is DM, instead of the normal matter made from SM particles~\cite{Planck2015AA, Zwicky1937APJ, Rubin1970APJ, Massey2007}.
The most popular class of DM candidates is that of weakly interacting massive particles (WIMPs), which
lies in the mass range of $1-1000$~GeV~\cite{Lee1977PRL, Hut1977PLB}.
These particles decouple from the thermal bath as the early Universe expands and cools and finally reach the 
appropriate relic density as observed now.

One of the simplest models to trigger a strong first-order EWPT is to introduce a new bosonic degree of freedom to the scalar potential of SM.
Various Higgs portal models with (or without) a DM candidate have been widely studied~
\cite{Ma2008PLB, Barger2009PRD, Gonderinger2012PRD, Yaguna2009JHEP, Profumo2010PRD, Arina2011, Mambrini2011PRD, Cline2013PRD, 
Bhattacharya2017PRD, Casas2017JHEP, GAMBIT2017EPJC, Jiang2016PRD, Chiang2018PRD, Baek2012JHEP, Fairbairn2013JHEP, Li2014JHEP, 
Beniwal2017JHEP, Beniwal2019JHEP, Hashino2018JHEP, Chiang2017PLB, Ahriche2019PLB, Alves2019JHEP, Breitbach2019JCAP, Duch2015JHEP, 
Chao2017JCAP, Huang2018JHEP, Kang2018JHEP, Kannike2019, He2016JHEP, Ghorbani2018, Dev2019JCAP, Paul2019JCAP, Alanne2019}.
The extension of SM with a complex singlet scalar for the above-mentioned purpose was proposed and studied in Refs.~\cite{Ma2008PLB, Barger2009PRD}, followed by extensive researches on its implications in subsequent works~\cite{Barger2009PRD, Gonderinger2012PRD, Jiang2016PRD, Chiang2018PRD}.
In the Higgs-portal model with a real singlet scalar as the DM candidate, the scalar is stabilized by a $\mathbb{Z}_2$ symmetry and cannot have a vacuum expectation value (VEV)~\cite{Profumo2007JHEP}.  DM phenomena related to this model have been studied in Refs.~\cite{Yaguna2009JHEP, Profumo2010PRD, Arina2011, Mambrini2011PRD, Cline2013PRD, Bhattacharya2017PRD, Casas2017JHEP, GAMBIT2017EPJC}. 
A previous study~\cite{Espinosa2008PRD} also shows that such a model cannot trigger a sufficiently strong first-order EWPT except when a large number of singlet scalars ($\sim 10$) are introduced. 
In the model with a global $U(1)$ group explicitly broken down to $\mathbb{Z}_2$ symmetry, Ref.~\cite{Kannike2019PRD} shows that all phase transitions leading to the correct vacuum are of the second order.  Furthermore,
Ref.~\cite{Kurup2017PRD} shows that a large portion of the parameter space for the $\mathbb{Z}_2$ symmetry model will be ruled out by the required bubble nucleation condition, $S(T_{n})/T_{n}\sim 140$.

By now, no obvious evidence for WIMPs has been observed in both DM direct detections and collider searches. In particular, the recent constraints from LUX~\cite{LUX2017PRL}, PandaX-II~\cite{PandaX-II2017PRL}, 
and XENON1T~\cite{XENON1T2018PRL} tell us that the DM interaction cross section with protons and neutrons is extraordinarily tiny, less than about $10^{-45}-10^{-46}~\rm{cm^{2}}$, low enough to rule out most of WIMP DM models. 
Refs.~\cite{Barger2008PRD,Barger2009PRD,Gonderinger2012PRD,Jiang2016PRD,Chiang2018PRD,Cai2011PRD,Gross2017PRL,He2009PRD,Cline2013JCAP} show that there is a novel mechanism for suppressing the direct detection cross section
if the global $U(1)$ symmetry of Higgs portal model is softly broken to $\mathbb{Z}_2$ symmetry by a mass term.
Such a suppression is due to a cancellation in the DM-nucleon scattering amplitude between the SM-like and new Higgs boson contributions at zero momentum transfer. Although the cancellation is spoiled at loop level, the scattering contribution from one loop is trivial and the conclusion remains practically unchanged~\cite{Azevedo2019JHEP, Ishiwata2018JHEP}.

In this work we are concerned with one of the simplest extensions of SM with a complex singlet scalar.  In the model, the $U(1)$ symmetry is softly broken by a cubic term to $\mathbb{Z}_3$ symmetry (see also recent works~\cite{Kang2018JHEP, Kannike2019}).
The real part of the complex scalar acquires a VEV, while the imaginary component (pseudo-Goldstone boson) 
plays the role of DM candidate due to a residual $\mathbb{Z}_2$ symmetry in the scalar potential. 
As shown in a previous study~\cite{Profumo2007JHEP}, such a cubic term could generally induce a potential barrier at tree level and trigger a strong first-order EWPT. 
We find that in the two-step EWPT scenario and for the heavy scalar mass falling in the range of $1-2$~TeV, the model with a Higgs mixing angle satisfying $\left | \theta \right |\gtrsim 0.2$ ($11.5^{\circ}$) can induce a strong first-order EWPT and the stochastic gravitational wave (GW) signals produced from the phase transition are potentially discoverable by future space-based interferometers, such as LISA~\cite{LISA2017}, DECIGO~\cite{Sato2017JPCS} and BBO~\cite{Crowder2005PRD}.

This work is presented as follows. In sections~\ref{sec:model} and \ref{sec:ewpt}, we describe our model and study detailed properties of the model at tree level. In section~\ref{sec:ss}, we present our search scheme and results for 
the strength of the EWPT from a comprehensive scan of the parameter space.
We study the bounds of vacuum stability and perturbativity in the parameter space in section~\ref{sec:VPbounds}.
We take into account the constraints from electroweak precision observables and Higgs searches at the colliders in section~\ref{sec:expbounds}.
The DM phenomenology of the pseudo-Goldstone boson $\chi$ is studied in section~\ref{sec:dmphen}.
The discussions of GW production along with the first-order cosmological phase transition and its detection by 
future space-based interferometers are given in section~\ref{sec:gw}.
In section~\ref{sec:gaugedepend}, we study the effects of gauge dependence in our results and conclusions.
We summarize our findings in section~\ref{sec:summary}.
Some detailed formulas and parameters are collected in the appendices: 
appendix~\ref{apd:fdms} gives the field-dependent mass matrices of the particles at finite temperature; 
appendix~\ref{apd:RGE} provides the renormalizable group equations of the parameters in the scalar potential, and shows the dependence of the critical temperature and the corresponding vacuum expectation value on the renormalization scale;
appendix~\ref{apd:ccp} provides the coefficients of counter-terms in the scalar potential; 
appendix~\ref{apd:dw} gives the partial decay widths of the SM-like Higgs and heavy scalar; 
and appendix~\ref{apd:ssbi} discusses the sensitivities of space-based interferometers.

\section{The model with $\mathbb{Z}_3$ symmetry}
\label{sec:model}

We consider an extension of the SM with just a complex gauge-singlet scalar field $S$ which transforms under a global $\mathbb{Z}_3$ transformation as $S\to \exp(i2\pi/3)S$. Imposing the $\mathbb{Z}_3$ symmmetry associated with $S$, we can write down the most general renormalizable scalar potential with softly $U(1)$ symmetry breaking:
\begin{equation}
  \label{eq:poten:1}
  V(H, S)=-\mu_{h}^{2}|H|^{2}+\lambda_{h}|H|^{4}+\frac{1}{2} \lambda_{m}|H|^{2}|S|^{2}+\frac{1}{2} \mu_{s}^{2}|S|^{2}+
  \frac{1}{6} \mu_{3}\left(S^{3}+S^{* 3}\right)+\frac{1}{4} \lambda_{s}|S|^{4},
\end{equation}
where $H$ denotes the SM Higgs doublet and $\mu_3$ is assumed to be real. The symmetry of global $U(1)$ transformation $S\to \exp(i\vartheta )S$ ($\vartheta$ is an arbitrary phase) is softly broken by the $\mu_3$ term to the $\mathbb{Z}_3$ symmetry, i.e., the potential remains unchanged only under those 
transformations with a rotation angle $\vartheta=2n\pi/3$, where $n$ is an integer. 
Notice that the Hermiticity of the potential implies a symmetry under the transformation $S\to S^{\ast}$, which turns into a $\mathbb{Z}_2$ symmetry for the imaginary component $\chi$ of $S$ ($\chi \to - \chi$). This ensures the stability of $\chi$ and makes it a DM candidate. 
It is worth mentioning that the usual $\mathbb{Z}_2$ symmetry models do not prohibit a term proportional 
to $|H|^2S^2$, which could also softly break the $U(1)$ symmetry but at the same time spoil the cancellation mechanism~\cite{Gross2017PRL}.
Such a term is not allowed by the $\mathbb{Z}_3$ symmetry in our model, which shows an advantage over the $\mathbb{Z}_2$ symmetry models.
Furthermore, this tree-level potential boasts analytical solutions which may explicitly reveal some of properties of the model.

We note in passing that the breakdown of a discrete symmetry during EWPT in the early universe can potentially lead to the problematic 
EW-scale cosmic domain walls~\cite{Zeldovich1974ZETF} whose gravitational effects may 
result in unacceptable anisotropy in the cosmic microwave background (CMB) radiation~\cite{Kibble1976JPA}.
Depending on the stability and evolution of such domain walls, several mechanisms have been proposed to avoid these 
quandaries~\cite{Kibble1976JPA, Sikivie1982PRL, Vilenkin1985RP, Gelmini1989PRD, Preskill1991NPB, Abel1995NP, Panagiotakopoulos1999PLB}.
One of the mechanisms proposed in Refs.~\cite{Kibble1976JPA, Sikivie1982PRL, Vilenkin1985RP, Gelmini1989PRD} was to assume that the discrete symmetry was not exact but approximate.
One can introduce a so-called ``bias'' term to the scalar potential to explicitly break the discrete symmetry. This term lifts the degenerate vacua and induces a difference 
in the energy density between these vacua. This difference in the energy density has effects on the wall as a volume pressure and finally leads
to the decay of the wall when the pressure becomes comparable to the surface tension of the wall~\cite{Hiramatsu2014JCAP}.
In a recent paper~\cite{Zhou2020}, the authors took our model and applied the approximate symmetry mechanism to solve the domain wall problem.
It was shown that there were two peaks in the GW spectrum, one from the first-order EWPT and the other from the domain wall decay.

\subsection{Field-dependent masses}

The Higgs and singlet scalar can be expanded around their classical backgrounds as
\begin{equation}
H=\left(\begin{array}{c}{G^{+}} \\ {\frac{1}{\sqrt{2}}\left(h+i G^{0}\right)}\end{array}\right),~~S=s+i\chi,
\end{equation}
where $G^{\pm}$, $G^{0}$, and $\chi$ are the Goldstone bosons after spontaneous symmetry breaking.
At zero temperature, the VEVs for the two scalars are $v \equiv \left \langle H \right \rangle|_{T=0}$ and 
$w \equiv \left \langle S \right \rangle|_{T=0}$, respectively.
We immediately obtain the tree-level potential in terms of the fields $h$ and $s$
\begin{equation}
  V(h, s)=-\frac{1}{2} \mu_{h}^{2} h^{2}+\frac{1}{4} \lambda_{h} h^{4}+\frac{1}{4} \lambda_{m} h^{2} s^{2}+\frac{1}{2} 
  \mu_{s}^{2} s^{2}+\frac{1}{3} \mu_{3} s^{3}+\frac{1}{4} \lambda_{s} s^{4}.
\end{equation}
Then the field-dependent mass matrix of the scalar bosons is given by
\begin{align}
\begin{split}
  \mathcal{M}^2(h,s)=&
  \begin{pmatrix}
    \mathcal{M}_{hh}^2 & \mathcal{M}_{hs}^2\\ 
    \mathcal{M}_{sh}^2 & \mathcal{M}_{ss}^2
  \end{pmatrix},
~\mbox{with}~
\begin{cases}
  \mathcal{M}_{hh}^2
  =-\mu_{h}^{2}+3 \lambda_{h} h^{2}+\frac{1}{2}\lambda_{m}s^2,
  \\
  \mathcal{M}_{ss}^2
  =\mu_{s}^{2}+3 \lambda_{s} s^{2}+2 \mu_{3} s+\frac{1}{2} \lambda_{m} h^{2},
  \\
  \mathcal{M}_{hs}^2
  =\mathcal{M}_{sh}^2(h,s)=\lambda_{m}hs.
\end{cases}
\end{split}
  \label{eq:mass1}
\end{align} 
The tree-level potential can now be rewritten as 
\begin{eqnarray}
  \label{eq:poten:3}
  V(h,s)=\frac{1}{2}\Phi^{\dagger }\mathcal{M}^{2}(h,s)\Phi,
\end{eqnarray} 
where $\Phi^{\dagger }=(h,s)$. 
With an orthogonal rotation
\begin{align}
  \begin{pmatrix}
    \mathcal{H}\\\mathcal{S}
   \end{pmatrix}=
   \begin{pmatrix}
   \cos\theta  &-\sin\theta  \\ 
   \sin\theta  & \cos\theta 
   \end{pmatrix}\begin{pmatrix}
   h\\ s
   \end{pmatrix}
\end{align}
from the $(h,s)^T$ basis to the physical basis $(\mathcal{H},\mathcal{S})$, one finds the physical masses given by
\begin{align}
\begin{split}
  M_{\mathcal{H}}^2(h,s)&=
  \frac{1}{2}\left(\mathcal{M}_{hh}^{2}+\mathcal{M}_{ss}^{2}-\sqrt{\left(\mathcal{M}_{hh}^{2}-
  \mathcal{M}_{ss}^{2}\right)^{2}+4 \mathcal{M}_{hs}^{2} \mathcal{M}_{sh}^{2}}\right),
  \\
  M_{\mathcal{S}}^2(h,s)&=
  \frac{1}{2}\left(\mathcal{M}_{hh}^{2}+\mathcal{M}_{ss}^{2}+\sqrt{\left(\mathcal{M}_{hh}^{2}-
  \mathcal{M}_{ss}^{2}\right)^{2}+4 \mathcal{M}_{hs}^{2} \mathcal{M}_{sh}^{2}}\right).
\end{split}
\end{align} 
The field-dependent mass of pseudo-Goldstone boson $\chi$ is
\begin{eqnarray}
  \label{eq:dmmass}
M_{\chi}^2=\mu_s^2+\lambda_s s^2+\frac{1}{2}\lambda_m h^2-2\mu_{3} s.
\end{eqnarray} 
Other field-dependent masses of the SM particles are given in appendix~\ref{apd:fdms}.

\subsection{Stationary points}
\label{sec:sp}

We now search for the local minima of the tree-level scalar potential.  An interesting scenario is when the potential has two local minima: the EW symmetry broken one located at $(v,~w)$ and the EW symmetric one located at $(0,~w_0)$. 
The tadpole conditions of the potential are
\begin{eqnarray}
  \label{eq:11}
  \frac{\partial V}{\partial h}&=&0~~ \Rightarrow ~~\left\{h=0,~~{\rm or}~~ h^{2}=\frac{1}{2 \lambda_{h}}
  \left(2 \mu_{h}^{2}-\lambda_{m} s^{2}\right)\right\},\\
  \label{eq:22}
  \frac{\partial V}{\partial s}&=&0~~ \Rightarrow ~~\left\{s=0, ~~{\rm or}~~ h^{2}=-\frac{2}{\lambda_{m}}
  \left(\mu_{s}^{2}+\mu_{3} s+\lambda_{s} s^{2}\right)\right\}.
\end{eqnarray} 

Besides, vacuum stability demands the following bounds on parameters in the potential~\cite{Espinosa2012NPB, Kanemura2016NPB}
\begin{equation}
  \label{eq:vacstab}
  \lambda_{h}>0,~~ \lambda_{s}>0,~~ \lambda_{h} \lambda_{s}>\frac{1}{4} \lambda_{m}^{2}.
\end{equation}

\subsubsection{Stationary points along $h$-axis}

The stationary point along the $h$-axis is given by
\begin{equation}
  s=0,~~ h_{\pm}=\pm \sqrt{\frac{\mu_{h}^{2}}{\lambda_{h}}}.
\end{equation}
The condition for a physical vacuum, i.e., $\mu_{h}^{2}/\lambda_{h}>0$, can be easily satisfied since
$\mu_{h}^{2}>0$ holds for most of the parameter space and $\lambda_{h}>0$ is guaranteed by the vacuum stability.  However, a zero VEV for the singlet scalar will lead to a vanishing DM mass,
which is of no interest to us in this work. To avoid such stationary points from being local minima, we can 
demand $\partial^2 V/\partial s^2<0$ at these points, giving
\begin{equation}
  \label{eq:cond1}
  \mu_{s}^{2}+\frac{\lambda_{m} \mu_{h}^{2}}{2 \lambda_{h}}<0.
\end{equation}
As we will see below, this condition also ensures a stationary point at $(h\neq 0,~s\neq 0)$.

\subsubsection{Stationary points along $s$-axis}

The stationary points along the $s$-axis is given by
\begin{equation}
  \label{eq:symmv}
  h=0,~~ s_{\pm}=\frac{-\mu_{3} \pm \sqrt{\mu_{3}^{2}-4 \lambda_{s} \mu_{s}^{2}}}{2 \lambda_{s}}.
\end{equation}
The required physical condition is $\mu_{3}^{2}-4 \lambda_{s} \mu_{s}^{2}>0$. We find that for most of the parameter space having sufficiently
strong first-order EWPT, the condition $\mu_{s}^{2}<0$ always holds. 
As we will see below, for the stationary points sitting on the $s$-axis, the potential located at $s_+$ is always lower than the potential located at $s_-$ if $\mu_3<0$ (i.e., $V(0,~s_+)<V(0,~s_-)$).  On the other hand, $V(0,~s_+)>V(0,~s_-)$ if $\mu_3>0$.

\subsubsection{Stationary points at $(h\neq 0,~s\neq 0)$}

There are also solutions off the $h$-axis and $s$-axis, given by
\begin{equation}
  \label{eq:brokv}
  h=v,~~ s_{\pm}=\frac{-\mu_{3} \pm \sqrt{\mu_{3}^{2}-4\left(\lambda_{s}-\frac{\lambda_{m}^{2}}
  {4 \lambda_{h}}\right)\left(\mu_{s}^{2}+\frac{\lambda_{m} \mu_{h}^{2}}{2 \lambda_{h}}\right)}}
  {2\left(\lambda_{s}-\frac{\lambda_{m}^{2}}{4 \lambda_{h}}\right)}.
\end{equation}
The condition for these solutions to be physical is 
\begin{equation}
  \label{eq:cond3}
  \mu_{3}^{2}-4\left(\lambda_{s}-\frac{\lambda_{m}^{2}}{4 \lambda_{h}}\right)\left(\mu_{s}^{2}+
  \frac{\lambda_{m} \mu_{h}^{2}}{2 \lambda_{h}}\right)>0.
\end{equation}
With vacuum stability condition \eqref{eq:vacstab}, we have $\lambda_{s}-\frac{\lambda_{m}^{2}}{4 \lambda_{h}}>0$.
As mentioned above, we can further impose the condition \eqref{eq:cond1} on the potential if we demand that 
the stationary points along the $h$-axis be unstable.
Hence, the condition \eqref{eq:cond3} is satisfied for most of the parameter space of interest.

Using Eq.~\eqref{eq:22}, we obtain
\begin{eqnarray}
  F&\equiv& V(h,~s_+)-V(h,~s_-)
  =-\frac{1}{6}\mu_3(s_+^3-s_-^3)-\frac{1}{4}\lambda _s(s_+^4-s_-^4).
\end{eqnarray}
One can show that $F<0$ is always true under the assumption of $\mu_3<0$.  This is in fact equivalent to the condition
\begin{eqnarray}
  \label{eq:cond4}
  \lambda_s>\frac{2a(b^2-ac)}{3(b^2-2ac)},
~~\mbox{with}~
  a=\lambda_s-\frac{\lambda_m^2}{4\lambda_h},~b=\mu_3,~{\rm and}~c=\mu_s^2+\frac{\lambda_m\mu_h^2}{2\lambda_h}.
\end{eqnarray}
Therefore, we finally reach the conclusion that with the assumption $\mu_3<0$, the tree-level potential
located at $(h,s_+)$ is always lower than the one located at $(h,s_-)$.
Similarly, We can also prove that $F>0$ is always true provided $\mu_3>0$.
We summarize our conclusion as follows:
\begin{eqnarray}
\left\{\begin{matrix}
  F<0~~{\rm and} ~~ w\equiv s_+>0~~{\rm if}~~ \mu_3<0,\\
  F>0~~{\rm and} ~~ w\equiv s_-<0~~{\rm if}~~ \mu_3>0.
\end{matrix}\right.
\end{eqnarray}
One can easily verify that these conclusions are also established for the case of local minima along the $s$-axis.

From Eq.~\eqref{eq:dmmass}, we see that the pseudo-Goldstone DM mass is 
\begin{eqnarray}
  m_{\chi}^2=-3\mu_3w.
\label{eq:dmmass2}
\end{eqnarray}
To avoid a tachyonic mass for the DM candidate, the signs of $\mu_3$ and $w$ should be opposite.
Without loss of generality, the singlet scalar's VEV of is assumed to be positive, and $\mu_3$ should thus have a negative value.

\subsection{Parameters}

Using Eqs.~\eqref{eq:mass1}, \eqref{eq:11}, and \eqref{eq:22} we have
\begin{eqnarray}
\label{eq:muh2}
\mu _{h}^2&=&\frac{1}{2}\mathcal{M}_{hh}^2(v,w)+\frac{w}{2v}\mathcal{M}_{hs}^2(v,w)\\
\label{eq:lambdah}
\lambda _h&=&\frac{\mathcal{M}_{hh}^2(v,w)}{2v^2},\\
\label{eq:lambdam}
\lambda_m&=&\frac{\mathcal{M}_{hs}^2(v,w)}{vw},\\
\label{eq:mus2}
\mu_s^2&=&-\frac{1}{2}\mathcal{M}_{ss}^2(v,w)+\frac{1}{6}m_{\chi}^2-\frac{v}{2w}\mathcal{M}_{hs}^2(v,w),\\
\label{eq:mu3}
\mu_3&=&-\frac{m_{\chi}^2}{3w},\\
\label{eq:lambdas}
\lambda_s&=&\frac{1}{2w^2}\left (\mathcal{M}_{ss}^2(v,w)+\frac{1}{3}m_{\chi}^2  \right ).
\end{eqnarray}
The reason for employing Eqs.~\eqref{eq:11} and \eqref{eq:22} is to ensure the existence of a local minimum at $(v,~w)$ for any choice of parameters.
The above parameters can be related to three physical parameters, the masses of two Higgs bosons $m_{\mathcal{H}}$ and $m_{\mathcal{S}}$ and the mixing angle $\theta$, with the relations
\begin{align}
\begin{split}
  \mathcal{M}_{hh}^2(v,w)&= \cos^2\theta m_{\mathcal{H}}^2+\sin^2\theta m_{\mathcal{S}}^2,
  \\
  \mathcal{M}_{ss}^2(v,w)&= \sin^2\theta m_{\mathcal{H}}^2+\cos^2\theta m_{\mathcal{S}}^2,
  \\
  \mathcal{M}_{hs}^2(v,w)&= \cos\theta \sin\theta (m_{\mathcal{S}}^2-m_{\mathcal{H}}^2),
\end{split}
\end{align}
where $m_{\mathcal{H}}=125$~GeV and $v=246$~GeV.
Thus, we take
$\left \{ w,~m_{\mathcal{S}},~m_{\chi},~\theta \right \}$ as the input parameters of the model.

\section{Effective potential}
\label{sec:ewpt}

At the one-loop level, the total effective potential is given by
\begin{equation}
  V_{\rm eff}(h, s, T)=V(h, s)+V_{\rm CW}(h, s)+V_{T}(h, s, T)+V_{\rm CT}(h, s),
\end{equation}
where the tree-level potential $V(h, s)$ has been given above, and the other components are discussed below.


At zero temperature, the one-loop corrections to the potential is given 
using the $\overline{\rm MS}$ renormalization scheme~\cite{Coleman1973PRD} as
\begin{equation}
  V_{\rm CW}(h, s)=\frac{1}{64 \pi^{2}} \sum_{i} N_{i} M_{i}^{4}(h, s)\left[\log \frac{M_{i}^{2}(h, s)}{\mu^{2}}-C_{i}\right],
\end{equation}
where the subscript $i=\left\{\mathcal{H}, G, \mathcal{S}, \chi, Z_{\rm T}, Z_{\rm L}, W_{\rm T}, W_{\rm L}, t,b\right\}$ denote respectively the SM-like Higgs boson $\mathcal{H}$, 
SM Nambu-Goldstone bosons, heavy scalar $\mathcal{S}$, pseudo-Goldstone DM $\chi$, transverse and longitudinal components of SM gauge bosons, and
top and bottom quarks, and $N_{i}=\left \{ 1,3,1,1,2,1,4,2,-12,-12 \right \}$.
The constant $C_i=1/2$ for gauge boson transverse modes and $3/2$ for all the other particles.  The renormalization scale $\mu$ is set to be $v$ in this work. 
Appendix~\ref{apd:RGE} gives the renormalization group equations (RGEs) for the parameters in the scalar potential, 
using which we calculate the renormalization group-improved (RGI) potential and show the dependence of the critical 
temperature and the corresponding VEV on the renormalization scale.

Due to plasma damping, the validity of the perturbative expansion of the effective potential breaks down
at high temperatures.  A remedy to this problem is to resum the daisy diagrams to all orders,
which results in an additional contribution to the bosonic masses~\cite{Carrington1992PRD}.
Thus, we replace the field-dependent bosonic masses at finite temperatures by
\begin{equation}
M_i^2(h,s)\to M_i^2(h,s,T)=M_i^2(h,s)+\Pi_i(T),
\end{equation}
where the thermal corrections $\Pi_i(T)$ are given in appendix~\ref{apd:fdms}.
The one-loop potential becomes gauge-dependent when thermal corrections of bosons' masses are introduced~\cite{Jackiw1974PRD, Patel2011JHEP}, 
leading to gauge-dependent critical temperature and GW spectrum produced from phase transition~\cite{Wainwright2011PRD, Chiang2017PLB}. 
To focus on our topic, in this work we take the Landau gauge with a vanishing gauge-fixing parameter ($\xi=0$) for the effective potential.
In section~\ref{sec:gaugedepend} we will scrutinize the gauge dependence issue. There we will show that our main conclusions 
made in the following sections with the $\xi=0$ effective potential are not changed when compared with the analyses made by using a gauge-independent effective potential proposed in Ref.~\cite{Katz2015PRD}.


At the one-loop level, the finite-temperature contributions to the effective potential are given by~\cite{Dolan1974PRD, Kanemura2016NPB}
\begin{align}
\begin{split}
  V_{\mathrm{T}}(h,s,T)=&
  \frac{T^{4}}{2 \pi^{2}} \sum_{i} N_{i} J_{B,F}\left(M_{i}^{2}(h,s,T) / T^{2}\right),
  \\
  \mbox{where}~~&
  J_{B,F}\left(z^{2}\right)=\int_{0}^{\infty} d x x^{2} \ln \left(1\mp e^{-\sqrt{x^{2}+z^{2}}}\right),
  \label{eq:jbf}
\end{split}
\end{align}
with the $-$ sign for bosons and $+$ for fermions.


To maintain the main properties of the tree-level potential derived above, we add the following counter-terms to the potential at zero temperature 
\begin{equation}
  \label{eq:cterm}
  V_{\rm CT}(h, s)=-\frac{1}{2} \delta \mu_{h}^{2} h^{2}+\frac{1}{4} \delta \lambda_{h} h^{4}+\frac{1}{4} 
  \delta \lambda_{m} h^{2} s^{2}+\frac{1}{2} \delta \mu_{s}^{2} s^{2}+\frac{1}{3} \delta u_{3} s^{3}+\frac{1}{4} \delta \lambda_{s} s^{4}.
\end{equation}
The coefficients of the counter-term potential are given in appendix~\ref{apd:ccp}.

\begin{figure}
  \centering
  \includegraphics[width=75mm,angle=0]{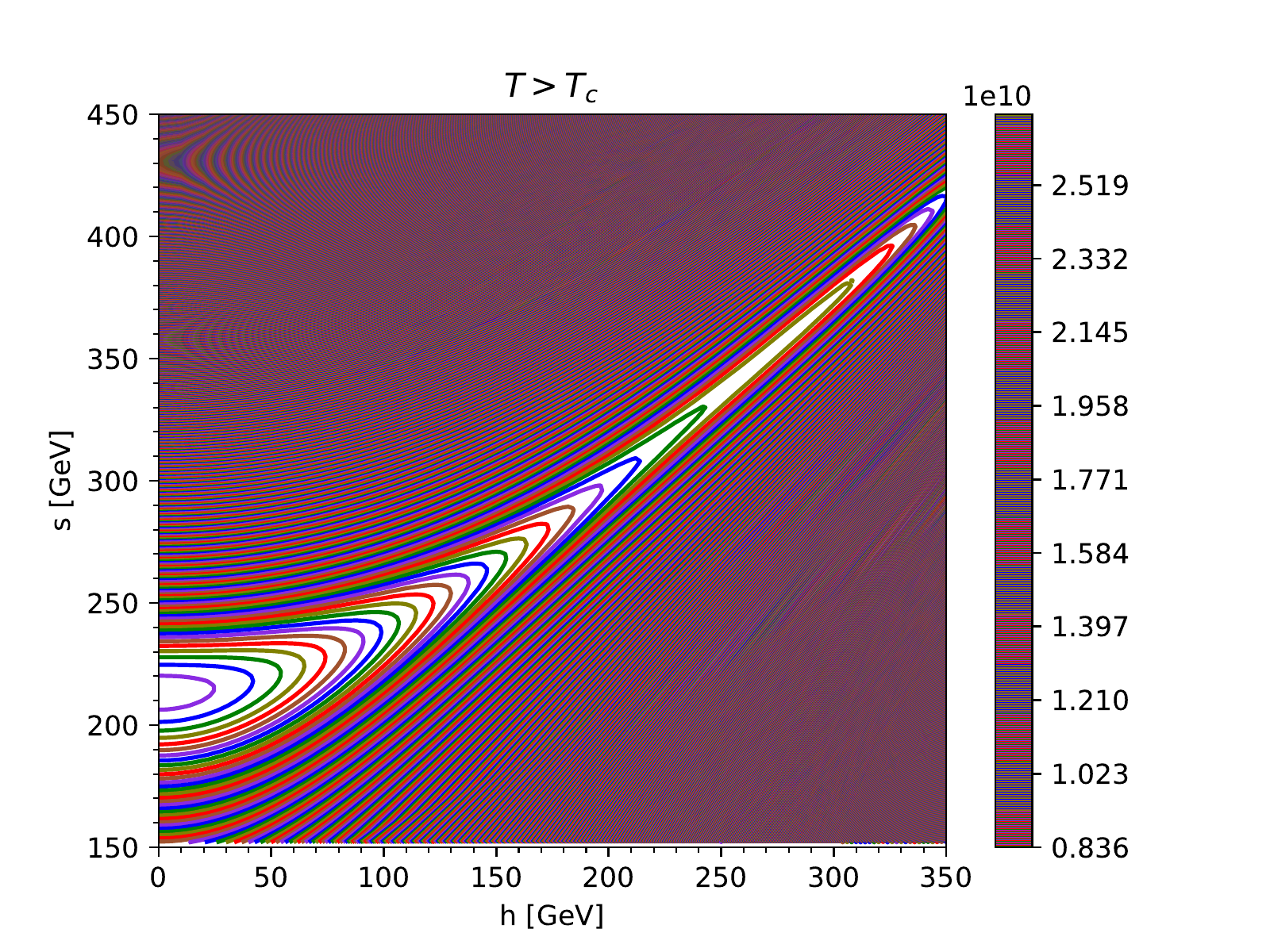}\\
  \includegraphics[width=75mm,angle=0]{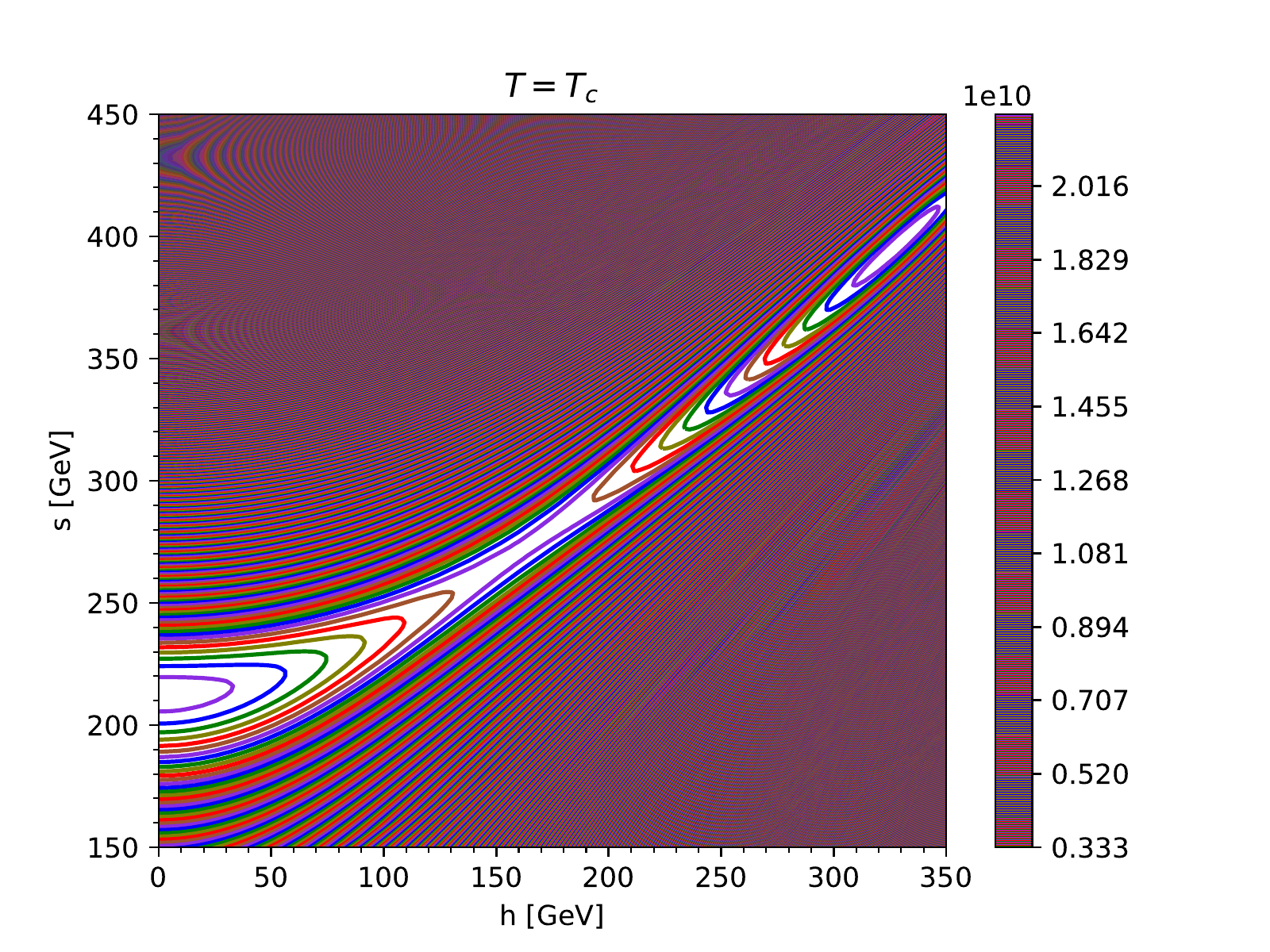}
  \includegraphics[width=75mm,angle=0]{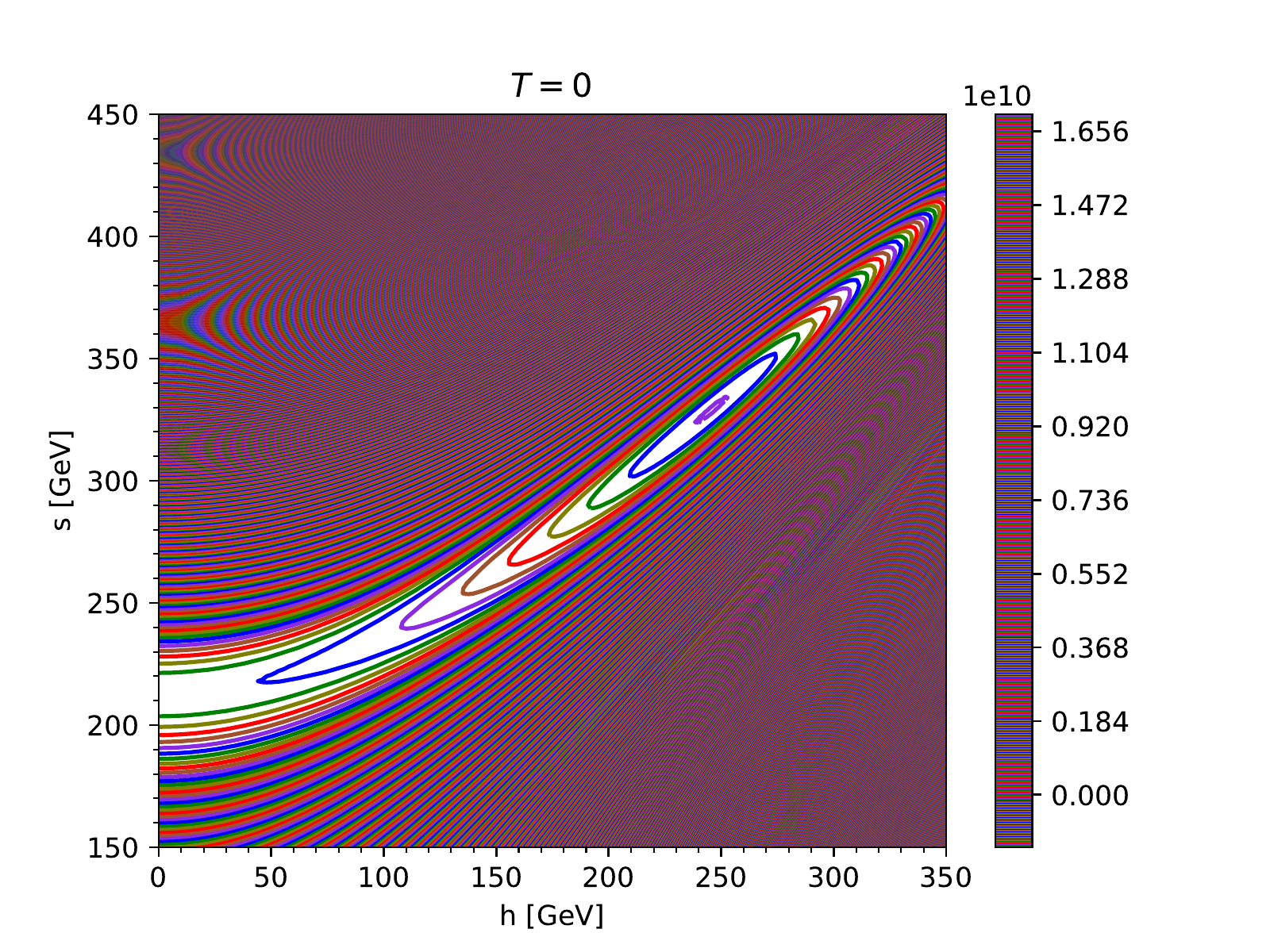}
  \caption{Contours of total effective potential in the $h-s$ plane, with parameters $w=329.2$ GeV, $m_{\mathcal{S}}=702.0$ GeV, $m_{\chi}=137.1$ GeV, 
  and $\theta=-0.64$. From upper plot to lower right plot, the potential is evaluated at temperature $T>T_c$, $T=T_c$, and $T=0$.
  }
  \label{fig:potential1}
\end{figure}

\section{Parameter space for electroweak phase transition}
\label{sec:ss}

In this section, we scan the parameter space for viable sample points for a sufficiently strong first-order phase transition.  We also show the distributions of the model parameters and physical parameters based upon our scan results.

\subsection{Two-step phase transition}
\label{sec:tspt}

A strong first-order EWPT could occur if there is a sufficiently high and wide potential barrier separating the two degenerate vacua of the thermal effective potential at critical temperature.
Introducing an extra bosonic degree of freedom could enhance the barrier and thus make the EWPT stronger.
In our model, there are two contributions to the barrier in the effective potential~\cite{Li2014JHEP}: one is the tree-level barrier coming from the cubic term of the tree-level potential; the other one arises from the bosonic thermal corrections to the potential at the one-loop level.  To see the latter, one can expand the integrations \eqref{eq:jbf} in the high temperature limit, i.e., $z\equiv M_i^2(h,s)/T^2\ll 1$, and find that there is a term proportional to $z^{3/2}$ that leads to terms cubic in both $h$ and $s$.
Most importantly, the barrier term cubic in $h$ is proportional to the Higgs portal coupling $\lambda_m$, 
and thus to the mixing angle $\theta$ (see Eq.~\eqref{eq:lambdam})~\cite{Profumo2007JHEP,Kozaczuk2019,Carena2019,Michael2019}.
Therefore a large value of mixing angle could strengthen the EWPT by boosting the potential barrier. 
The potential barrier is shallow in the SM and its EWPT is confirmed to be a crossover~\cite{Onofrio2014PRL}.

In this work, we focus on the so-called two-step phase transition as shown in Fig.~\ref{fig:potential1}. At very high temperatures, both scalar fields have no VEV.  When the temperature drops to a certain value $T_s$, (above the critical temperature $T_c$, at which two degenerate vacua exist concurrently) 
the scalar potential along the $s$ direction develops a global minimum at $(0,~w_0(T_s))$.  We call it the symmetric phase, as shown in the upper plot of Fig.~\ref{fig:potential1}.  As the temperature further lowers to the critical temperature, another local minima located at $(v(T_c),~w(T_c))$, designating the symmetry-broken phase, appears and becomes degenerate with the symmetric phase $(0,~w_0(T_c))$, as shown in the lower left plot of Fig.~\ref{fig:potential1}.  Meanwhile, a tunneling path, along which is a lowest barrier between the two degenerate minima, opens up.
The decrease of potential at the broken phase is much faster than that at the symmetric phase as the temperature approaches zero.
As a result, the broken phase moves to $(v,~w)$ and becomes a global minimum, as shown by the lower right plot of Fig.~\ref{fig:potential1}.

\subsection{Searching scheme}
\label{sec:seachsch}

The strength of the EWPT is measured according to the order parameter $v_c/T_c$, 
where $v_c\equiv v(T_c)$ is the VEV of SM-like Higgs boson at the critical temperature.  For a successful baryogenesis, 
the first-order EWPT should be strong enough so that the sphaleron process in the broken phase is sufficiently suppressed to avoid baryon asymmetry washout \cite{Cline2006}.
This gives the conventional criterion for a sufficiently strong EWPT:
\begin{equation}
  \label{eq:criterion}
  \frac{v_c}{T_c}\gtrsim \zeta,
\end{equation}
where $\zeta $ is a criterion value usually around unity.
We note that a theoretical ambiguity of this criterion may arise due to the gauge dependence and uncertainty from higher-order calculations, as studied in Ref.~\cite{Patel2011JHEP}.  Nevertheless, we will still use it as a useful guidance to the relevant regions in the parameter space.
For the left-hand side (LHS) of Eq.~\eqref{eq:criterion}, the ratio $v_c/T_c$ obtained by using our procedure provided below is gauge-dependent.
As mentioned above, the full one-loop effective potential is gauge-dependent due to the thermal corrections. 
With the leading-order high-temperature expansion of the effective potential and the implementation of Nielsen's identity, 
Ref.~\cite{Patel2011JHEP} provided a gauge-independent determination of $v_c$, $T_c$ and thus $v_c/T_c$.
To estimate the impacts of gauge dependence in criterion \eqref{eq:criterion}, we will adopt the gauge-independent effective potential, which is obtained by truncating the one-loop effective potential at second order in the EW gauge couplings, to calculate
$v_c/T_c$ in section \ref{sec:gaugedepend}. We will show that in comparison with the full one-loop potential, using the gauge-independent potential generally reduces the critical temperature
while increasing $v_c/T_c$. Thus, the samples satisfying criterion \eqref{eq:criterion} in the Landau gauge can also satisfy the gauge-independent version of criterion \eqref{eq:criterion}.
For the RHS of criterion \eqref{eq:criterion}, there exist several sources of theoretical uncertainties in obtaining this quantity (a summary of these uncertainties can be found in Ref.~\cite{Patel2011JHEP} and references therein).
One of the uncertainties is the lower bound on the ``washout factor,'' $S>e^{-X}$, where $S$ is the ratio of the baryon densities after and before the phase transition. Taking $X\simeq 10$ could lead to the conventional criterion value, $\zeta\simeq 1.0$. However, for a certain scenario, uncertainties in the value of $X$ could arise from the efficiency of the CP violation mechanism and 
the duration of phase transition. A more realistic treatment of the criterion is to replace unity by a range  determined by an appropriate choice of $X$ and the other theoretical inputs~\cite{Patel2011JHEP}. 
In this work, we will simply take the conventionally used criterion value $\zeta\simeq 1.0$ for the determination of a sufficiently strong phase transition.


Aiming at the search of parameter space giving a sufficiently strong EWPT, we make a random scan of the parameters in the following ranges: 
\begin{align}
\begin{split}
\label{eq:ranges}
&100{~\rm GeV}\leq  w \leq  2000{~\rm GeV},~~150{~\rm GeV}\leq  m_{\mathcal{S}} \leq  2000{~\rm GeV},
\\
&10{~\rm GeV}\leq  m_{\chi} \leq  1500{~\rm GeV},~~-\frac{\pi}{4}\leq \theta \leq \frac{\pi}{4}.
\end{split}
\end{align}
In addition to the above enforced restrictions, the parameters are also subject to the constraints discussed in section~\ref{sec:sp}, 
including vacuum stability \eqref{eq:vacstab} and the conditions to ensure the existence of 
stationary points at the symmetric phase $(0,~w_0)$ and the broken phase $(v,~w)$.  Further constraints come from the requirement of perturbative unitarity~\cite{Pruna2013PRD, Kanemura2016NPB}
\begin{equation}
  \label{eq:pertu}
  \lambda_{h}<4 \pi,~ \lambda_{s}<4 \pi,~|\lambda_{m} |<8 \pi,~
  3 \lambda_{h}+2 \lambda_{s}+\sqrt{\left(3 \lambda_{h}-2 \lambda_{s}\right)^{2}+2 \lambda_{m}^{2}}<8 \pi.
\end{equation}
We also require that $V(v,~w)<V(0,~w_0)$ and $V(v,~w)<V(0,~0)$ at zero temperature, so that the broken phase is a global minimum.

To search for the critical temperature where two degenerate vacua coexist, 
we start from an initial temperature $T$ between a minimum value of temperature $T_{\rm min}$ and a maximum value of temperature $T_{\rm max}$, we then search between these two temperatures for the local minima of the potential around the positions $(0,~w_0)$ and $(v,~w)$, which can be determined using the 
analytical formulas \eqref{eq:symmv} and \eqref{eq:brokv} for given parameters.
If the local minimum at the symmetric phase $(0,~w_0(T))$ is found to be larger (smaller) than the one at the broken phase $(v(T),~w(T))$, the temperature is 
increased (decreased) in the next trial.  We conclude that no electroweak phase transition for given 
parameters occurs if the following two cases are met:
\begin{itemize}
  \item Case 1. The local minimum at the symmetric phase $(0,~w_0(T_{\rm max}))$ is larger than 
  the one at the broken phase $(v(T_{\rm max}),~w(T_{\rm max}))$.
  \item Case 2. The local minimum at the symmetric phase $(0,~w_0(T_{\rm min}))$ is less than 
  the one at the broken phase $(v(T_{\rm min}),~w(T_{\rm min}))$.
\end{itemize}
As obvious, the lower and upper temperature limits are critical for our parameters searches.
In our preliminary scan of the parameter space, we find that no points with $v_{c}/T_{c}\gtrsim 1$ can be found at a temperature higher than about 350~GeV. 
We thus restrict the scan range of temperature to $(10-350)~{\rm GeV}$.

\subsection{Parameter distributions}
\label{sec:paramsdist}

We generate one million random floats uniformly for each of the input parameters, among which about 
1.8\% are found to be able to trigger a sufficiently strong phase transition while fulfilling the other basic criteria mentioned above.
We show the distributions of various physical parameters in Fig.~\ref{fig:dist1}, and summarize our observations from the figures as follows:

\begin{figure}
  \centering
  \includegraphics[width=75mm,angle=0]{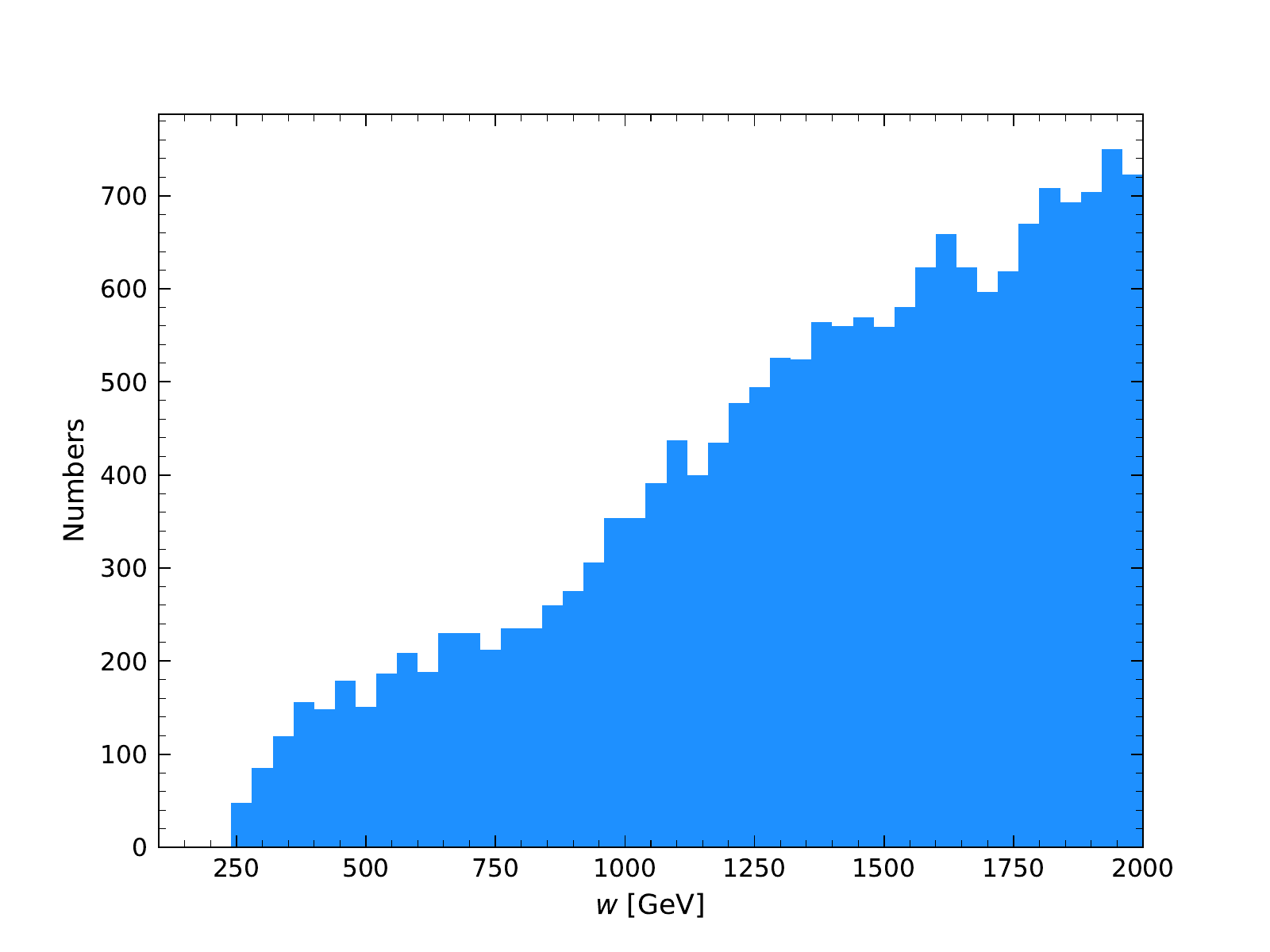}
  \includegraphics[width=75mm,angle=0]{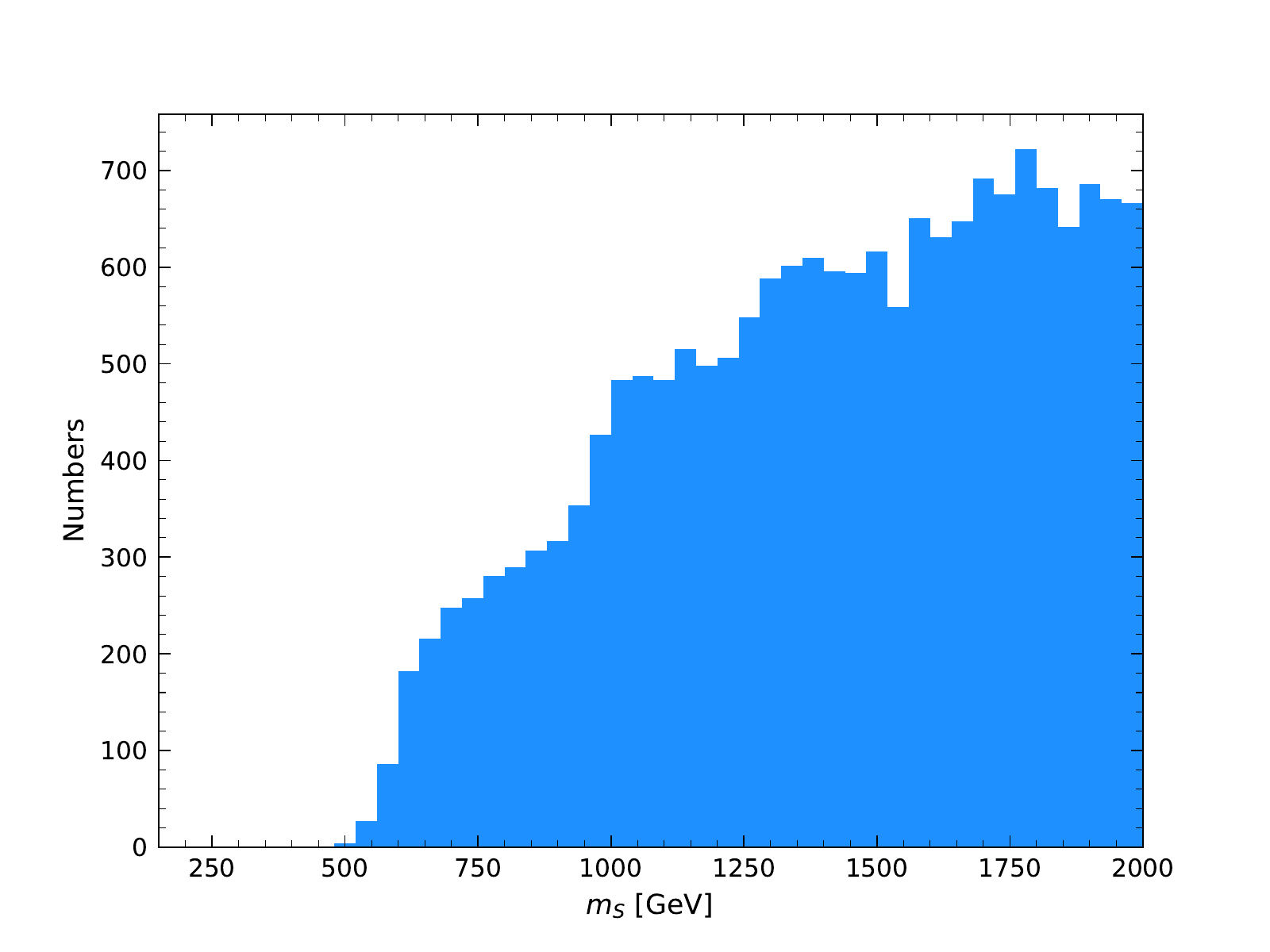}\\
  \includegraphics[width=75mm,angle=0]{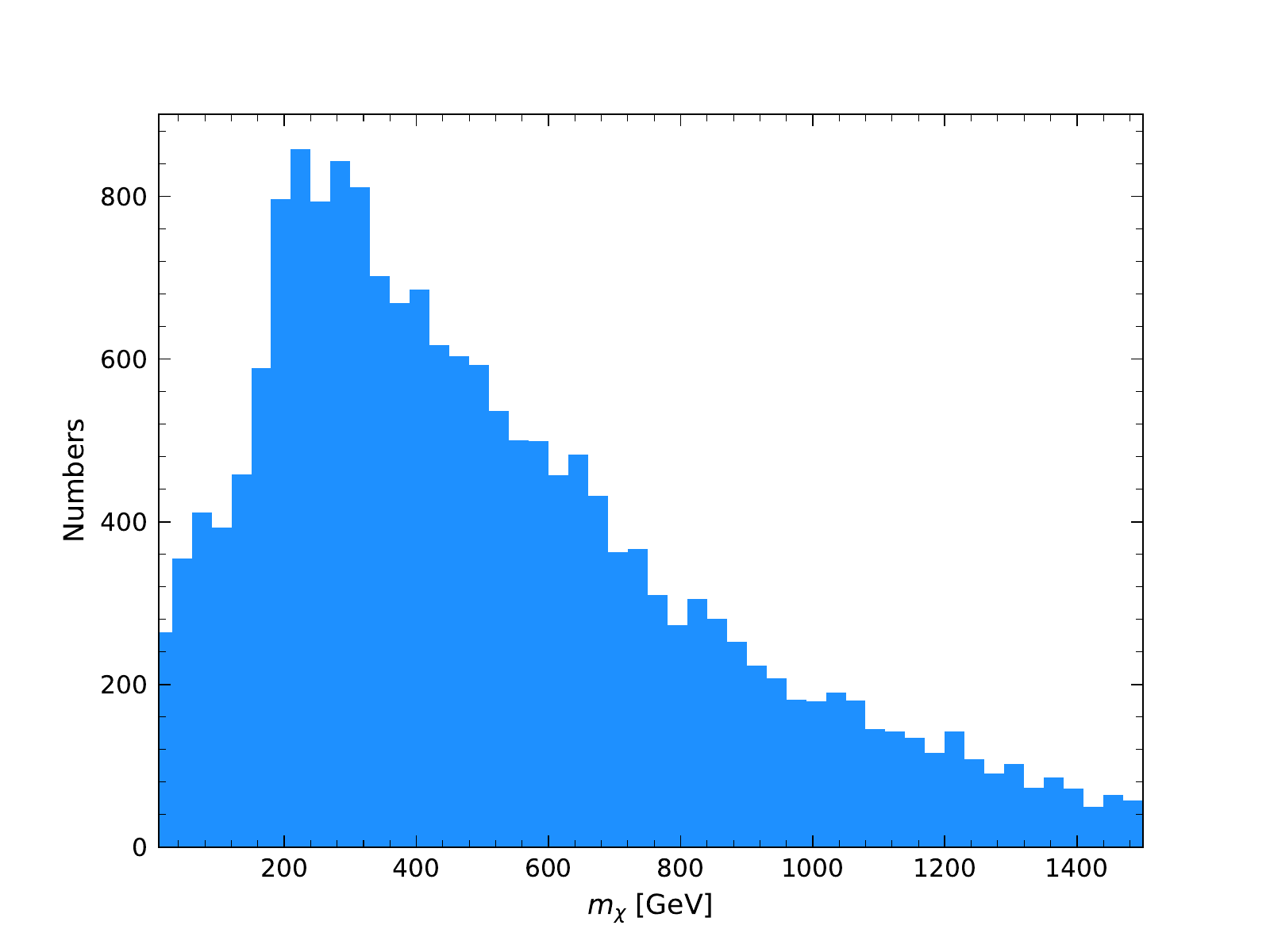}
  \includegraphics[width=75mm,angle=0]{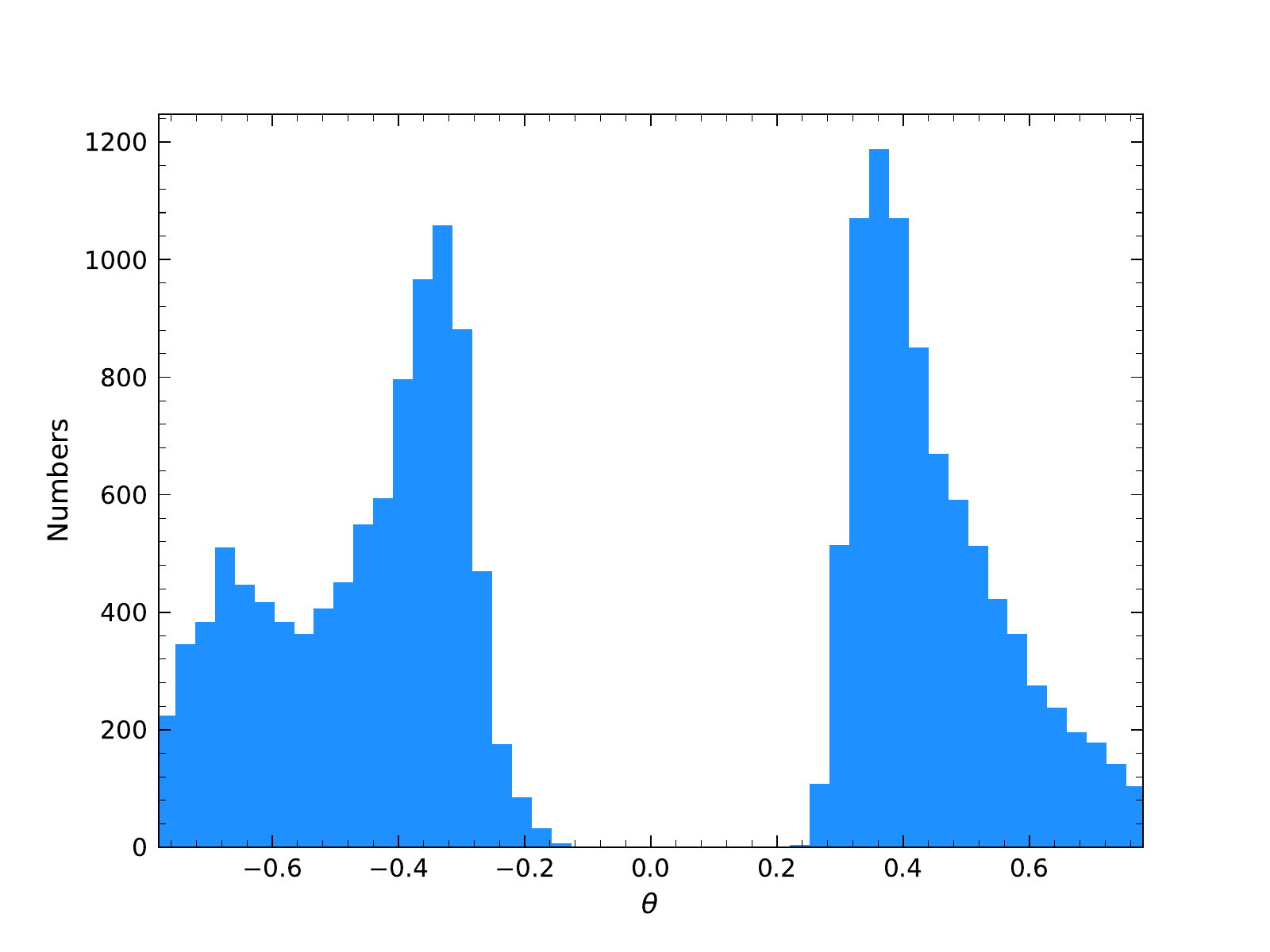}
  \caption{Distributions of parameters that can generate a sufficiently strong first-order EWPT.  For each fixed value of the parameter on the horizontal axis, all the other parameters are scanned in their respective full ranges.}
  \label{fig:dist1}
\end{figure}

\begin{figure}
  \centering
  \includegraphics[width=75mm,angle=0]{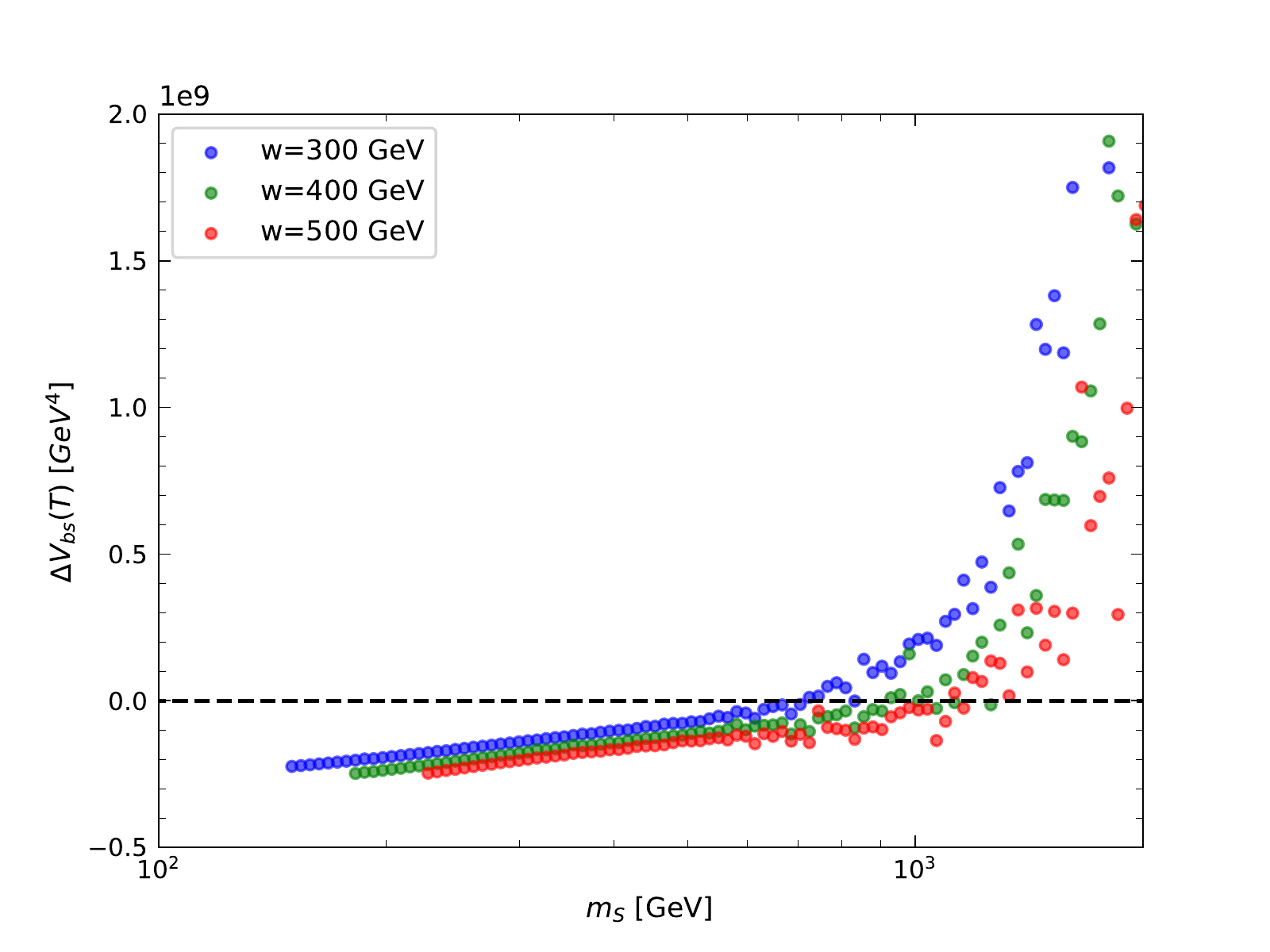}
  \includegraphics[width=75mm,angle=0]{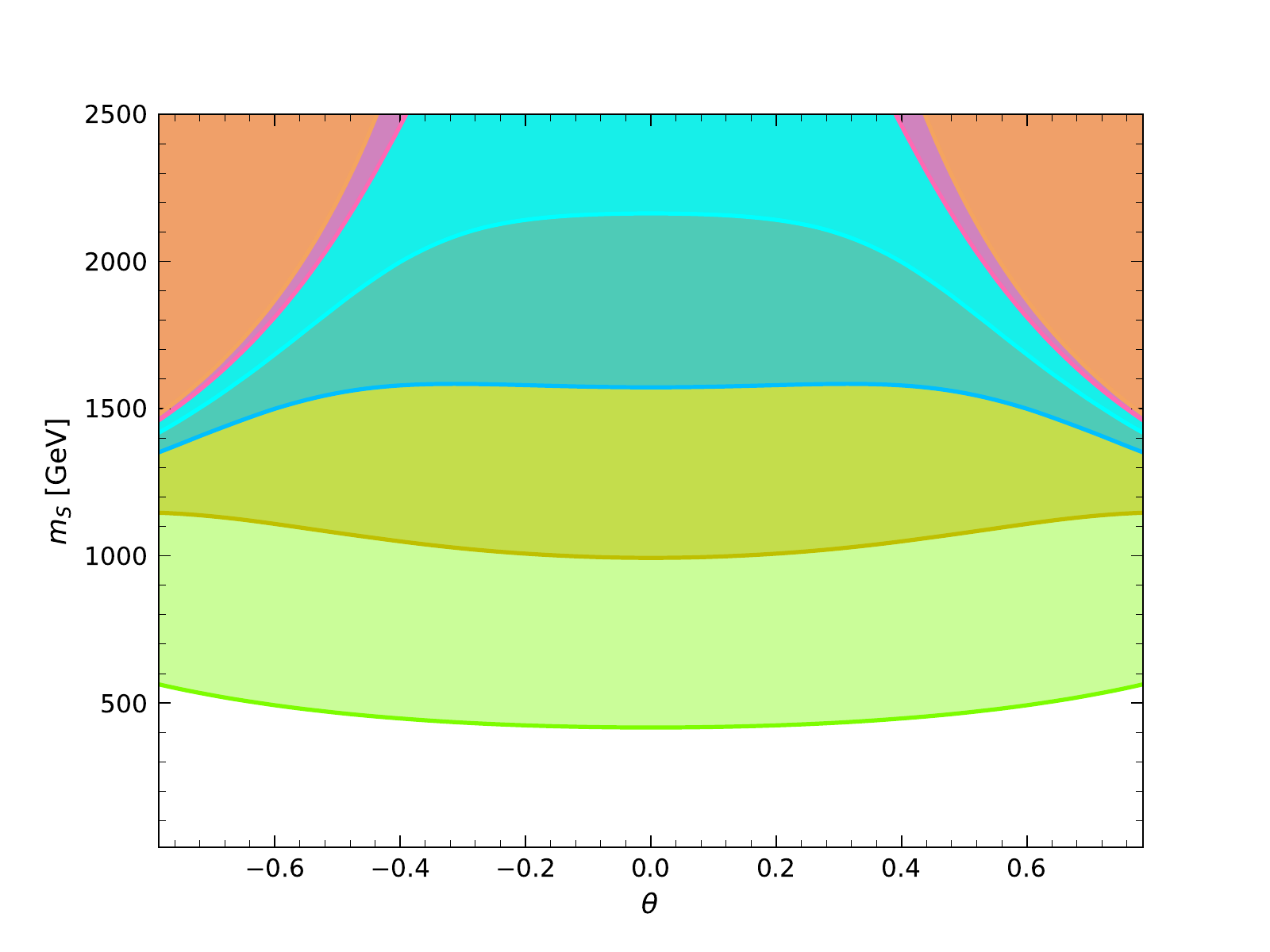}\\
  \includegraphics[width=75mm,angle=0]{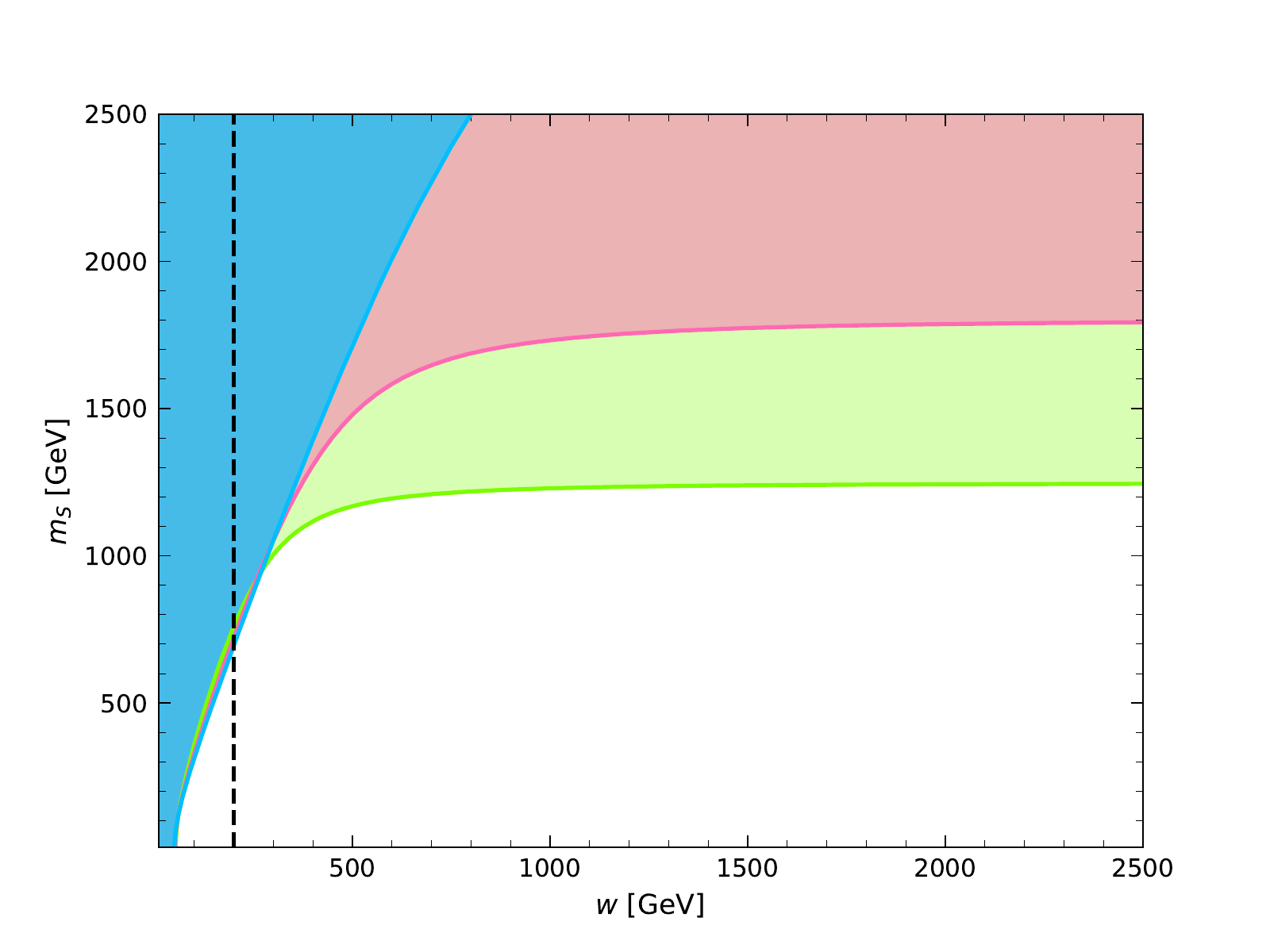}
  \includegraphics[width=75mm,angle=0]{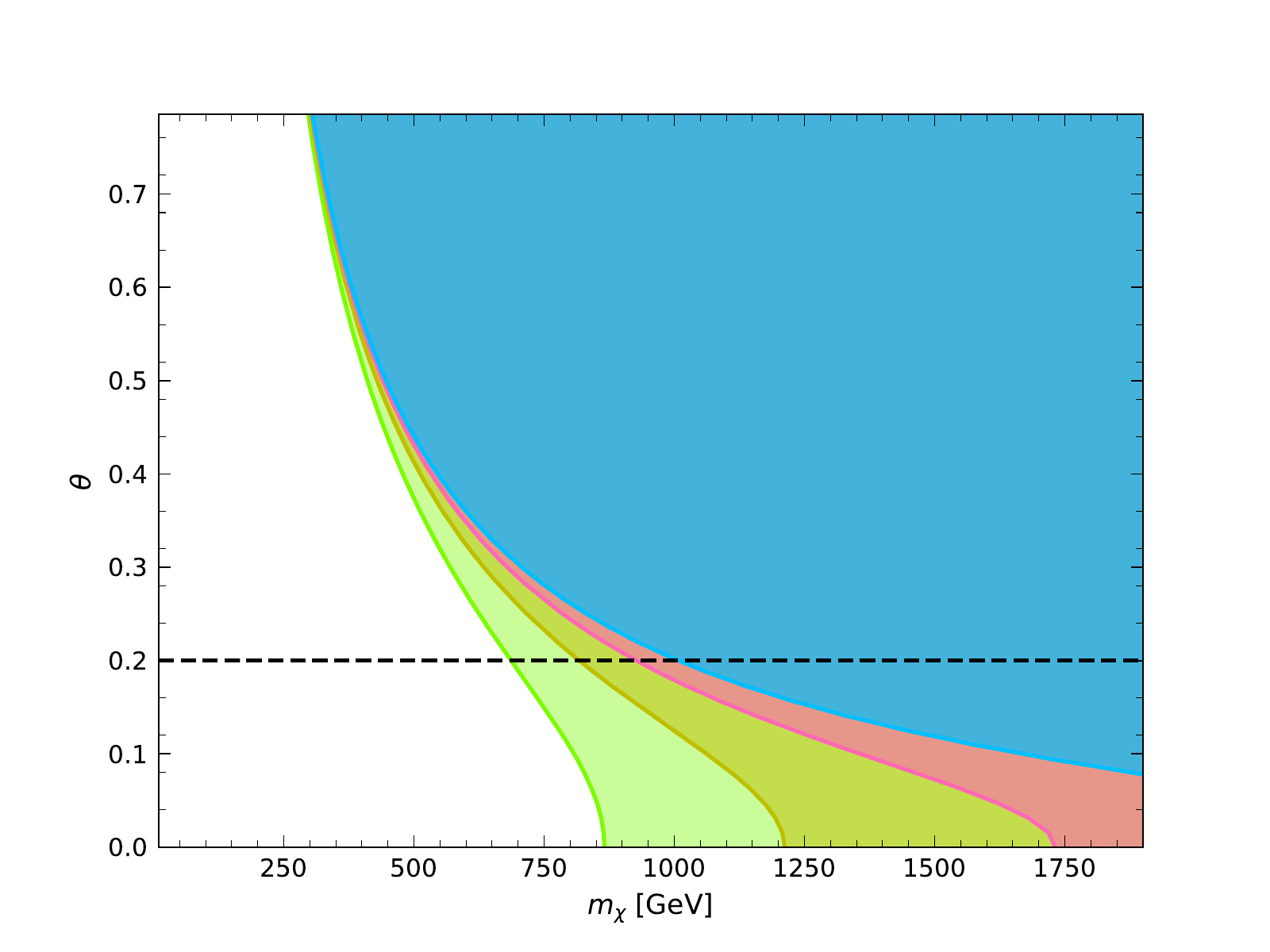}
  \caption{Upper left plot: $\Delta V_{bs}(T)$ as a function of $m_{\mathcal{S}}$, with $\theta=0.4$, $m_{\chi}=300$~GeV, and $T=300$~GeV. The blue, green, and red points represent respectively the results with $w=300$~GeV, 400~GeV, and 500~GeV.
   Upper right plot: Unitarity bounds in the $\theta$-$m_{\mathcal{S}}$ plane, with $m_{\chi}=300$~GeV. Colored regions in this figure are excluded by various constraints. The values of $w$ in this plot are 100~GeV (green), 200~GeV (yellow), 300~GeV (deep blue), 400~GeV (light blue), 600~GeV (pink), and 1000~GeV (sandy-brown) respectively.
   Lower left plot: Unitarity bounds in the $w$-$m_{\mathcal{S}}$ plane, with $\theta=0.2$ (blue), 0.4 (pink), and 0.6 (green), respectively.
   Lower right plot: Condition~\eqref{eq:cond1} in the $m_{\chi}$-$\theta$ plane, with $w=1000$~GeV.  The green, yellow, pink, and blue regions represent the bounds with $m_{\mathcal{S}}=$ 500~GeV, 700~GeV, 1000~GeV, and 1500~GeV, respectively.
   }
  \label{fig:cst}
\end{figure}

\begin{figure}
  \centering
  \includegraphics[width=75mm,angle=0]{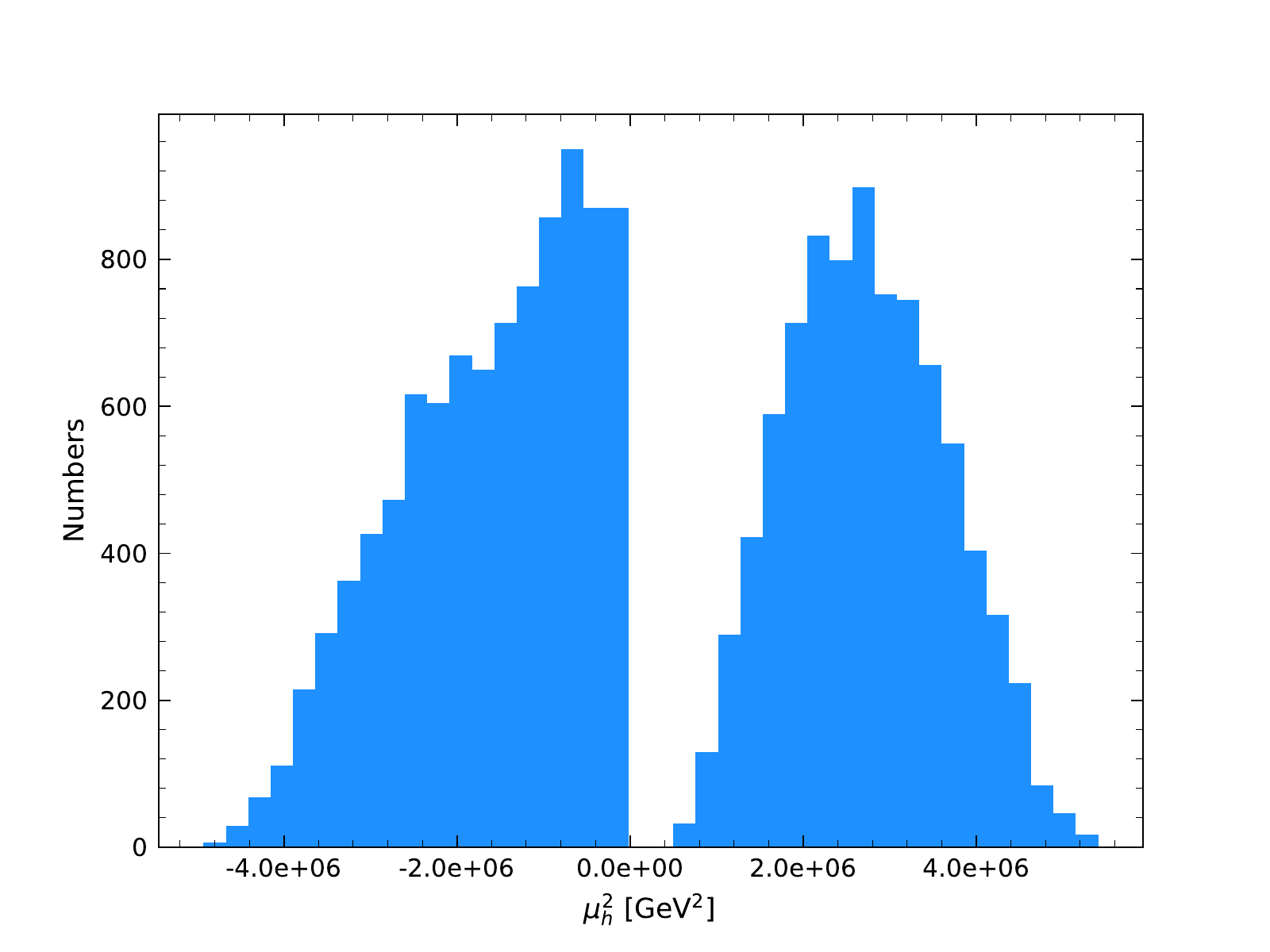}
  \includegraphics[width=75mm,angle=0]{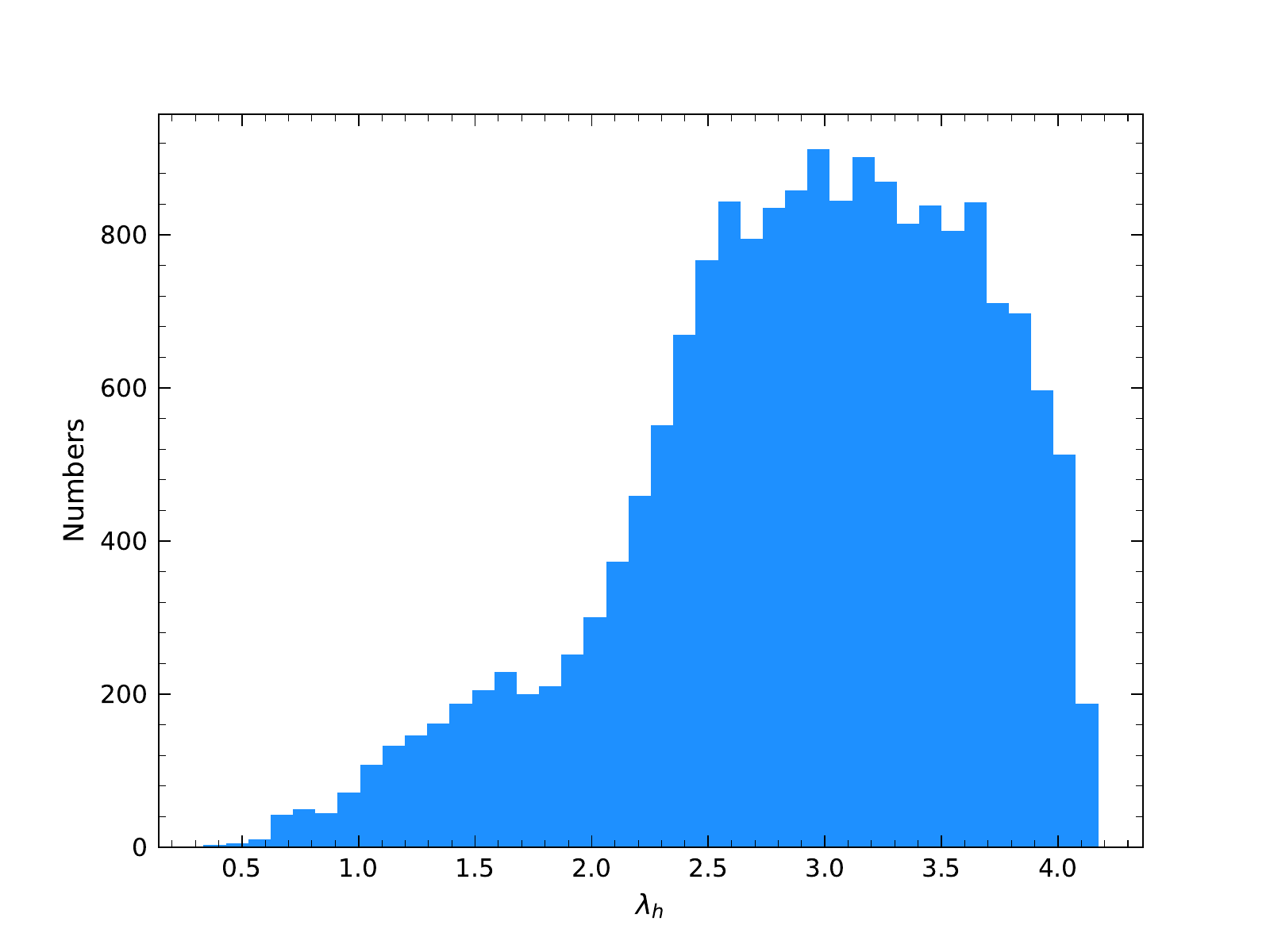}\\
  \includegraphics[width=75mm,angle=0]{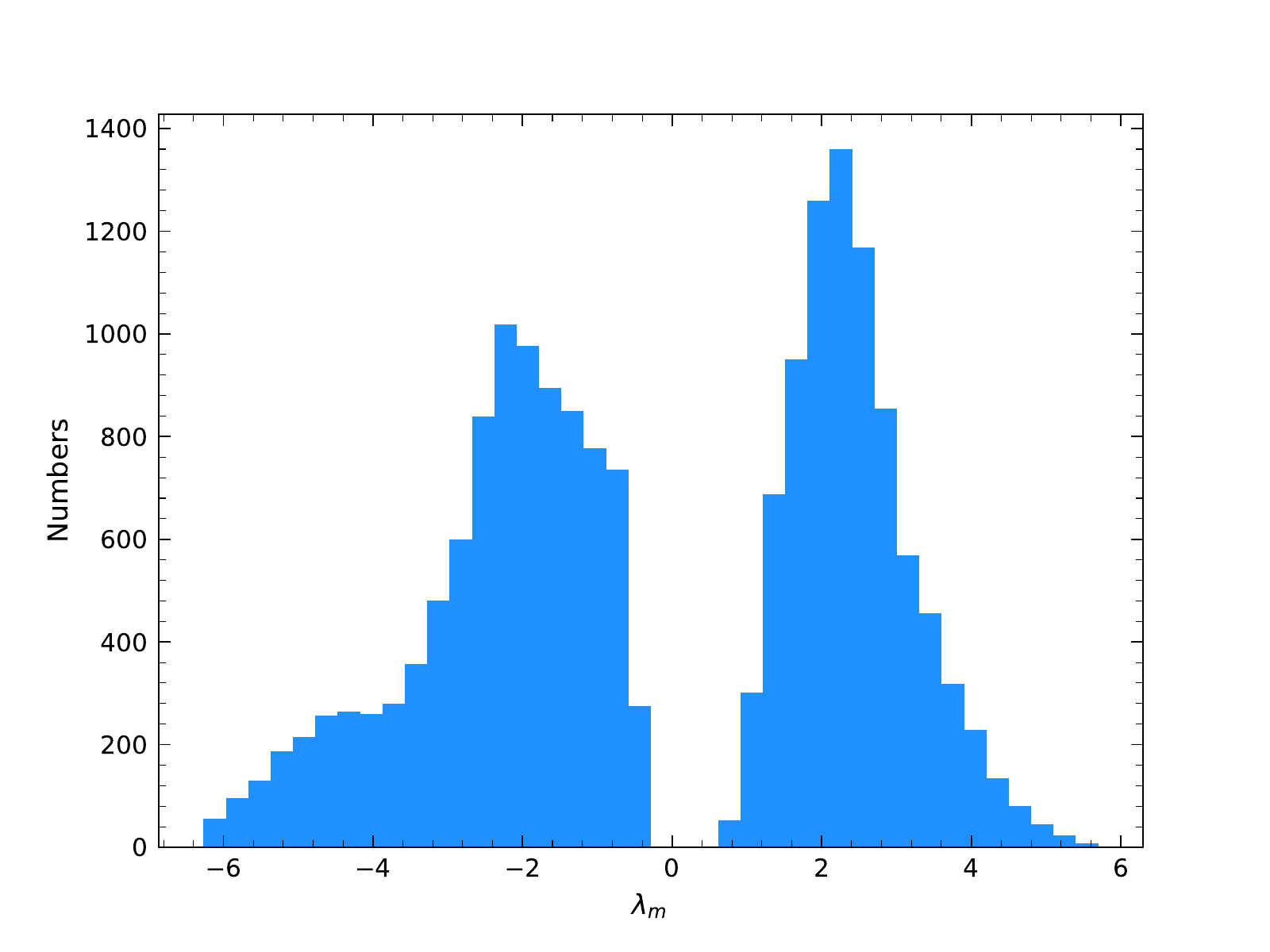}
  \includegraphics[width=75mm,angle=0]{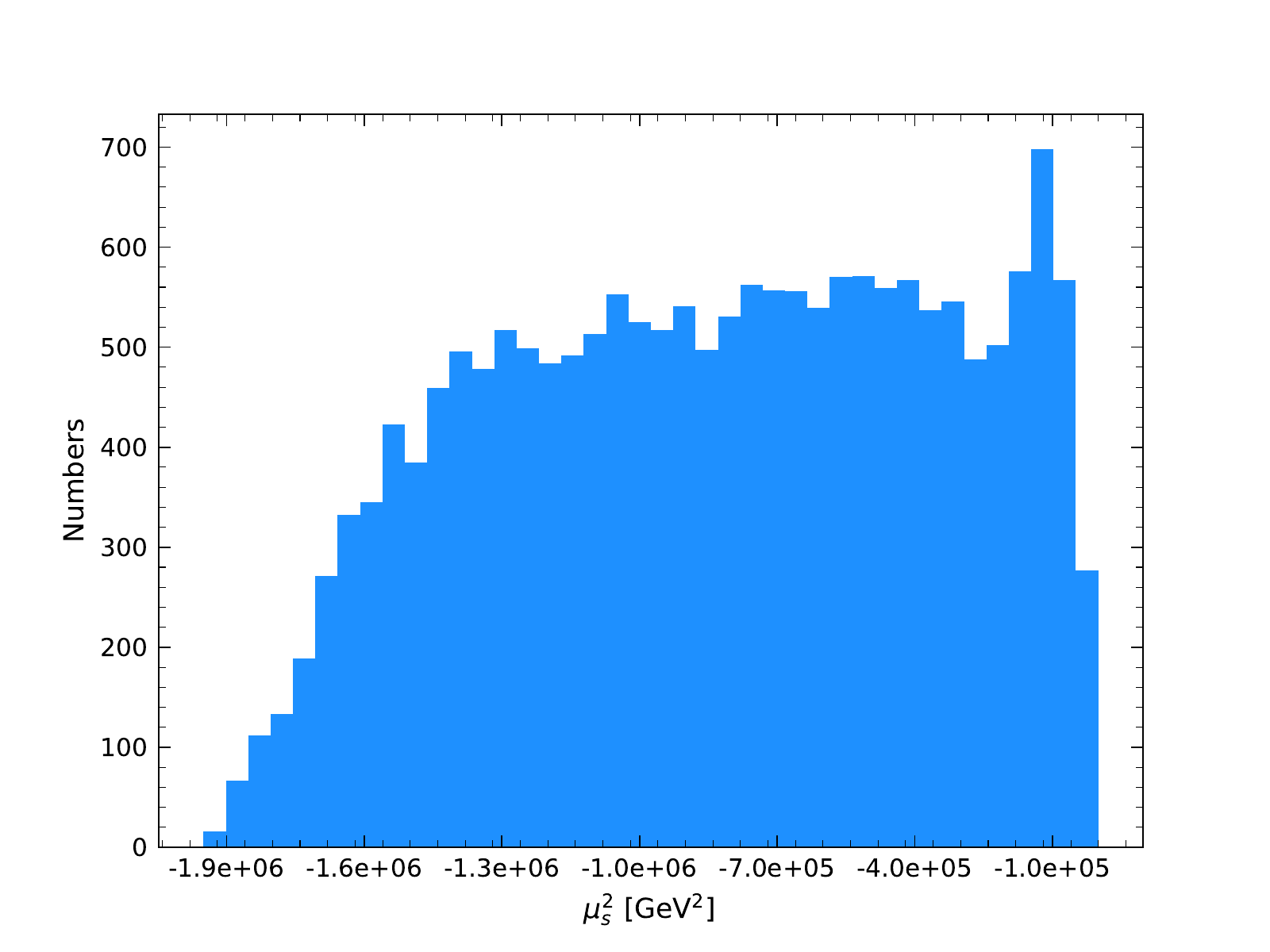}\\
  \includegraphics[width=75mm,angle=0]{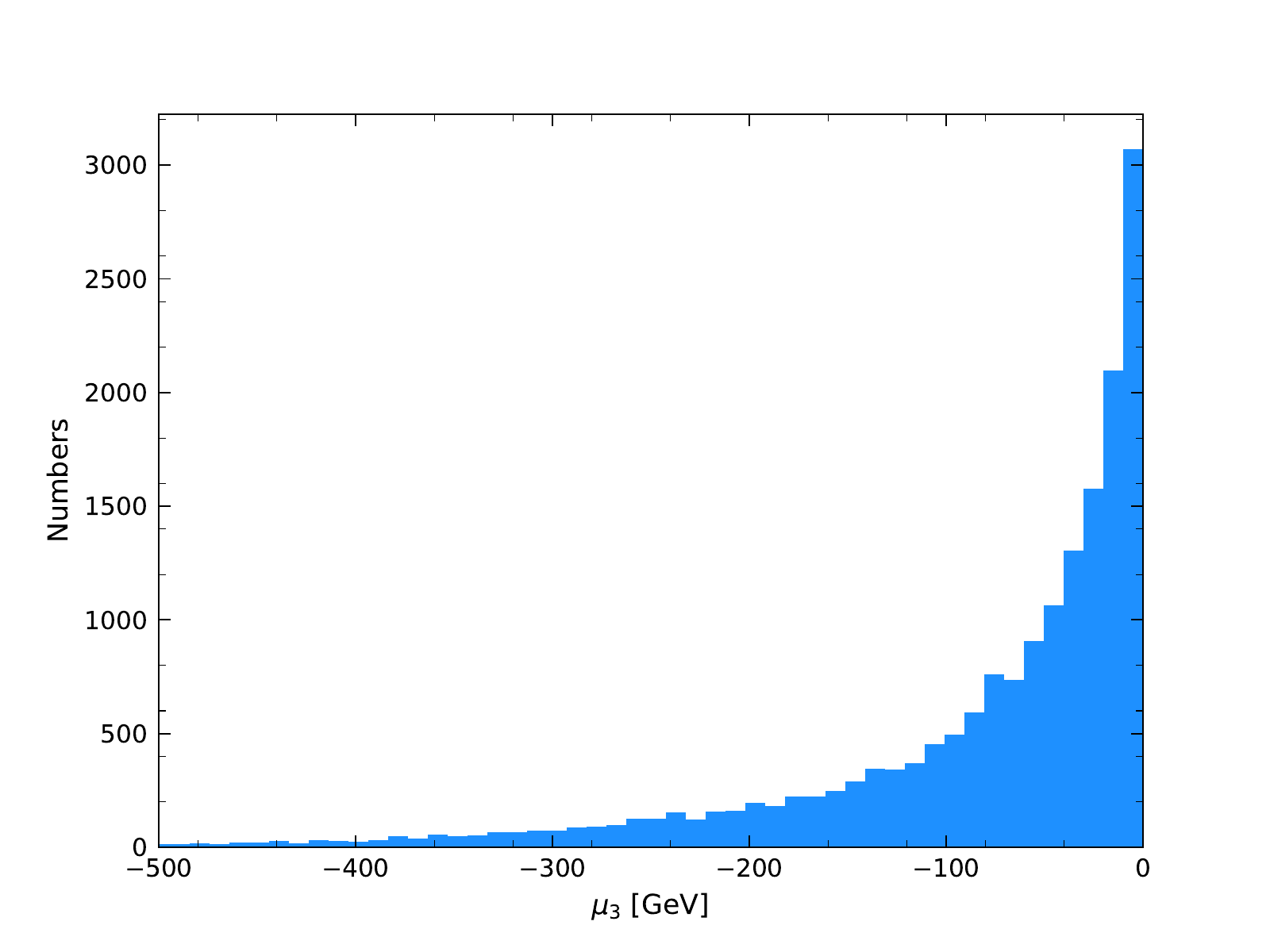}
  \includegraphics[width=75mm,angle=0]{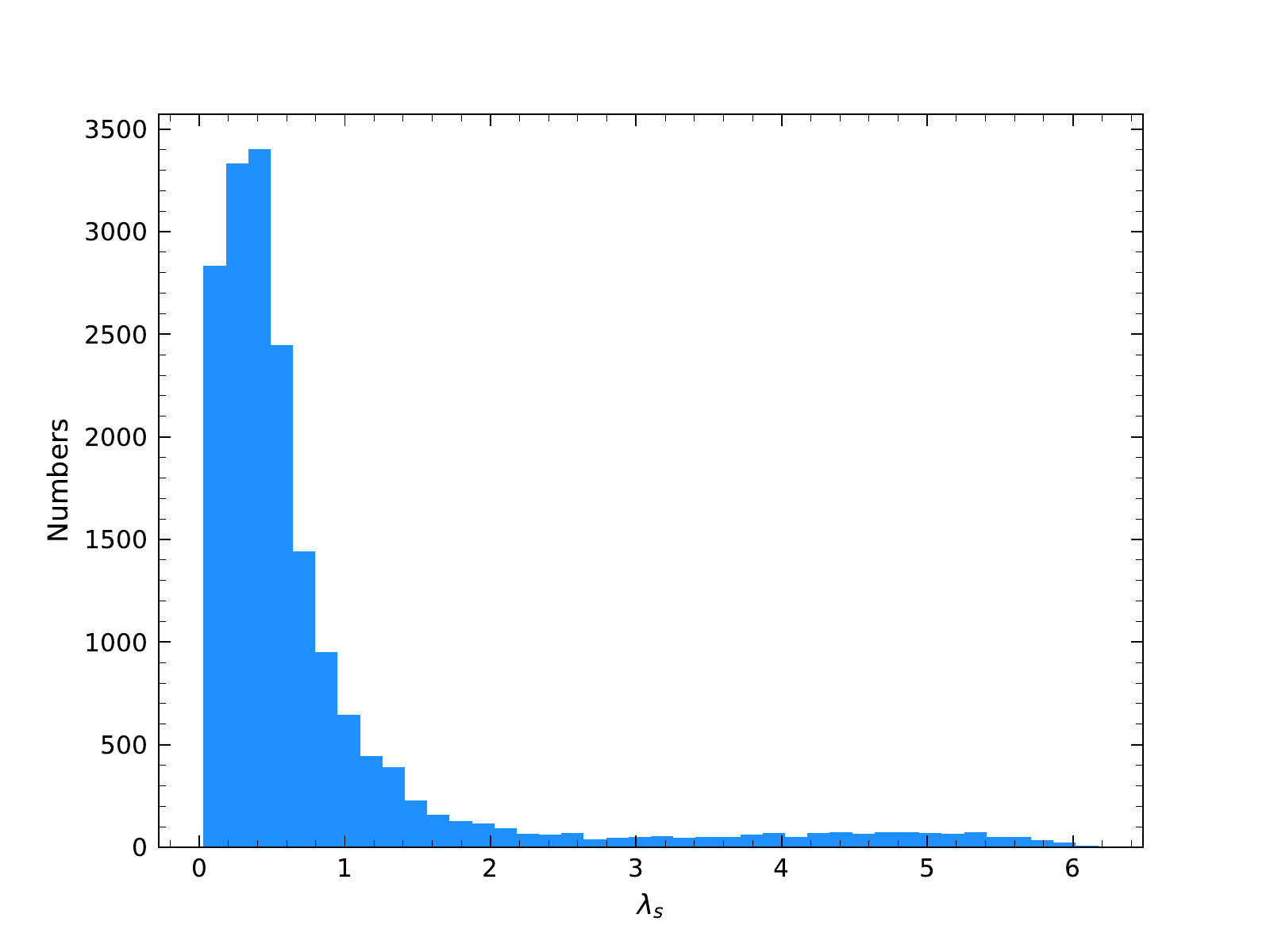}\\
  \includegraphics[width=75mm,angle=0]{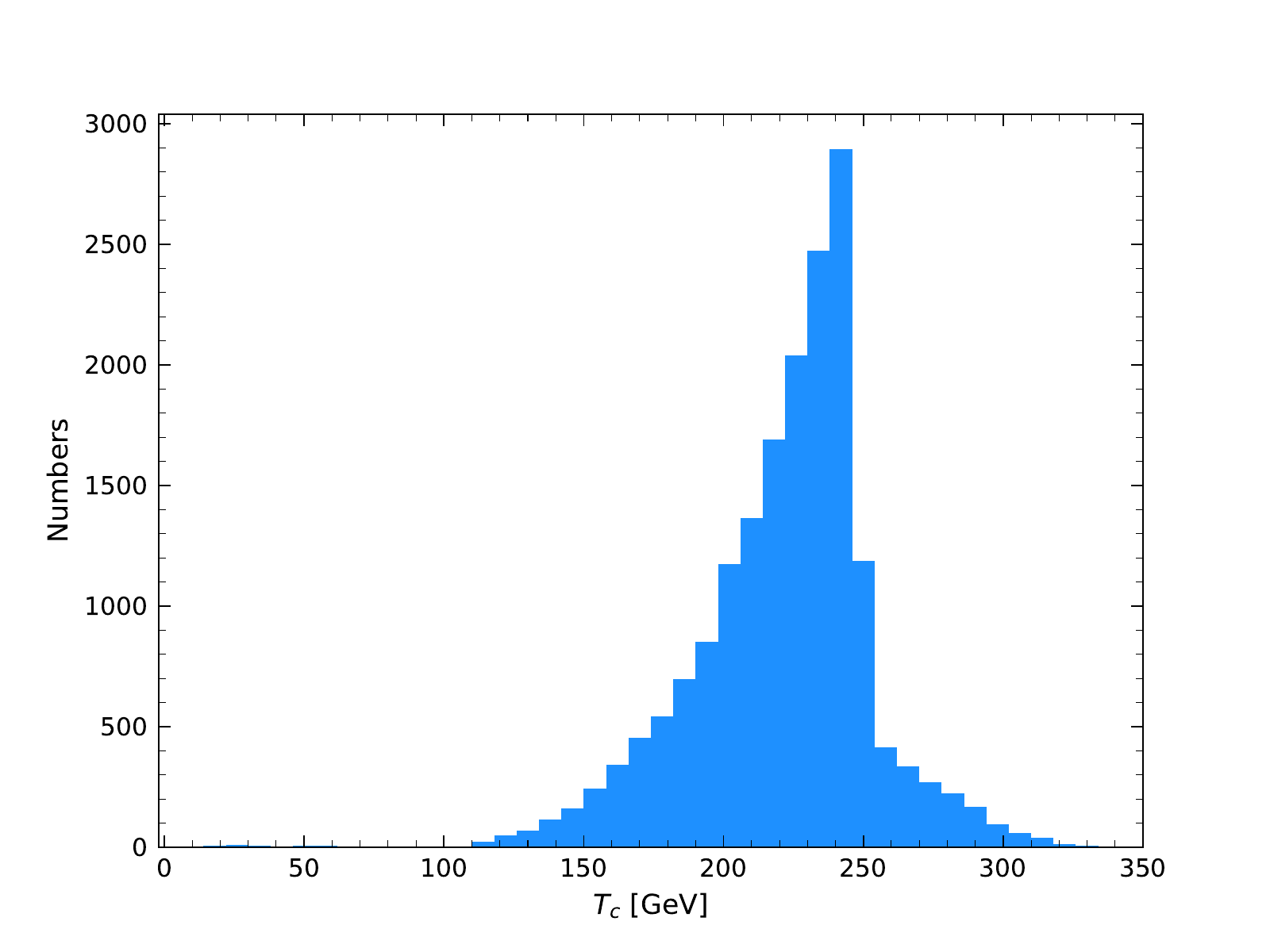}
  \includegraphics[width=75mm,angle=0]{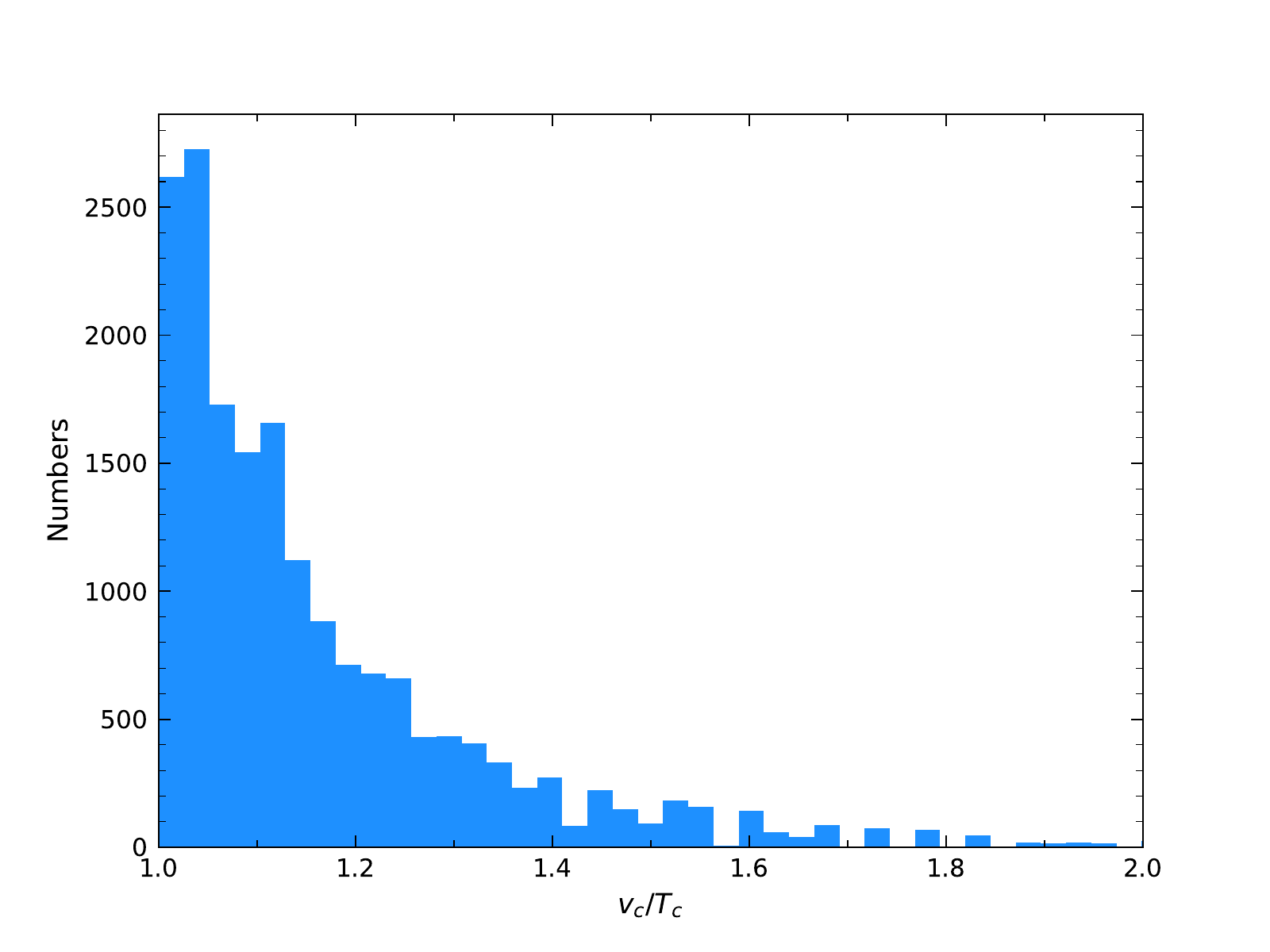}
  \caption{Distributions of parameters in the effective potential that can generate a sufficiently strong first-order EWPT. 
  The lowest two plots give the distributions of $v_c/T_c$ and $T_c$.}
  \label{fig:distributions2}
\end{figure}

\begin{itemize}
\item[1.] There is a lower bound in the distribution of $m_{\mathcal{S}}$, i.e., $m_{\mathcal{S}}\gtrsim 500$~GeV. This is because in order to trigger a phase transition, the local minimum at the broken phase $(v(T),~w(T))$ should be 
larger than the local minimum at the symmetric phase $(0,~w_0(T))$ when the temperature is higher than critical temperature.  To see this, we show in the upper left plot of Fig.~\ref{fig:cst} the potential difference $\Delta V_{bs}(T)\equiv V_{\rm eff}(v(T),~w(T),~T) - V_{\rm eff}(0,~w_0(T),~T)$ 
at temperature $T=300$~GeV (the distribution of critical temperature can be found in Fig.~\ref{fig:distributions2}) as a function of $m_{\mathcal{S}}$, with $\theta=0.4$ and $m_{\chi}=300$~GeV. The blue, green, and red points in Fig.~\ref{fig:cst} represent the results with $w=300$~GeV, 400~GeV, and 500~GeV, respectively.  As shown in the drawing, the potential difference slowly increases with $m_{\mathcal{S}}$ and remains
negative for $m_{\mathcal{S}}\lesssim 500$~GeV. At larger values of $m_{\mathcal{S}}$, the potential difference sharply increases to a positive value.
Therefore, the lower bound $m_{\mathcal{S}}\gtrsim 500$~GeV is to guarantee a successful phase transition.

\item[2.] There is a lower bound in the distribution of $\left | \theta \right |$, i.e., $\left | \theta \right |\gtrsim 0.2$. 
Such a relatively large lower bound on the absolute value of the mixing angle is straightforwardly derived from the requirement of a sufficiently strong first-order EWPT. 
As shown in Eq.~\eqref{eq:lambdam}, the mixing angle $\left | \theta \right |$ directly controls the Higgs-portal interacting strength $\left | \lambda_m \right |$ (whose distribution can be found in the left plot of the second row of Fig.~\ref{fig:distributions2}).
For $\left | \theta \right |\lesssim 0.2$ (or, equally, $\left | \lambda_m \right |\lesssim 1$), 
the Higgs boson $h$ and singlet scalar $s$ have nearly no mixing and the phase transition takes place mostly along the $h$ direction. 
A larger $|\theta|$ would induce a significant mixing between the $h$ and $s$ fields and curve the tunneling path to be along a linear combination of the two fields (see Fig.~\ref{fig:potential1}) and finally lead to a strong EWPT (see also Ref.~\cite{Cline2009JHEP} for similar conclusions).

\item[3.] The lower bound on $w$ and the decrease in the distribution of $\left | \theta \right |$ at larger values are from the bound of unitarity (last condition of Eq.~\eqref{eq:pertu}). 
We plot this bound condition in various planes in Fig.~\ref{fig:cst}, the colored regions are excluded by the constraints. 
In the upper right plot of Fig.~\ref{fig:cst}, we show the unitarity bounds in the $\theta$-$m_{\mathcal{S}}$ plane, fixing $m_{\chi}=300$~GeV.
The values of $w$ in this plot are 100~GeV (green), 200~GeV (yellow), 300~GeV (blue), 400~GeV (light blue), 
600~GeV (pink), and 1000~GeV (sandy-brown), respectively. As shown in this plot, when $w=100$~GeV, the heavy scalar mass $m_{\mathcal{S}}$ is restricted to lie below about 500~GeV.  As one increases $w$, more parameter space opens up.
Sufficient sample points for strong EWPT are available for $w\gtrsim 200$~GeV.
These indicate that the unitarity bound on large values of $m_{\mathcal{S}}$ can be avoided by increasing $w$.
In the lower left plot of Fig.~\ref{fig:cst}, we plot the unitarity bounds in the $w$-$m_{\mathcal{S}}$ plane, 
with $\theta=0.2$ (blue), 0.4 (pink), and 0.6 (green), respectively. As shown in this plot, the available parameter space
is largely reduced with the increase mixing angle, which explains the decreasing behavior in the $\left | \theta \right |$ distribution at large values.  When $m_{\mathcal{S}}\gtrsim 500$~GeV, nearly all of the parameter space is excluded by the unitarity bound for $w\lesssim 200$~GeV. The constraints become independent of $m_{\mathcal{S}}$ for $w\gtrsim 500$~GeV and $\theta\gtrsim 0.4$.

\item[4.] The distribution of pseudo-Goldstone DM mass $m_{\chi}$ has a peak around $(200-400)$~GeV and then decreases at larger values.  The decreasing behavior of $m_{\chi}$ distribution involves a few contributions. One is the unitarity bound; it is easy to check that the unitarity bound becomes tighter for a larger DM mass $m_{\chi}$.  Another reason is the condition \eqref{eq:cond1}, which is to ensure that the stationary point located along the $h$ axis cannot be a local minimum.  We show this constraints in the lower right plot of Fig.~\ref{fig:cst}, 
the colored regions are excluded by the constraints.  In this plot, the value of $w$ is fixed at 1000~GeV, 
the green, yellow, pink, and blue regions represent the bounds with $m_{\mathcal{S}}=$ 500~GeV, 700~GeV, 1000~GeV, and 1500~GeV, respectively. 
We see that for $\theta\gtrsim 0.2$, the parameter space with $m_{\chi}\gtrsim 1000$~GeV is excluded. However, we should note that this bound is imposed only when there is a stationary point along $h$ axis, which requires $\mu_h^2>0$.
From the distribution of $\mu_h^2$ shown in the upper left plot of Fig.~\ref{fig:distributions2}, we find that only about half of the total sample points have a stationary point along $h$ axis.  We also note that such a requirement is somehow too stringent because of the loop corrections on the potential.
Further constraint may arise from the requirement of $d\Delta V_{bs}(T)/dT^2>0$ near the critical temperature~\cite{Espinosa2012NPB}, we find this bound is relatively weak and do not give a detailed account here.

\item[5.] We see that within the parameter ranges considered here, the distributions of $w$ and $m_{\mathcal{S}}$ do not decrease with increasing values. Such distributions are consistent with a recent global fit performed in Ref.~\cite{Beniwal2019JHEP}.
\end{itemize}

In Fig.~\ref{fig:distributions2} we show the distributions of the various parameters in the scalar potential, which are determined from the four physical parameters above using Eqs.~\eqref{eq:muh2}-\eqref{eq:lambdas}.
The Higgs self coupling $\lambda_h$ takes a value between 1 and 4 most of the time.
The mixing coupling $\lambda_m$ can take either sign, but cannot be too small, as demanded by $|\theta| \gtrsim 0.2$. 
Due to our choice of a positive $w$, $\mu_3$ takes only negative values in order to have a positive DM mass, given in Eq.~\eqref{eq:dmmass2}.
The self coupling parameter of the $S$ field is usually less than 1, though larger values are sometimes allowed as well.
The distributions of the critical temperature $T_c$ and the ratio $v_c/T_c$ are also depicted in the figure, with $T_c$ falling mostly between 150 and 300~GeV.

\section{Vacuum stability and perturbativity}
\label{sec:VPbounds}

In this work, we restrict ourselves in the case of absolute vacuum stability, which requires that the EW vacuum of the scalar effective potential at zero temperature be a global minimum below the energy scale for which the model is valid. 

For our model, the tree-level vacuum stability is implemented in Eq.~\eqref{eq:vacstab}. For the vacuum stability
at one-loop level, we follow the gauge-invariant treatment used in Ref.~\cite{Gonderinger2012PRD} by requiring
\begin{equation}
  \label{eq:vacstablp}
  \lambda_{h}(\mu)>0,~~ \lambda_{s}(\mu)>0,~~ \lambda_{h}(\mu) \lambda_{s}(\mu)>\frac{1}{4} \lambda_{m}(\mu)^{2}, {~~\rm for~~} \mu<\Lambda,
\end{equation}
where $\Lambda$ is a cutoff energy scale where new physics comes in or perturbation breaks down. These extend the requirements of tree-level vacuum stability on the couplings to any energy scale 
below the cutoff scale. These conditions ensure a positive potential for large field values. 
The large-field behavior of the potential is dominated by the quartic terms. With the relations given in Eq.~\eqref{eq:muh2}, the quartic part of the 
potential is given by \cite{Espinosa2012NPB}
\begin{eqnarray}
  V_4=\frac{1}{8\mathcal{M}_{h}^2v^2}[(\mathcal{M}_{h}^2h^2+\lambda_mv^2s^2)^2+4\lambda^2v^4s^4],
\end{eqnarray}
where $\lambda^2\equiv \lambda_h\lambda_s-\lambda_m^2/4$. We see that if $\lambda_m<0$, the first term in the square brackets vanishes along a particular direction.  In this case, the last condition of Eq. \eqref{eq:vacstablp} ensures the stability of the EW vacuum.  However, in the case of $\lambda_m>0$, the condition $\lambda^2>0$ could be over-restrictive since the potential is already bounded from below for large values of the fields along all directions with the first two conditions of Eq.~\eqref{eq:vacstablp}.

\begin{figure}
  \centering
  \includegraphics[width=75mm,angle=0]{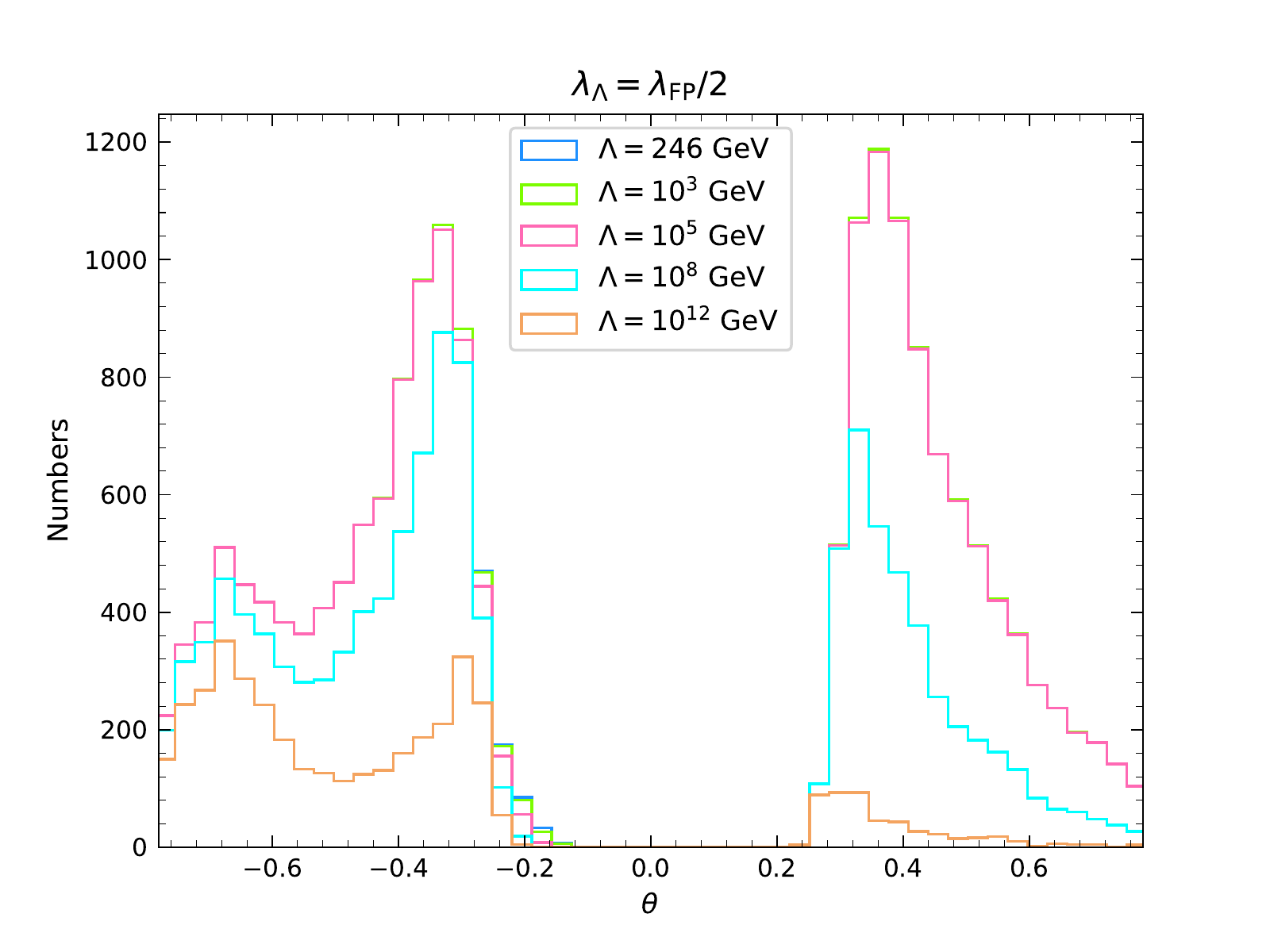}
  \includegraphics[width=75mm,angle=0]{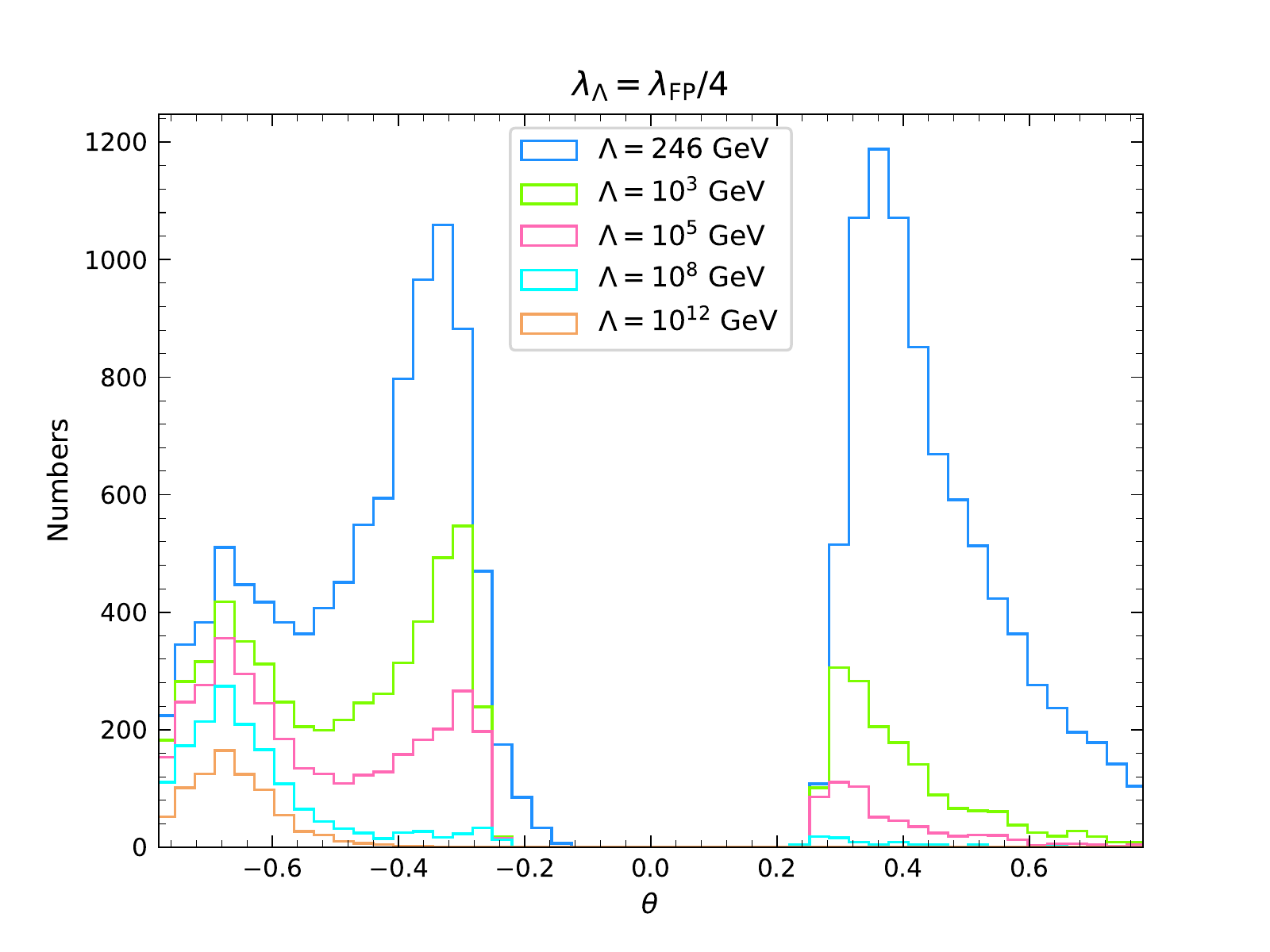}\\
  \includegraphics[width=75mm,angle=0]{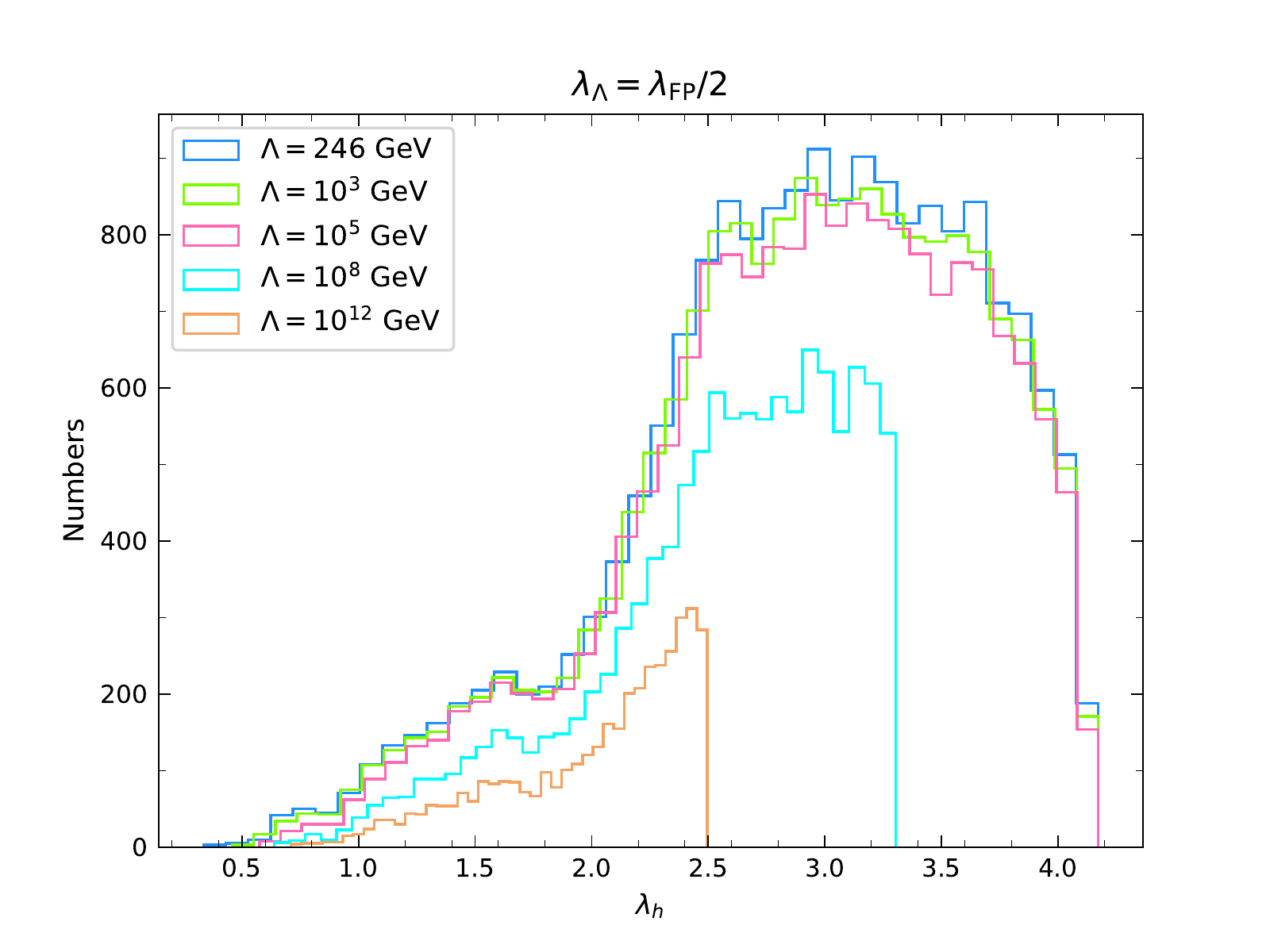}
  \includegraphics[width=75mm,angle=0]{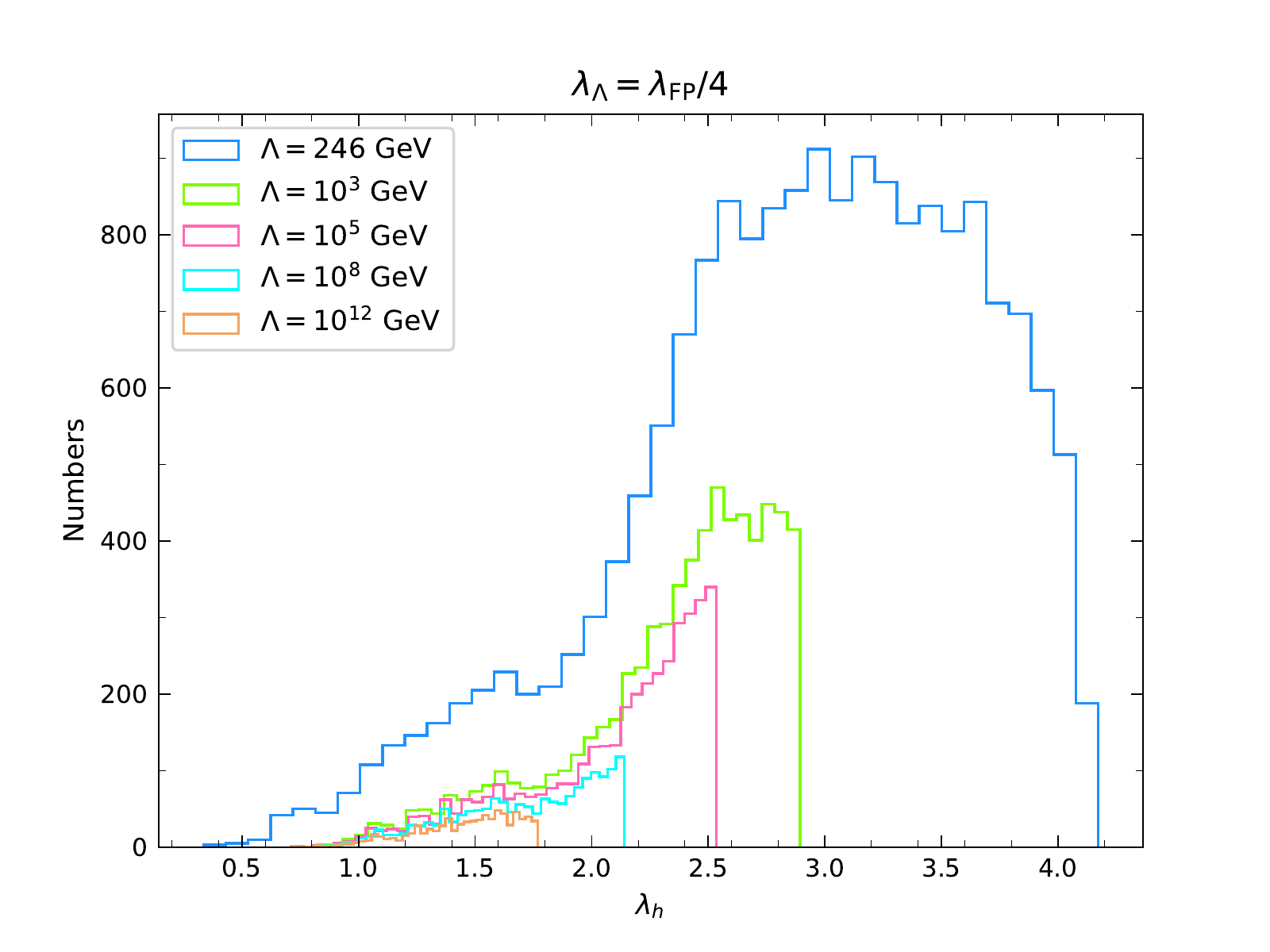}\\
  \includegraphics[width=75mm,angle=0]{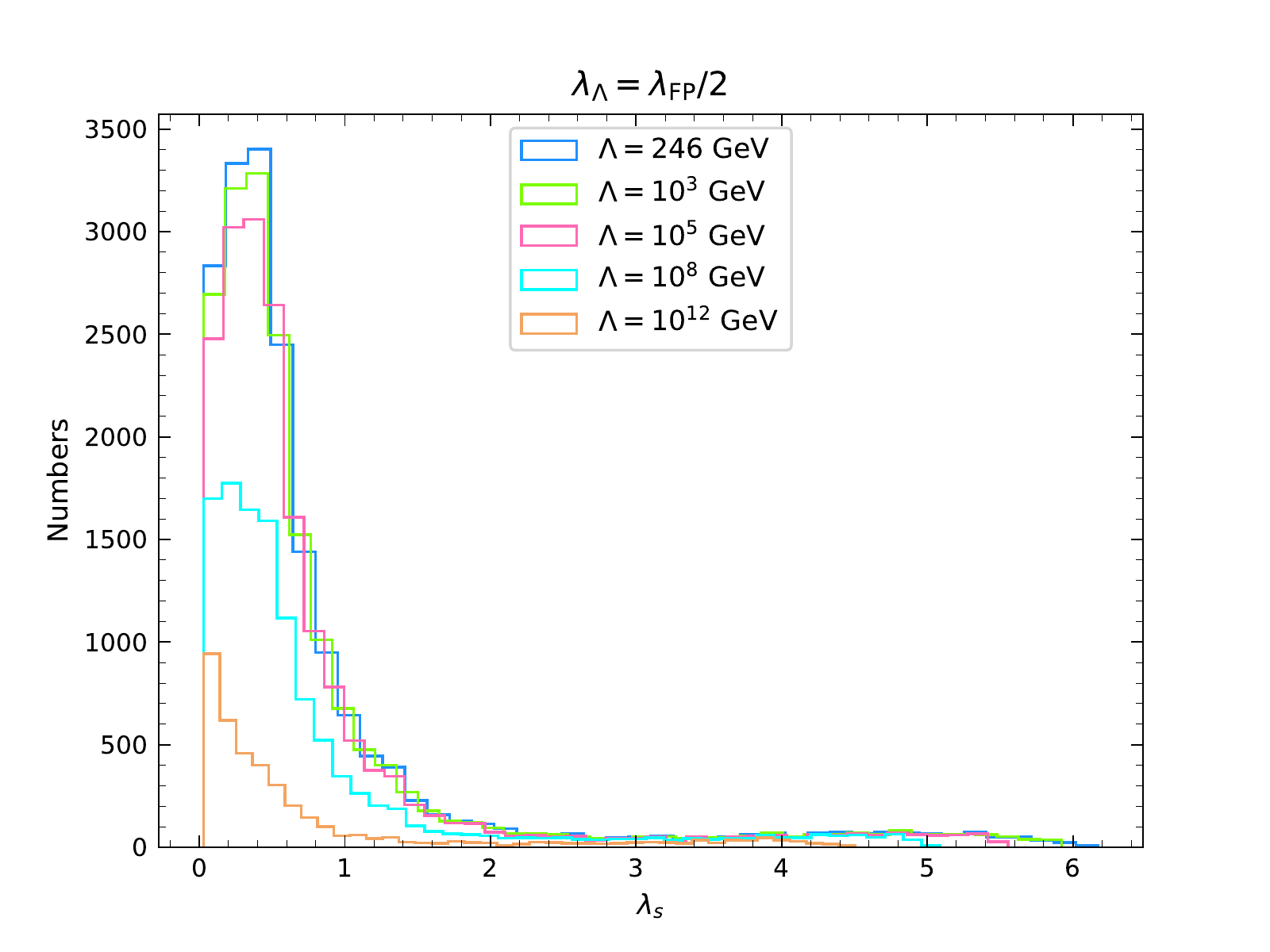}
  \includegraphics[width=75mm,angle=0]{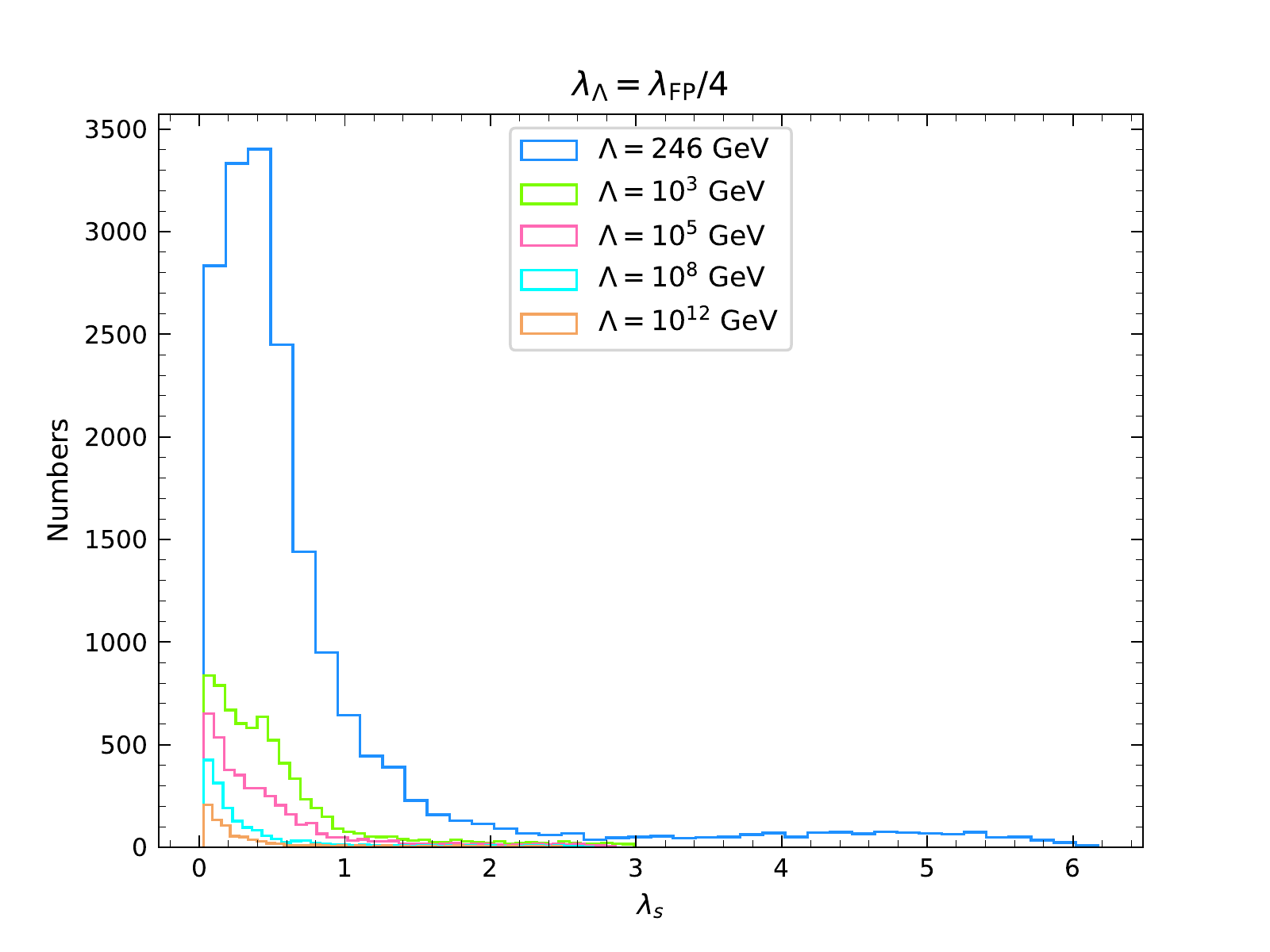}
  \caption{Distributions of parameters after taking into account the constraints from vacuum stability and perturbativity at the one-loop level.
  The blue, green, pink, light blue, and yellow histograms represent the bounds with $\Lambda$ = 246 GeV, $10^3$ GeV, $10^5$ GeV, $10^8$ GeV, 
  and $10^{12}$ GeV, respectively. $\lambda_{\Lambda}=\lambda_{\rm FP}/2$ for the left plots and $\lambda_{\rm FP}/4$ for the right plots.
  }
  \label{fig:VPbounds}
\end{figure}

We now turn to the perturbativity of the scalar potential. At the one-loop level, a large Higgs quartic coupling at EW scale may give 
rise to a positive $\beta_{\lambda_h}$ function, which results in a monotonic increase of $\lambda_{h}(\mu)$ with $\mu$ and eventually develops a Landau pole at some scale $\Lambda_{\rm L}$.
However, the perturbativity of theory has already been invalid long before $\mu$ reaches $\Lambda_{\rm L}$.
At the two-loop level, $\lambda_{h}(\mu)$ may approach an ultraviolet fixed point where $\beta_{\lambda_h}\to 0$ and $\lambda_{h}(\mu)\to \lambda_{\rm FP}$~\cite{Hambye1997PRD}. For a given cutoff scale $\Lambda$, perturbation theory is expected to be reliable for the value of 
$\lambda_{\Lambda}\equiv \lambda_{h}(\Lambda)$ in the range of $\lambda_{\rm FP}/4-\lambda_{\rm FP}/2$~\cite{Riesselmann1997PRD}.
Following Ref.~\cite{Gonderinger2012PRD}, the approximate perturbativity bounds on the couplings are given by
\begin{equation}
  \label{eq:perturbativity}
  \lambda_{h}(\mu)<\lambda_{\Lambda},~~ \lambda_{s}(\mu)<\lambda_{\Lambda},~~ \lambda_{m}(\mu)<\lambda_{\Lambda}, {~~\rm for~~} \mu<\Lambda.
\end{equation}

Given the cutoff scale $\Lambda$, we evolve the running couplings from the EW scale up to $\Lambda$ with the RG equations.
We pick out those samples simultaneously satisfying the constraints from both vacuum stability and perturbativity, and present the results 
in Fig.~\ref{fig:VPbounds}. In the figure, we show the distributions of $\theta$, $\lambda_{h}$, and $\lambda_{s}$ with various choices of 
$\Lambda$ and $\lambda_{\Lambda}$. The constraints, of course, become stronger with a larger $\Lambda$ and a lower $\lambda_{\Lambda}$.
We find that most of the samples satisfy the vacuum stability up to a very large scale, while perturbativity can impose a stricter bound on the parameters.
For the case of $\lambda_{\Lambda}=\lambda_{\rm FP}/2$, the constraints with $\Lambda=10^3$~GeV (light green histogram) are nearly negligible 
so that its distribution overlaps with the initial distribution (blue histogram). We see that in this case, most of the parameter
space of the theory remains valid for the energy scale below $\sim 10^8$ GeV. However, for the case of $\lambda_{\Lambda}=\lambda_{\rm FP}/4$,
most of the samples are excluded for the scale above $\sim 10^5$ GeV. We note that the constraints from vacuum stability and perturbativity are subjective due to the somewhat arbitrary choice of the cutoff scale and the uncertainties in the value of $\lambda_{\Lambda}$~\cite{Gonderinger2012PRD}.

\section{Experimental constraints}
\label{sec:expbounds}

In this section, we consider various constraints of collider experiments on our model, including the electroweak precision observable measurements and Higgs signal strength measurements.

\subsection{Electroweak precision observables}

We first consider the constraints arising from electroweak precision observables (EWPOs)~\cite{Peskin1990PRL, Peskin1992PRD}.
We introduced two scalars to the SM, one of them is a real singlet scalar which can mix with the SM Higgs boson.
Both the SM Higgs boson and the real singlet scalar will contribute to the SM gauge boson self-energies at loop-level and finally induce corrections to the oblique parameters $S$, $T$, and $U$.
The other one is a pseudo-Goldstone boson $\chi$ which becomes a DM candidate due to an accidental $\mathbb{Z}_2$ symmetry hidden in our $\mathbb{Z}_3$ symmetry model. The pseudo-Goldstone boson do not mix with SM particles and thus do not affect the oblique parameters.  Constraints from the EWPO's can alter the distribution of $m_{\chi}$ via the parameters $m_{\mathcal{S}}$ and $\theta$.

The contributions of the heavy scalar to the oblique parameters 
$\Delta \mathcal{O}_{i} \equiv \mathcal{O}_{i}-\mathcal{O}_{i}^{\rm SM}$ 
(here $i$ denotes $T$, $S$, or $U$) can be quantified as follows~\cite{Profumo2007JHEP, Baek2012JHEP}:
\begin{align}
\begin{split}
\Delta T
&=  \frac{3}{16 \pi s_{W}^{2}}\left[\cos ^{2} \theta\left\{f_{T}\left(\frac{m_{\mathcal{H}}^{2}}{m_{W}^{2}}\right)
-\frac{1}{c_{W}^{2}} f_{T}\left(\frac{m_{\mathcal{H}}^{2}}{m_{Z}^{2}}\right)\right\}+\sin ^{2} \theta\left\{f_{T}
\left(\frac{m_{\mathcal{S}}^{2}}{m_{W}^{2}}\right)\right.\right.
\\ 
& \qquad\qquad
\left.\left.-\frac{1}{c_{W}^{2}} f_{T}
\left(\frac{m_{\mathcal{S}}^{2}}{m_{Z}^{2}}\right)\right\}-\left\{f_{T}\left(\frac{m_{\mathcal{H}}^{2}}{m_{W}^{2}}\right)
-\frac{1}{c_{W}^{2}} f_{T}\left(\frac{m_{\mathcal{H}}^{2}}{m_{Z}^{2}}\right)\right\}\right],
\\
\Delta S
&= \frac{1}{2 \pi}\left[\cos ^{2} \theta f_{S}\left(\frac{m_{\mathcal{H}}^{2}}{m_{Z}^{2}}\right)+\sin ^{2} 
  \theta f_{S}\left(\frac{m_{\mathcal{S}}^{2}}{m_{Z}^{2}}\right)-f_{S}\left(\frac{m_{\mathcal{H}}^{2}}{m_{Z}^{2}}\right)\right],
\\
\Delta U
&= \frac{1}{2 \pi}\left[\cos ^{2} \theta f_{S}\left(\frac{m_{\mathcal{H}}^{2}}{m_{W}^{2}}\right)+\sin ^{2} \theta f_{S}
  \left(\frac{m_{\mathcal{S}}^{2}}{m_{W}^{2}}\right)-f_{S}\left(\frac{m_{\mathcal{H}}^{2}}{m_{W}^{2}}\right)\right]-\Delta S,
\end{split}
\end{align}
where $m_Z$ and $m_W$ are the SM weak gauge boson masses, the cosine of the weak mixing angle $c_{W}^{2}=m_{W}^{2}/m_{Z}^{2}$ and  $s_{W}^{2}=1-c_{W}^{2}$. The loop functions $f_T(x)$ and $f_S(x)$ are given by~\cite{Grimus2008NPB, Beniwal2019JHEP}
\begin{align}
\begin{split}
f_{T}(x)
&=
\frac{x \log x}{x-1},
\\
f_{S}(x)
&=
\begin{cases}
\displaystyle
{\frac{1}{12}\left[-2x^{2}+9x+\left((x-3)\left(x^{2}-4 x+12\right)+\frac{1-x}{x}\right) f_{T}(x)\right.} 
\\ 
\displaystyle
\qquad {\left.+2 \sqrt{(4-x) x}\left(x^{2}-4 x+12\right) \tan^{-1}\sqrt{\frac{4-x}{x}}\right], 
~ \mbox{for~} 0<x<4,} 
\\ 
\displaystyle
{\frac{1}{12}\left[-2 x^{2}+9 x+\left((x-3)\left(x^{2}-4x+12\right)+\frac{1-x}{x}\right) f_{T}(x)\right.} 
\\ 
\displaystyle
\qquad {\left.+\sqrt{(x-4) x}\left(x^{2}-4 x+12\right) \log \left(\frac{x-\sqrt{(x-4) x}}{x+\sqrt{(x-4) x}}\right)\right], 
~ \mbox{for~} x \geq 4.}
\end{cases}
\end{split}
\end{align}
In order to analyze the impacts of the EWPO's on the strong EWPT parameter distributions, 
we follow the procedure given in Refs.~\cite{Profumo2007JHEP, Cline2009JHEP, Barger2008PRD, Profumo2015PRD}
by defining a $\Delta \chi^{2}$ as 
\begin{equation}
  \Delta \chi^{2}_{\rm ewpo}=\sum_{i, j}\left(\Delta \mathcal{O}_{i}-\Delta \mathcal{O}_{i}^{\rm exp}\right)\left(\Sigma^{2}\right)_{i j}^{-1}
  \left(\Delta \mathcal{O}_{j}-\Delta \mathcal{O}_{j}^{\rm exp}\right),
\end{equation}
where $\Delta\mathcal{O}_{i}^{\rm exp}$ denotes the experimental measurements of the deviations of oblique parameters from its SM reference values.  We take the most recent analysis from the Gfitter Group~\cite{Haller2018EPJC}
\begin{equation}
  \label{eq:operr}
  \Delta S=0.04 \pm 0.11, \quad \Delta T=0.09 \pm 0.14, \quad \Delta U=-0.02 \pm 0.11.
\end{equation}
The covariance matrix $\Sigma_{i j}^{2} \equiv \sigma_{i} \rho_{i j} \sigma_{j}$ involves $\sigma_i$ as 
the various errors given in Eqs.~\eqref{eq:operr} and 
\begin{equation}
\rho_{i j}=\left(\begin{array}{ccc}{1} & {0.92} & {-0.68} \\ {0.92} & {1} & {-0.87} \\ {-0.68} & {-0.87} & {1}\end{array}\right)
\end{equation}
is the correlation matrix of the experiment.
The electroweak observables are governed by only two parameters: the heavy scalar mass $m_{\mathcal{S}}$ and the mixing angle $\theta$.
We consider the singlet scalar extended models to be consistent with the EWPO's if the oblique parameters lie within the 95\% confidence level (CL) ellipsoid, which corresponds to taking $\Delta \chi^{2}_{\rm ewpo} \le 5.99$
for the models with given $m_{\mathcal{S}}$ and $\theta$.

\begin{figure}
  \centering
  \includegraphics[width=120mm,angle=0]{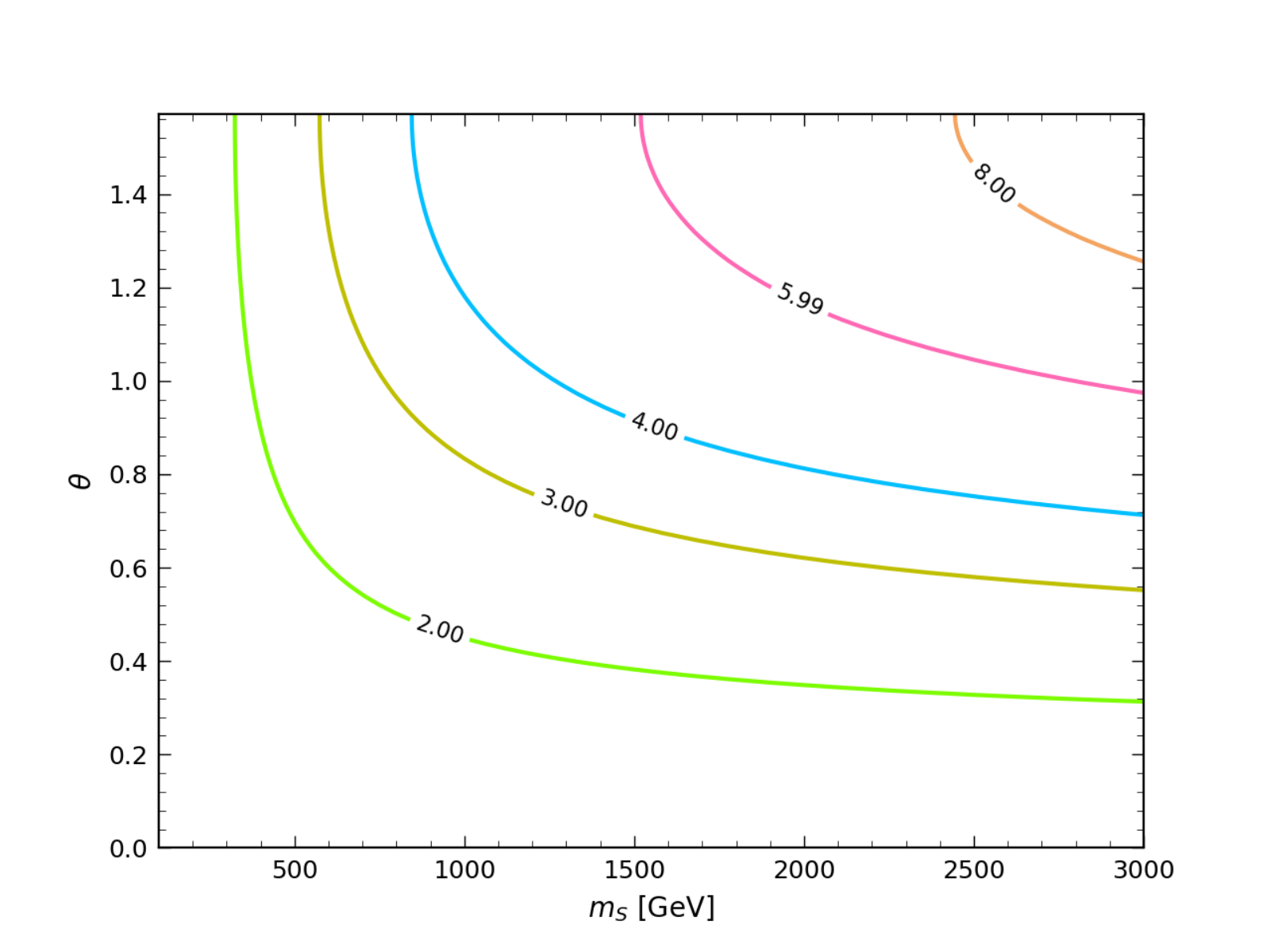}
  \caption{Contours of $\Delta \chi^{2}_{\rm ewpo}$ in the $m_{\mathcal{S}}$-$\theta$ plane.
  }
  \label{fig:chi2ewpo}
\end{figure}

We plot the contours of $\Delta \chi^{2}_{\rm ewpo}$ in Fig.~\ref{fig:chi2ewpo} and find that in the $m_{\mathcal{S}}$-$\theta$ space, the region with $m_{\mathcal{S}}\gtrsim 1500$ GeV and $\theta\gtrsim 1.0$ (to the right of the pink curve) is excluded at the 95\% CL.  The constraint is seen to be not very strong.

\subsection{Higgs signal strengths}


In our extension of the SM with a complex singlet scalar, the SM-like Higgs boson coupling strengths to other SM particles are modified by 
a common factor of $\cos \theta$.  This leads to the productions of $X\bar{X}$ (where $X$ can be a SM gauge boson or fermion) via the SM-like Higgs boson being suppressed by a factor $\cos^2\theta$, provided that no additional decay channel is permitted.  Note that the same factor of $\cos^2\theta$ is involved in all major production channels (gluon fusion, vector boson fusion and $t \overline{t} h$).
To take into account the published exclusion bounds from Higgs searches at the LEP, Tevatron and LHC experiments, we make use of the 
$\textsf{HiggsBounds}\_\textsf{v4.3.1}$~\cite{Bechtle2014EPJC} package, which calculates the predicted signal rates for 
the search channels considered in the experimental data. 
More information on search channels and experimental data used in $\textsf{HiggsBounds}$ can be found in Ref.~\cite{Bechtle2014EPJC}.
The $\textsf{HiggsBounds}$ package determines whether a point in the model parameter space is excluded at 95\% CL by comparing the predicted signal rates against the expected and observed cross section limits from the 
direct Higgs searches~\cite{Bechtle2014EPJC}.


Distinct signal strengths, defined as the production rate times the decay branching fraction relative to  the SM expectation, i.e., $\mu_{i} \equiv(\sigma \times \mathrm{BR})_{i} /(\sigma \times \mathrm{BR})_{i}^{\mathrm{SM}}$, in various decay channels including $\gamma\gamma$, $WW^{\ast}$, $ZZ^{\ast}$, $b\bar{b}$, and $\tau^+\tau^-$ have already been measured with high precision at the LHC.
From these signal strengths, one can obtain information on the couplings of the Higgs boson to SM particles 
and derive constraints on the extension models.
For the model considered in this work and in the narrow width approximation, the signal strength of $\mathcal{H}$ is~\cite{Baek2012JHEP}
\begin{equation}
  \mu_{\mathcal{H}}=\frac{\Gamma_{\mathcal{H}}^{\mathrm{SM}} \cos ^{4} \theta}{\Gamma_{\mathcal{H}}^{\mathrm{SM}} \cos ^{2} 
  \theta+\Gamma_{\mathcal{H} \rightarrow \chi\chi}+\Gamma_{\mathcal{H} \rightarrow \mathcal{SS}}},
\end{equation}
where $\Gamma_{\mathcal{H}}^{\rm SM}$ denotes the total decay widths of the SM-like Higgs boson with mass being set to $m_{\mathcal{H}}$, and $\Gamma_{\mathcal{H}\rightarrow XX}$ is the width of $\mathcal{H}$ decaying to a pair of $X$ ($=\chi, {\cal H}$), which can be found in appendix \ref{apd:dw}.
We see that the signal strengths of $\mathcal{H}$ is suppressed by two factors: $\cos^2\theta$
and the presence of new decay channels. Even when the new decay channels are kinematically forbidden, the signal strength is still reduced by $\cos^2\theta$. 
This means that a generic signature of the mixing of the SM Higgs boson with an extra singlet scalar boson can be derived from a reduced signal of the Higgs bosons at the LHC~\cite{Baek2012JHEP}.

We use the $\textsf{HiggsSignals}\_\textsf{v1.4.0}$ package~\cite{Bechtle2014EPJC-2} to estimate the $\chi^2$ of a given model and assess which sample points are allowed by the signal strength measurements.
The package assumes a Gaussian probability distribution and uses the peak-centered method for the calculation 
of $\chi^2=\chi_{\mu}^2+\chi_{m}^2$, where $\chi_{\mu}^2$ is evaluated by comparing the signal strength measurements for the peak to the model-predicted signal strengths and $\chi_{m}^2$ is evaluated by comparing the model-predicted Higgs boson mass and the observed one if a mass measurement is also available~\cite{Bechtle2014EPJC-2}.
The signal strength measurements used in the $\textsf{HiggsSignals}$ analysis are summarized in table~\ref{tab:i}.

We find that $\chi_{\rm min}^2$, the minimum of $\chi^2$, is obtained in the model with the mixing angle $\theta=0$, which means no mixing for the SM Higgs boson.
Totally 89 observables are used in each fit, and $\chi^2_{\rm min} = 102.02$.
We conclude that a sample point in the parameter space is not excluded by the experimental results at 95\% CL if $\Delta \chi^2_{\rm hs}=\chi^2-\chi^2_{\rm min} < 9.49$ (for four free parameters).

\begin{figure}
  \centering
  \includegraphics[width=120mm,angle=0]{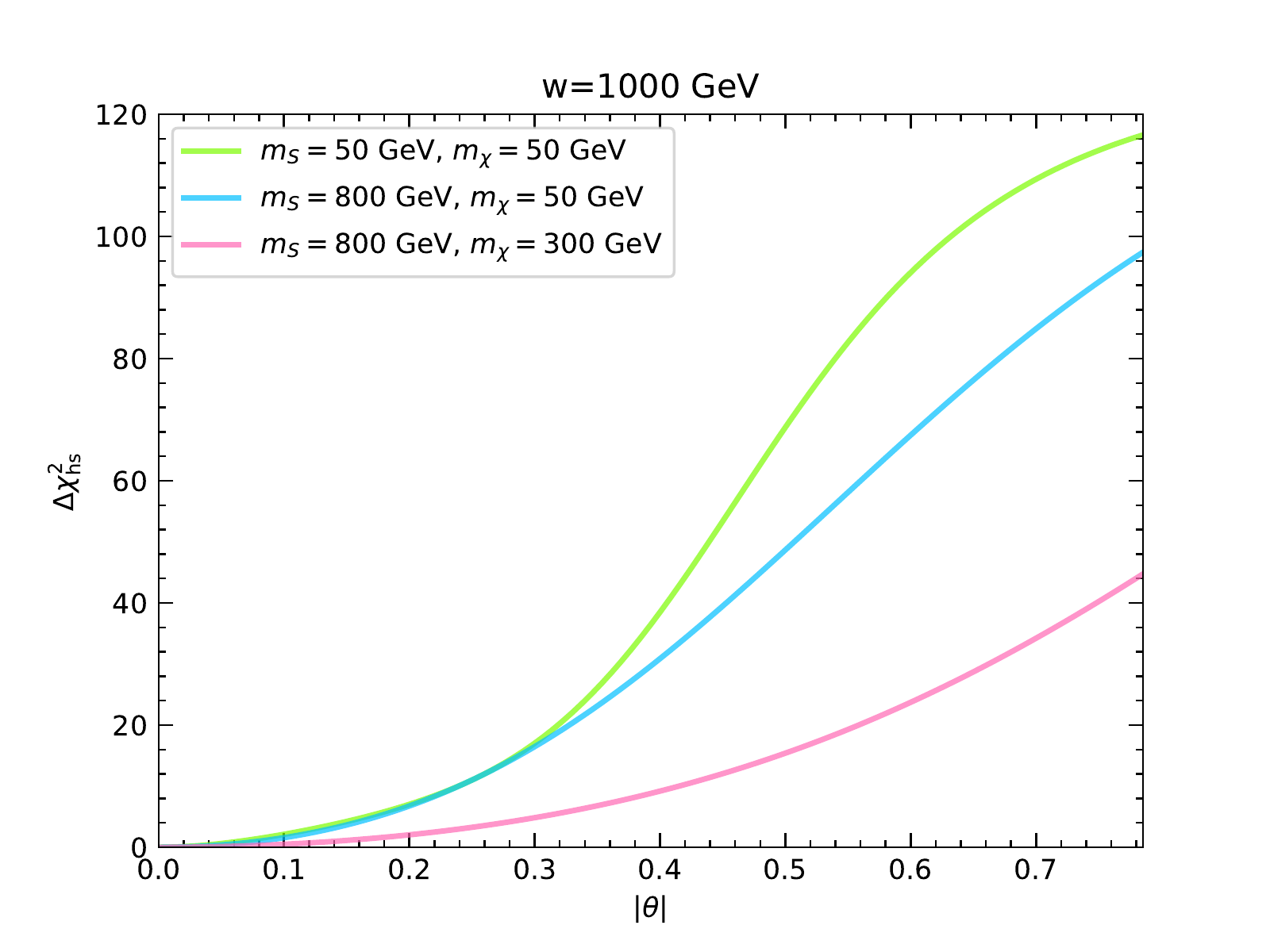}
  \caption{$\Delta \chi^{2}_{\rm hs}$ as a function of $|\theta|$, with $w=1000$~GeV.
  The green, blue, and pink curves represent the results with $(m_{\mathcal{S}},~m_{\chi})=$ (50, 50)~GeV, (800, 50)~GeV,
  and (800, 300)~GeV respectively.
  }
  \label{fig:chi2higgs}
\end{figure}

We plot $\Delta \chi^2_{\rm hs}$ as a function of the mixing angle in Fig.~\ref{fig:chi2higgs}, with
the pink, blue, and green curves denoting the results for $(m_{\mathcal{S}},~m_{\chi})=$ (800, 300)~GeV, (800, 50)~GeV, and (50, 50)~GeV, respectively.
As discussed above, when $m_{\mathcal{S}},~m_{\chi}>m_{\mathcal{H}}/2$, the SM Higgs signal strength is reduced by a factor of $\cos^2\theta$, and thus $\Delta \chi^2_{\rm hs}$ increases with $|\theta|$. When $m_{\chi}<m_{\mathcal{H}}/2$, the SM Higgs invisible decay $\mathcal{H}\to\chi\chi$ is 
kinematically allowed, leading to an additional suppression in the SM Higgs signal strength, as is indicated by the blue curve.  One can also see that the opening of new decay channels will play a dominant role in suppressing the SM Higgs signal strength.  
Furthermore, if $m_{\mathcal{S}}<m_{\mathcal{H}}/2$ the decay channel $\mathcal{H}\to \mathcal{SS}$ opens up and dominates $\Delta \chi^2_{\rm hs}$ for $|\theta|\gtrsim 0.3$, as shown by the green curve.

In summary, the constraints from Higgs signal strength measurements can be divided into two cases:
\begin{itemize}
  \item Case 1. If all of the extra particles are heavier than half of the SM Higgs boson mass, the Higgs signal strength scales as $\cos^2\theta$.
  The mixing angle of the SM Higgs boson with an extra singlet scalar should be $\lesssim 0.4$ ($23^{\circ}$) at 95\% CL.
  \item Case 2. If there is at least one extra particle is lighter than half of the SM Higgs boson mass, the Higgs signal strength will 
  receive an additional suppression, the constraint on the mixing angle becomes more rigorous than Case 1.
\end{itemize}

\begin{table}[tbp]
  \centering
  \renewcommand\arraystretch{1.5}
  \begin{tabular}{|l|c|c|c|}
  \hline
  \multirow{2}*{Channel} & 
  \multicolumn{2}{c|}{Signal strength} &
  \multirow{2}{*}{refs.} \\
  \cline{2-3}
    & ATLAS & CMS & \\
  \hline 
  $h\to \gamma\gamma$ & $1.17_{-0.26}^{+0.28}$ & $1.14_{-0.23}^{+0.26}$  &~\cite{ATLAS2016EPJC},\cite{CMS2014EPJC} \\
  $h\to WW^{\ast}$ & $1.18_{-0.21}^{+0.24}$ & $0.72_{-0.18}^{+0.20}$ &~\cite{ATLAS2016EPJC},\cite{CMS2014JHEP}\\
  $h\to ZZ^{\ast}$ & $1.46_{-0.34}^{+0.40}$ & $0.93_{-0.25}^{+0.29}$ &~\cite{ATLAS2016EPJC},\cite{CMS2014PRD}\\
  $h\to b\bar{b}$ & $0.63_{-0.37}^{+0.39}$ & $1.00_{-0.50}^{+0.50}$ &~\cite{ATLAS2016EPJC},\cite{CMS2014NP}\\
  $h\to \tau^+\tau^-$ & $1.44_{-0.37}^{+0.42}$ & $0.78_{-0.27}^{+0.27}$ &~\cite{ATLAS2016EPJC},\cite{CMS2014NP}\\
  \hline
  \end{tabular}
  \caption{\label{tab:i} Experimental results of Higgs signal strength measurements in various channels
  that are used in the $\textsf{HiggsSignals}$ analysis~\cite{Bechtle2014EPJC-2}.}
\end{table}
\begin{figure}
  \centering
  \includegraphics[width=75mm,angle=0]{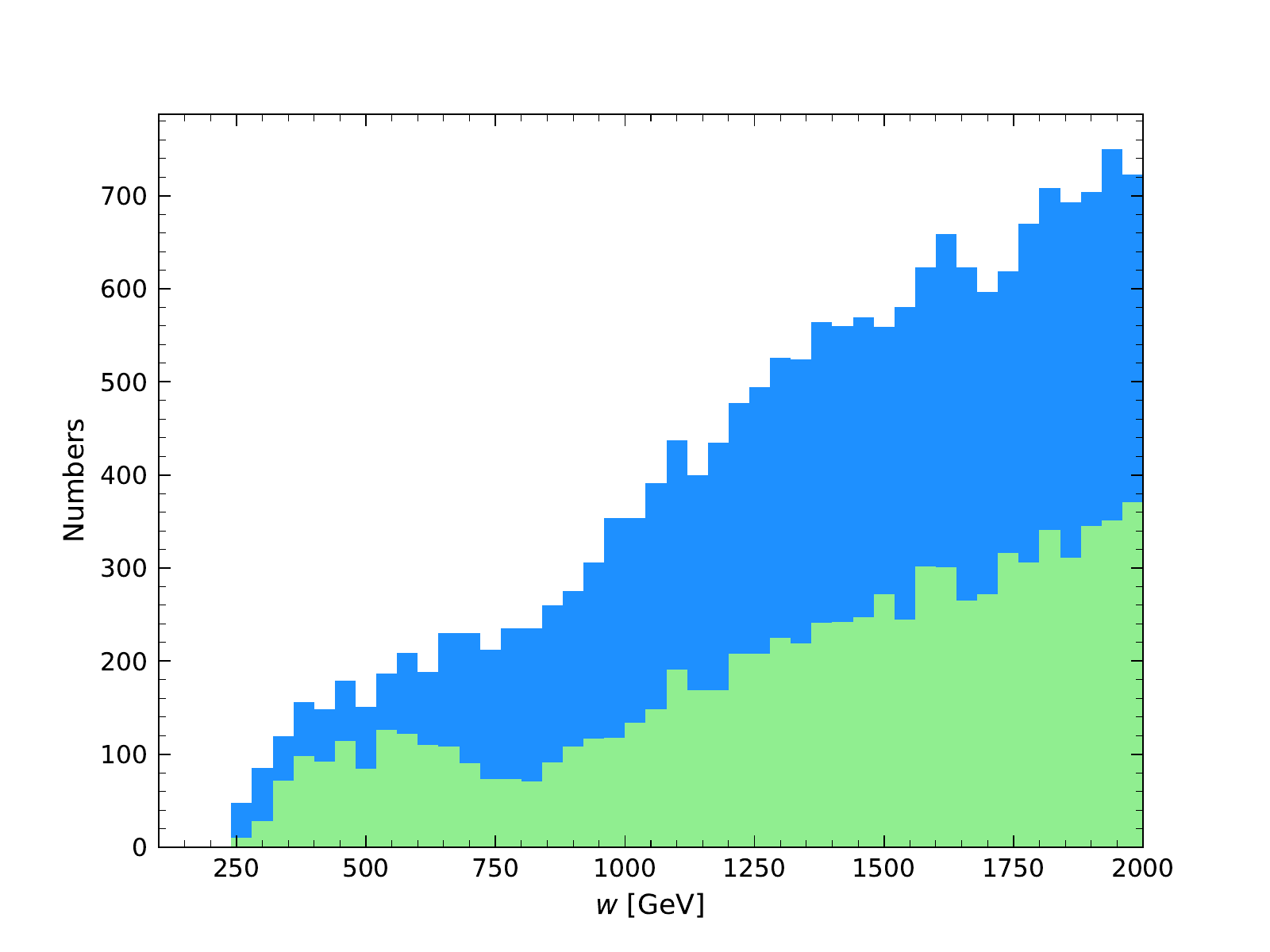}
  \includegraphics[width=75mm,angle=0]{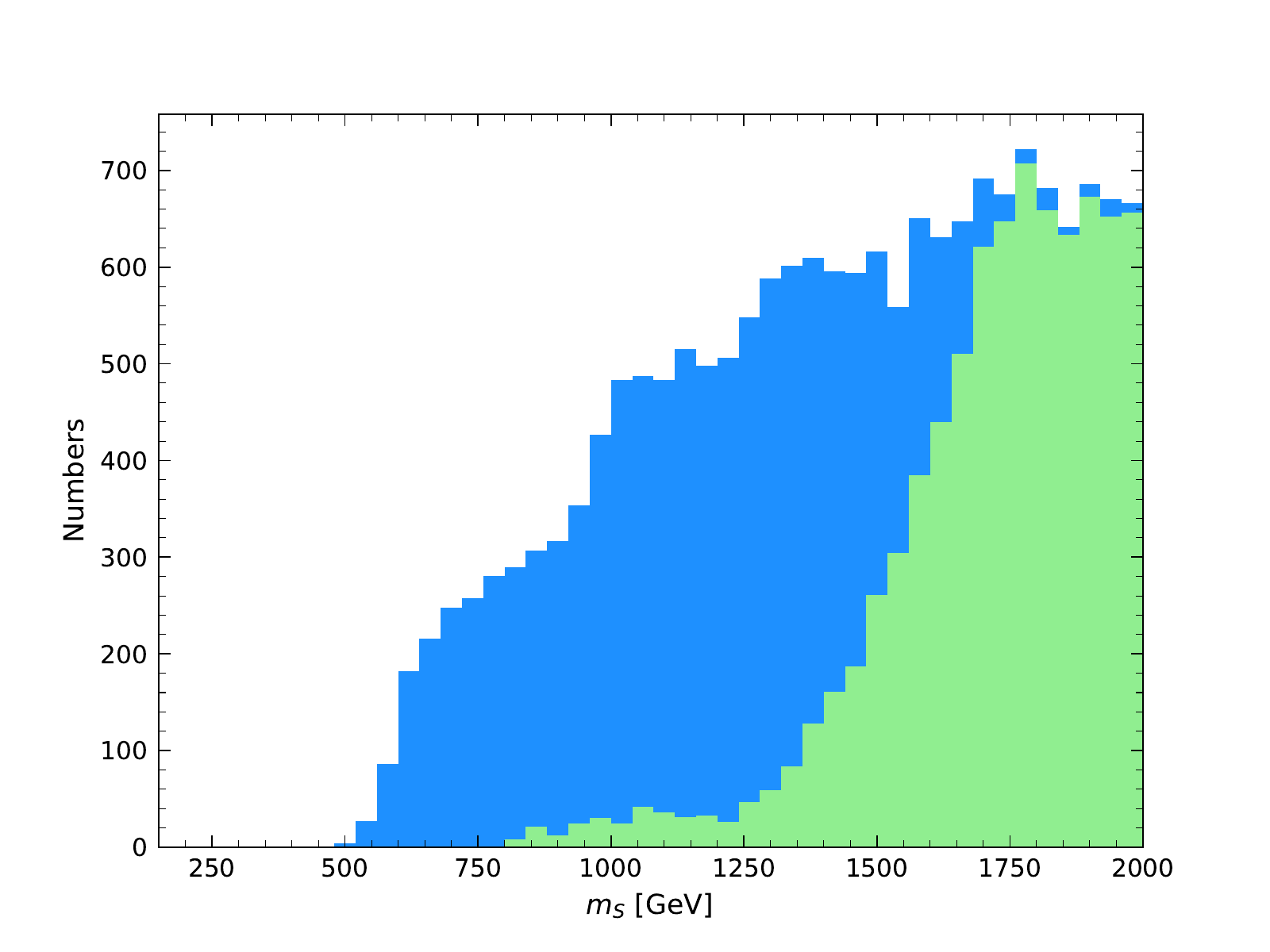}\\
  \includegraphics[width=75mm,angle=0]{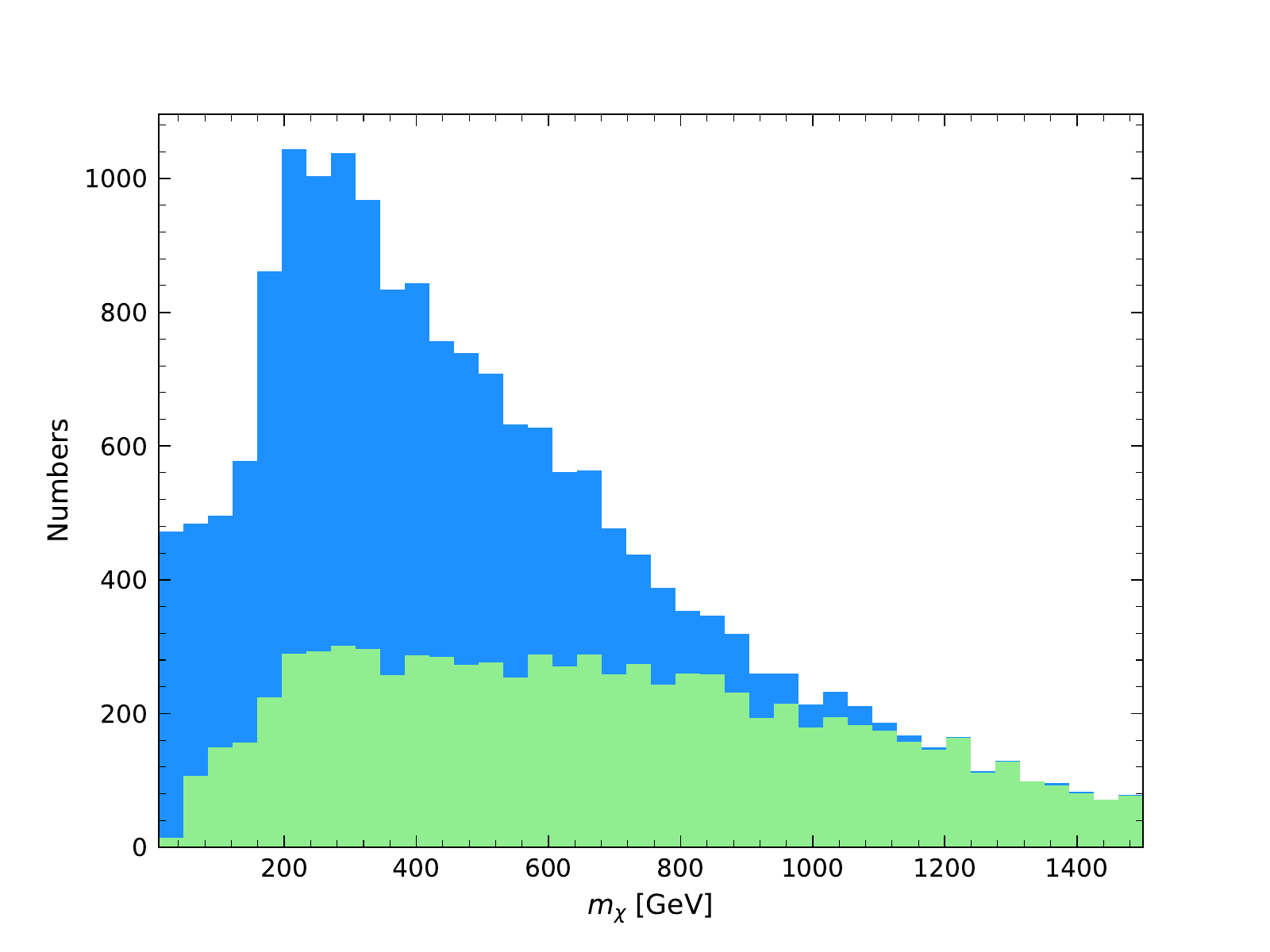}
  \includegraphics[width=75mm,angle=0]{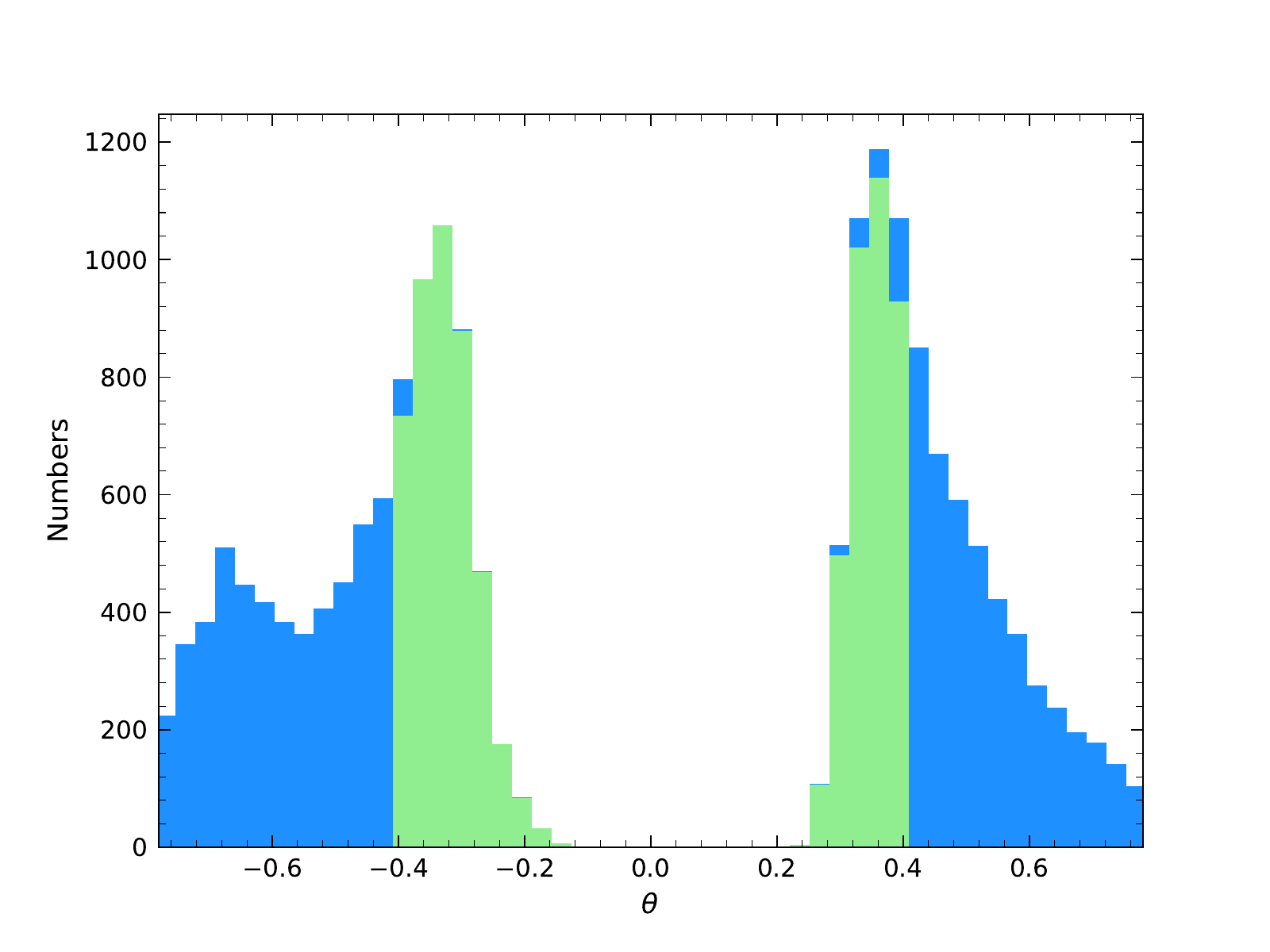}
  \caption{Distributions of the input parameters. The blue histogram represents those samples that can trigger a strong EWPT, the green histogram represents those samples that further survive the experimental constraints.}
  \label{fig:dist2}
\end{figure}

\subsection{Results}

We show our results in Fig.~\ref{fig:dist2} after combining all the collider experimental constraints discussed above.  We find that 8104 (in the green histogram) out of 18047 sample points (in the blue histogram) that can trigger a sufficiently strong EWPT survive the experimental bounds. As discussed above, the most stringent constraints come from the Higgs signal strength measurements.

As shown by the blue histogram, all of the $m_{\mathcal{S}}$ and most of the $m_{\chi}$ sample points that induce a strong EWPT distribute at masses larger than $m_{\mathcal{H}}/2=62.5$~GeV.
According to the above discussions, we see that there is a universal constraint on the mixing angle, 
$\left | \theta \right |\lesssim 0.4$, from the Higgs signal strength measurements.
Here we have chosen the scan range of $m_{\mathcal{S}}\geq 150$~GeV, apparently, the bound on the mixing angle can be much stronger if $m_{\mathcal{S}}$ and/or $m_{\chi}$
is lighter than $m_{\mathcal{H}}/2$. Anyway, the constraint $\left | \theta \right |\lesssim 0.4$ is not dependent on the assumed parameter ranges. Hence, as shown by the lower right plot of Fig.~\ref{fig:dist2}, there is a hard cut-off around $\pm 0.4$ in the distribution of $\theta$.

The peak around $200-400$~GeV in the distribution of $m_{\chi}$ is now removed, and it tends to a flater distribution in the 
range $100-1500$~GeV. While there seems to be no preferred range in the $w$ distribution by the signal strength constraints, the sample points with $m_{\mathcal{S}}\lesssim 1.2$ TeV are strongly disfavored,
as shown in the top two plots. For one thing, the signal strength bounds require a mixing angle 
$\left | \theta \right |\lesssim 0.4$.  Yet samples with large values of mixing angle are mainly associated with a lighter scalar mass $m_{\mathcal{S}}$ due to the perturbative unitarity.
For another reason, the scalar mass $m_{\mathcal{S}}$ should be large enough to induce a strong EWPT when the mixing angle is small.

We summarize our main conclusions obtained in this section below:
\begin{itemize}
  \item The Higgs signal strength measurements give a universal constraint on the mixing angle $\left | \theta \right |\lesssim 0.4$ ($23^{\circ}$).
  \item The mass of heavy scalar $\mathcal{S}$ should be larger than 1.2~TeV 
  from the combined constraints of Higgs signal strength measurements and perturbative unitarity and the requirement of strong EWPT.
  \item Our analysis supports $\left | \theta \right |\gtrsim 0.2$ ($11.5^{\circ}$) for the scalar mass $m_{\mathcal{S}}\lesssim 2$~TeV, which is the requirement of a sufficiently strong EWPT. 
  However, this conclusion depends on the scanning range of scalar mass $m_{\mathcal{S}}$ pre-assumed in this study.
  A strong EWPT for the mixing angle less than $0.2$ might be available if the scanned heavy scalar mass extends beyond 2~TeV (we leave this for future studies). 
\end{itemize}

\section{Dark matter phenomenology}
\label{sec:dmphen}

In this section, we discuss constraints on the model from the observed DM relic density and null direct search result.

\subsection{Dark matter relic density}

As mentioned above, the pseudo-Goldstone boson $\chi$ from the spontaneous symmetry breaking has a $\mathbb{Z}_2$ symmetry, ensuring the stability of the pseudo-Goldstone boson as a DM candidate.  In the standard freeze-out scenario, the DM 
particles are in chemical equilibrium with the other SM particles via annihilation-production reaction in the early Universe.
The DM population becomes nonrelativistic and the annihilations take over the thermal productions with the adiabatic expansion of the Universe.
At around the temperature that the reaction rate falls below the Universe expansion rate, the DM particles begin to decouple from the thermal bath.
The evolution of the DM number density is described by the Boltzmann equation
\begin{equation}
  \frac{d Y}{d T}=\sqrt{\frac{\pi g_{*}(T)}{45}} M_{\rm pl}\left\langle\sigma v_{\mathrm{rel}}\right\rangle
  \left[Y(T)^{2}-Y_{\mathrm{eq}}(T)^{2}\right],
\end{equation}
where the abundance $Y(T)$ denotes the ratio of the DM number density $n_{\chi}$ to the entropy density, 
$M_{\rm{pl}}=1.22 \times 10^{19}~\mathrm{GeV}$ is the Planck mass, $g_{\ast}$ is the effective number of 
relativistic degrees of freedom, $Y_{\rm eq}(T)$ is the thermal equilibrium abundance, 
and $\left\langle\sigma v_{\mathrm{rel}}\right\rangle$ is the relativistic thermally averaged
annihilation cross section. The resulting DM relic density is given by
\begin{equation}
  h^{2}\Omega_{\rm DM}=2.742 \times 10^{8}Y_0 \frac{m_{\chi}}{\mathrm{GeV}},
\end{equation}
where $Y_0$ is the abundance of DM in the present Universe.
In our numerical analysis, we make use of the $\textsf{MicrOMEGAs}\_\textsf{5.0.4}$ package~\cite{Barducci2018CPC} to calculate the DM relic density.

\begin{figure}
  \centering
  \includegraphics[width=75mm,angle=0]{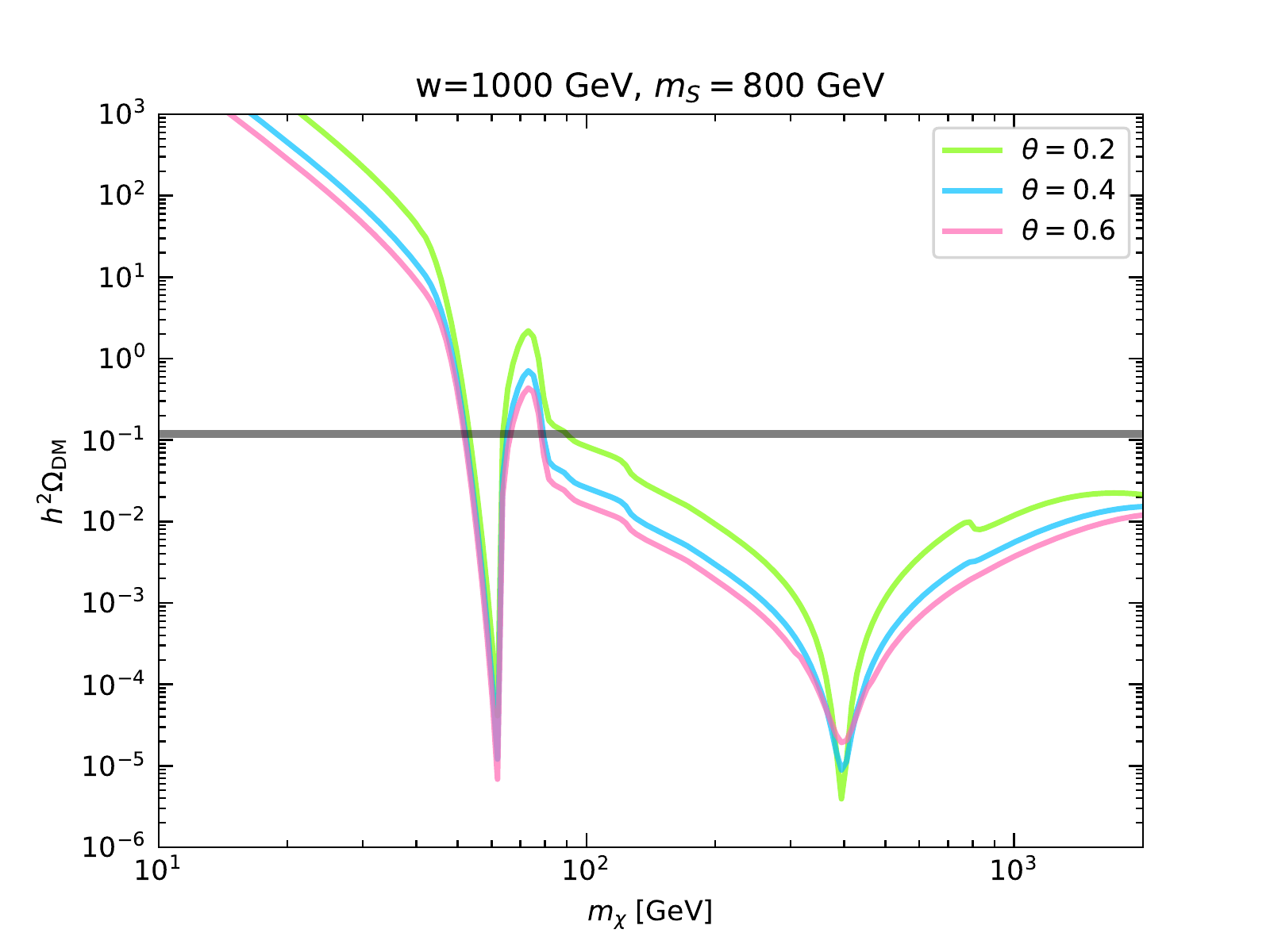}
  \includegraphics[width=75mm,angle=0]{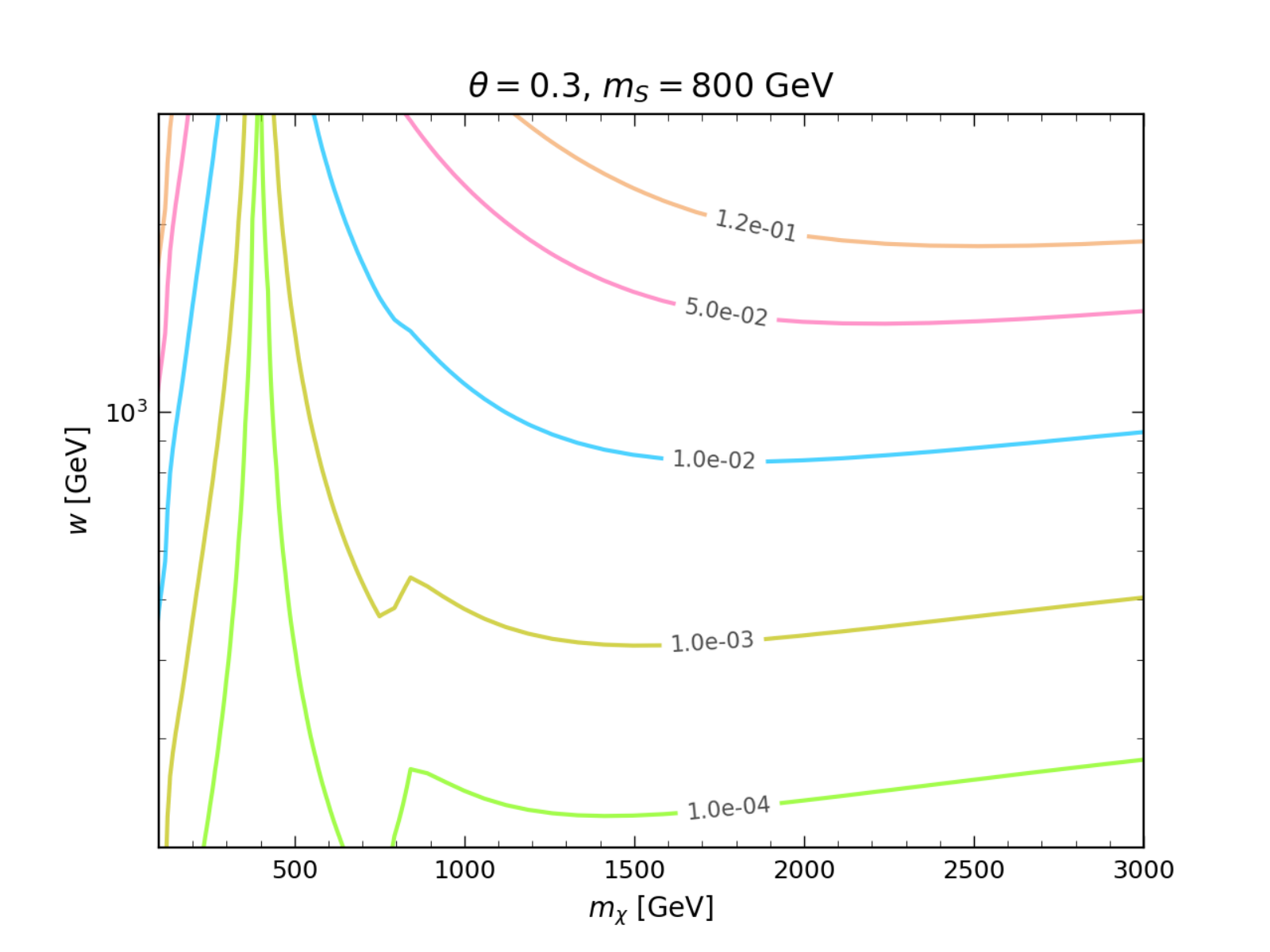}
  \caption{Left plot: DM thermal relic density as a function of the DM mass for $w=1000$~GeV, $m_{\mathcal{S}}=800$~GeV, and
  $\theta=$ 0.2 (green), 0.4 (light blue) and 0.6 (green). The black line denotes the DM relic density given by the Planck satellite's observation of the CMB radiation~\cite{Planck2016}
  Right plot: Contours of the DM thermal relic density in the $m_{\chi}$-$w$ plane for $\theta=0.3$ and
  $m_{\mathcal{S}}=800$~GeV.
  }
  \label{fig:dmrd}
\end{figure}

In the left plot of Fig.~\ref{fig:dmrd}, we plot the DM relic density $h^{2}\Omega_{\rm DM}$ as a function of the DM mass $m_{\chi}$, with the values of $w$ and $m_{\mathcal{S}}$ fixed at 1000~GeV and 800~GeV, respectively.
In our model, the DM annihilation to the SM particles are mediated by the SM-like Higgs boson $\mathcal{H}$ and heavy scalar $\mathcal{S}$.
When $m_\chi\lesssim m_{\mathcal{H}}/2$, its annihilation process is kinematically suppressed, leading to a large value of DM relic density.
The resonant DM annihilation occurs at $m_{\chi}\simeq m_{\mathcal{H}}/2$ and $m_{\mathcal{S}}/2$, which would result in a sharp decrease of the DM relic density, as shown by the two dips in the curves.
In the right plot of Fig.~\ref{fig:dmrd}, we plot the contours of DM relic density in the $m_{\mathcal{S}}$-$w$ plane.
We see that the DM relic density becomes larger as $w$ increases.

\begin{figure}
  \centering
  \includegraphics[width=75mm,angle=0]{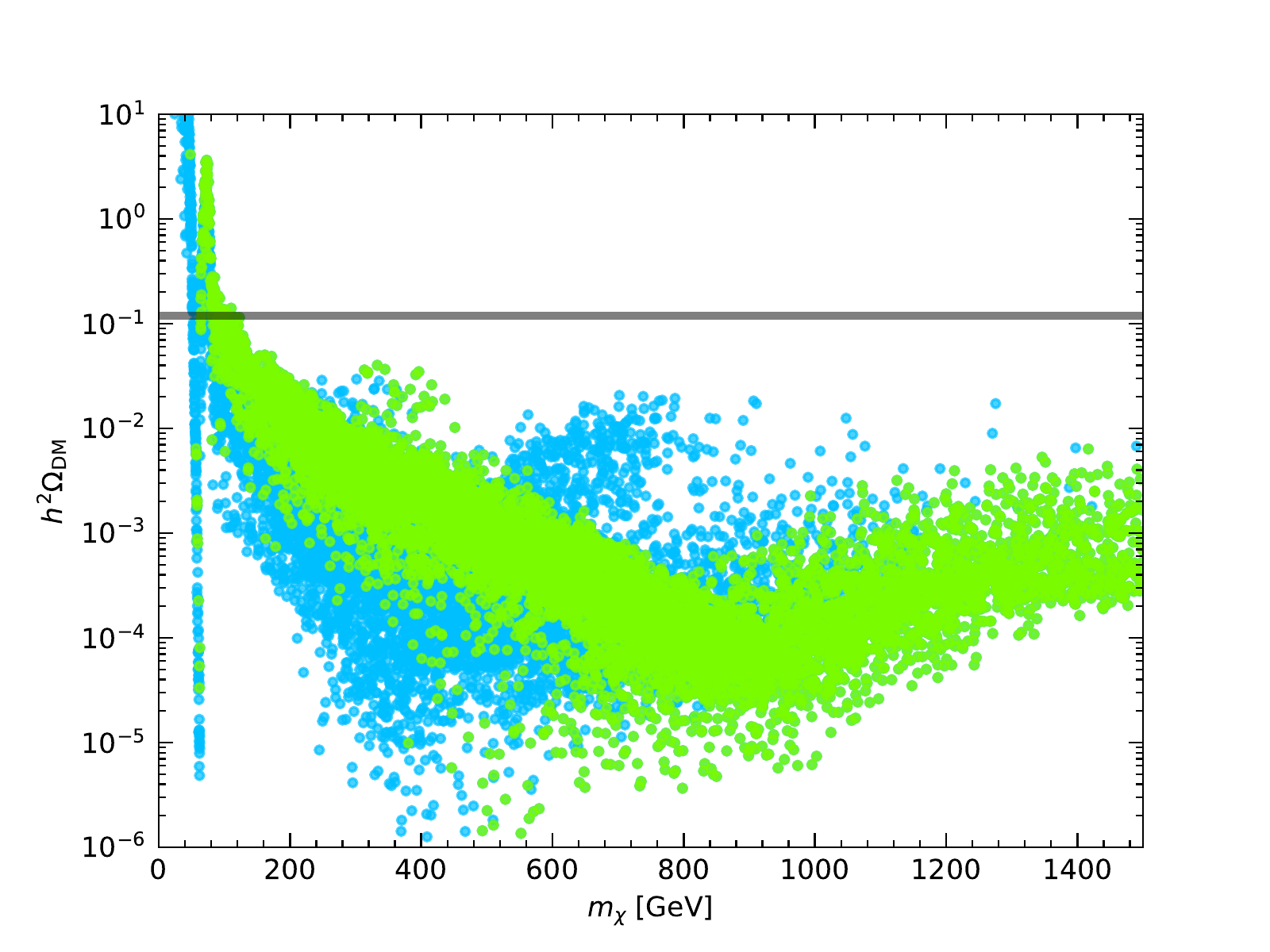}
  \includegraphics[width=75mm,angle=0]{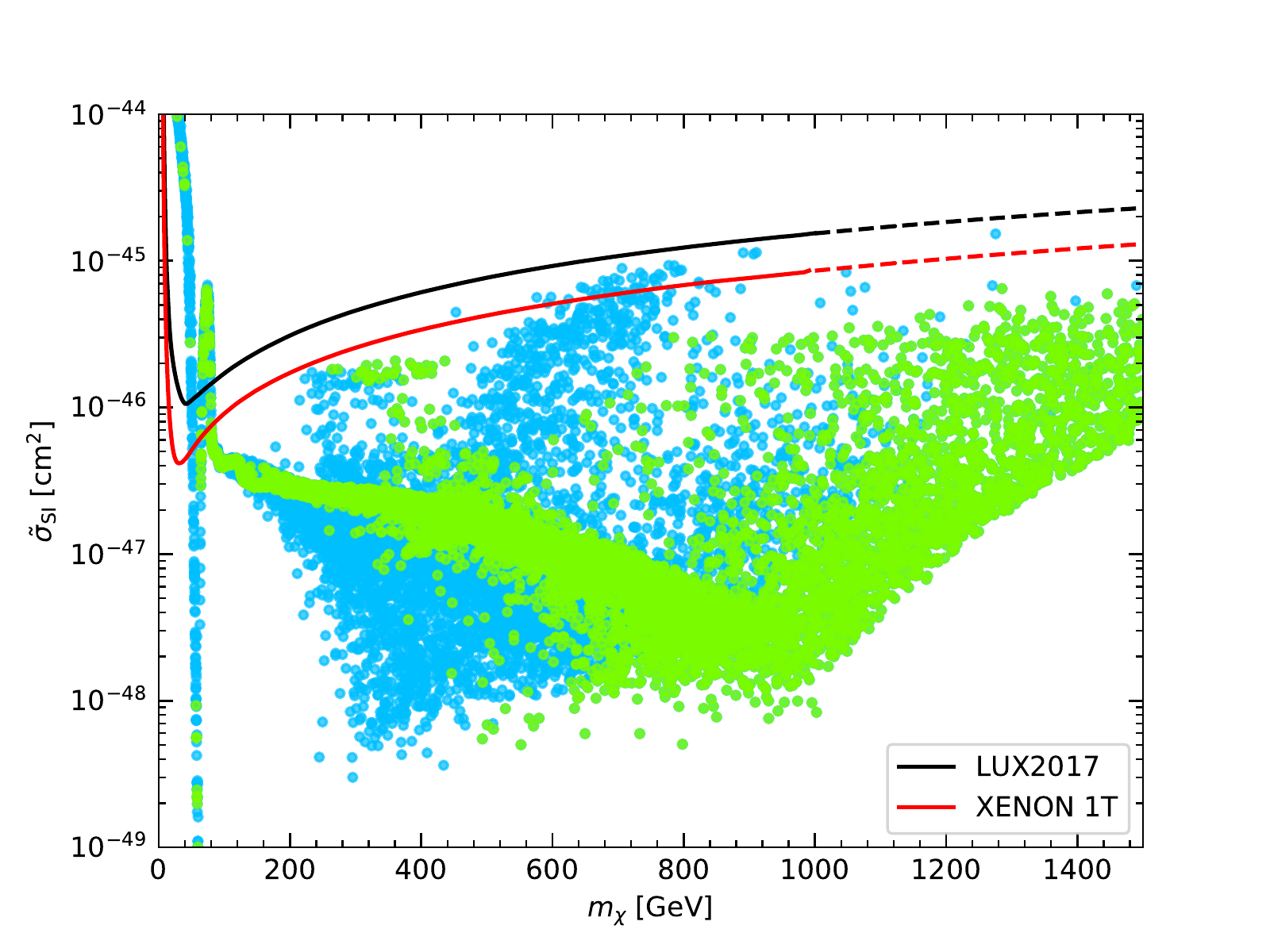}
  \caption{Left plot: DM thermal relic density as a function of the DM mass.
  Right plot: Effective spin-independent DM-nucleon elastic scattering cross section as a function of the DM mass. The black and red curves denote the upper limits on scattering cross section from LUX~\cite{LUX2017PRL} and XENON1T~\cite{XENON1T2018PRL} experiments, respectively.
 In both plots, the blue scatter points represent the samples that have a sufficiently strong EWPT and the green scatter points are those further surviving the experimental constraints.}
  \label{fig:dmphe}
\end{figure}

In the left plot of Fig.~\ref{fig:dmphe}, we calculate the DM relic density of the sample points.
The blue scatter points represent the samples that have a sufficiently strong EWPT and the green scatter points are 
the samples further survive all the experimental constraints. The black line denotes the DM relic density given 
by the Planck satellite's observation of the CMB radiation~\cite{Planck2016}
\begin{equation}
  h^{2}\Omega_{\mathrm{DM}}^{\rm obs}=0.1188 \pm 0.0010.
\end{equation}
It is seen that most of the samples have $h^{2}\Omega_{\mathrm{DM}}$ much below the observed result.  On one hand, 
these sample parameters can evade from a over-closed Universe.  On the other hand, only a small fraction of DM consists of the pseudo-Goldstone boson from spontaneous symmetry breaking in our model.

\subsection{Direct detection}

For DM direct detection, we use the $\textsf{MicrOMEGAs}$ package to compute the spin-independent DM-nucleon elastic scattering cross section
$\sigma_{\rm SI}$.  As shown above, for the parameter space considered here, only a small fraction of DM is made up of the pseudo-Goldstone bosons.
To compare with the experimental upper bounds, we have to scale the scattering cross section as 
\begin{equation}
  \label{eq:effsc}
  \tilde{\sigma}_{\rm SI}=f_{X}\sigma_{\rm SI},
  ~~\mbox{where}~
  f_{X}=\frac{h^2\Omega_{\rm DM}}{0.1188}.
\end{equation}
We simultaneously calculate the DM relic density $h^2\Omega_{\rm DM}$ and scattering cross section $\sigma_{\rm SI}$ with the 
help of $\textsf{MicrOMEGAs}$, then we obtain the effective scattering cross section $\tilde{\sigma}_{\rm SI}$ 
using Eq.~\eqref{eq:effsc}.  We plot our results in the right plot of Fig.~\ref{fig:dmphe}.
Again, the blue scatter points represent the samples that have a sufficiently strong EWPT, and the green scatter points are 
the samples further surviving all the collider constraints. The black curve denotes the upper limit on 
DM-nucleon elastic scattering cross section from the LUX~\cite{LUX2017PRL} experiment, and the red curve from the 
XENON1T~\cite{XENON1T2018PRL} experiment.  They are the most stringent constraints on the spin-independent DM-nucleon scattering cross section up to date. The scattering cross section in our model is suppressed by both a small mixing angle $\theta$ and a large value of $w$.  
It can be seen that our scanned sample points have sustained the most stringent constraints from DM direct detection experiments with most of the scattering cross sections being much below the upper limits from the direct detections.

\section{Gravitational waves}
\label{sec:gw}

A first-order cosmological phase transition can only occur in the presence of a scalar effective potential 
barrier separating the symmetry-broken and -unbroken vacua. 
Although the probability of tunneling from the metastable minimum to the stable one via the instantons is very tiny, 
the decay of the false vacuum can proceed through thermal fluctuations which help overcome the potential barrier.
The tunneling rate per unit volume and time element is approximately given by~\cite{Apreda2002NPB, Espinosa2008PRD}
\begin{eqnarray}
    \Gamma(T)=A(T)e^{-S_{3}/T},
\end{eqnarray}
where $A(T)\simeq [S_3/(2\pi T)]^{3/2}T^{4}$ and $S_3$ denotes the three-dimensional on-shell Euclidean action of a instanton.
Due to the supercooling effect, the onset of the bubble nucleations can be delayed to a temperature $T_{\rm n}$ smaller than 
the critical temperature $T_{\rm c}$.
The nucleation temperature $T_{\rm n}$ is defined to be at which the probability of nucleating one bubble per horizon volume 
is of order one, i.e., $p(T)\sim 1$, where the probability of bubble nucleations per Hubble volume is defined as
\begin{eqnarray}
    p(T)=\int _{T}^{T_{\rm c}}\frac{\Gamma(x)}{H^{4}(x)}\frac{dx}{x}\approx \left( \frac{T}{H} \right)^4e^{-S_{3}/T}.
\end{eqnarray}
In a radiation dominated Universe, the Hubble parameter is given by $H^2=8\pi^{3}g_{\ast}T^4/(90M_{\rm pl}^3)$
and $g_{\ast}\simeq 110$.
The condition $p(T)\sim 1$ guarantees the percolation of bubbles in the early Universe and can be
translated into the following criterion for determining the nucleation temperature~\cite{Espinosa2008PRD}
\begin{eqnarray}
  \label{eq:bnc}
  \frac{S_3(T_{\rm n})}{T_{\rm n}}\simeq 4\ln\left( \frac{T_{\rm n}}{H} \right)\simeq 142-4\log\left( \frac{T_{\rm n}}{246~\rm GeV} \right ).
\end{eqnarray}
For the phase transition at a characteristic temperature $T\sim \mathcal{O}(100~\rm GeV)$, 
the condition above is well approximated by $S_3(T_{\rm n})/T_{\rm n}\simeq 140$.
The successful bubble nucleations at EW scale are guaranteed by Eq.~\eqref{eq:bnc},
which requires a sufficiently large bubble nucleation rate to overcome the expansion rate.  On one hand,
a sufficiently strong EWPT ensures that the washout of baryon asymmetry through sphalerons is suppressed.
On the other hand, a successful bubble nucleation is the requirement of triggering baryogenesis in the EW broken phase.
The latter is typically a more stringent requirement on the model.

The Euclidean action for a spherical bubble configuration can be written as 
\begin{eqnarray}
    S_{3}(T)=4\pi\int dr~r^2\left [ \frac{1}{2}\left ( \frac{d\Phi}{dr} \right )^2+V_{\rm eff}(\Phi,T) \right ].
\end{eqnarray}
By extremizing the Euclidean action, we obtain the following differential equation
\begin{eqnarray}\label{ce}
    \frac{d^2\Phi}{dr^2}+\frac{2}{r}\frac{d\Phi}{dr}-\frac{dV}{d\Phi}=0,
\end{eqnarray}
with the boundary conditions
\begin{eqnarray}
    \left.\begin{matrix}\frac{d\Phi}{dr} \end{matrix}\right|_{r=0}=0,~\left.\begin{matrix}\Phi \end{matrix}\right|_{r\to\infty}=0.
\end{eqnarray}
If $\Phi(r)$ represents the profile of a particle in the potential $V$, then Eq.~(\ref{ce}) can be treated as the classical equation of motion, which can be solved by the traditional overshooting/undershooting method~\cite{Apreda2002NPB}. 
In this work, we employ the $\textsf{CosmoTransitions~2.0.2}$ package~\cite{Wainwright2012PLB} to perform the numerical calculations of 
the bubble profile and Euclidean action.  Afterwards, we use Eq.~\eqref{eq:bnc} to determine the nucleation temperature $T_{n}$.

\subsection{Gravitational wave parameters}

It has been shown that the stochastic gravitational waves (GW's) produced from a cosmological phase transition can be fully characterized by the 
knowledge of two primary parameters~\cite{Kamionkowski1994PRD}, which are defined as
\begin{eqnarray}
    \label{eq:ab}
    \alpha =\frac{\epsilon (T_{\ast})}{\rho _{\rm rad}(T_{\ast})}~ {\rm and}~\frac{\beta}{H_{\ast}}=T_{\ast}\frac{d}{dT}\left.
    \left ( \frac{S_3(T)}{T} \right )\right|_{T=T_{\ast}},
\end{eqnarray}
where $T_{\ast}$ is the GW generation temperature, $\rho_{\rm rad}=\pi^2g_{\ast}T^4/30$ is
the radiation energy density in the plasma, and the latent heat associated with the phase transition is given by
\begin{eqnarray}
    \epsilon (T)=T\frac{\partial \Delta V_{bs}(T)}{\partial T}- \Delta  V_{bs}(T),
\end{eqnarray}
where $\Delta V_{bs}(T)\equiv V_{\rm eff}(v(T),~w(T),~T)-V_{\rm eff}(0,~w_0(T),~T)$ is the potential difference between the 
broken phase and the symmetric phase at temperature $T$. 
Therefore, the parameter $\alpha$ is related to the maximum available energy budget for gravitational wave emissions.
The parameter $\beta$ represents the rate of time variation of the nucleation rate, whose inverse gives the duration of the bubble nucleation. 
Consequently, $\beta/H_{\ast}$ defines the characteristic frequency of the GW spectrum produced from the phase transition.

In addition to the GW parameters $\alpha$ and $\beta/H_{\ast}$ and the nucleation temperature $T_{n}$,
the GW spectrum also depends on the bubble wall velocity $v_{w}$, which is the expanding speed of the true vacuum.
It has been pointed out in Refs.~\cite{Espinosa2010JCAP, No2011PRD} that it is the relative velocity between the wall and the plasma in the front ($v_+$) rather than $v_w$ that should be used in the calculations of EW baryogenesis. For a strong EWPT, 
the condition $v_+\ll v_w$ could be achieved, making it possible to generate a viable EW baryogenesis and a loud GW signal 
in the same scenario. In this work, the bubble velocity $v_w$ is simply assigned to be around 1 for the calculations of GW spectra.

\begin{figure}
  \centering
  \includegraphics[width=75mm,angle=0]{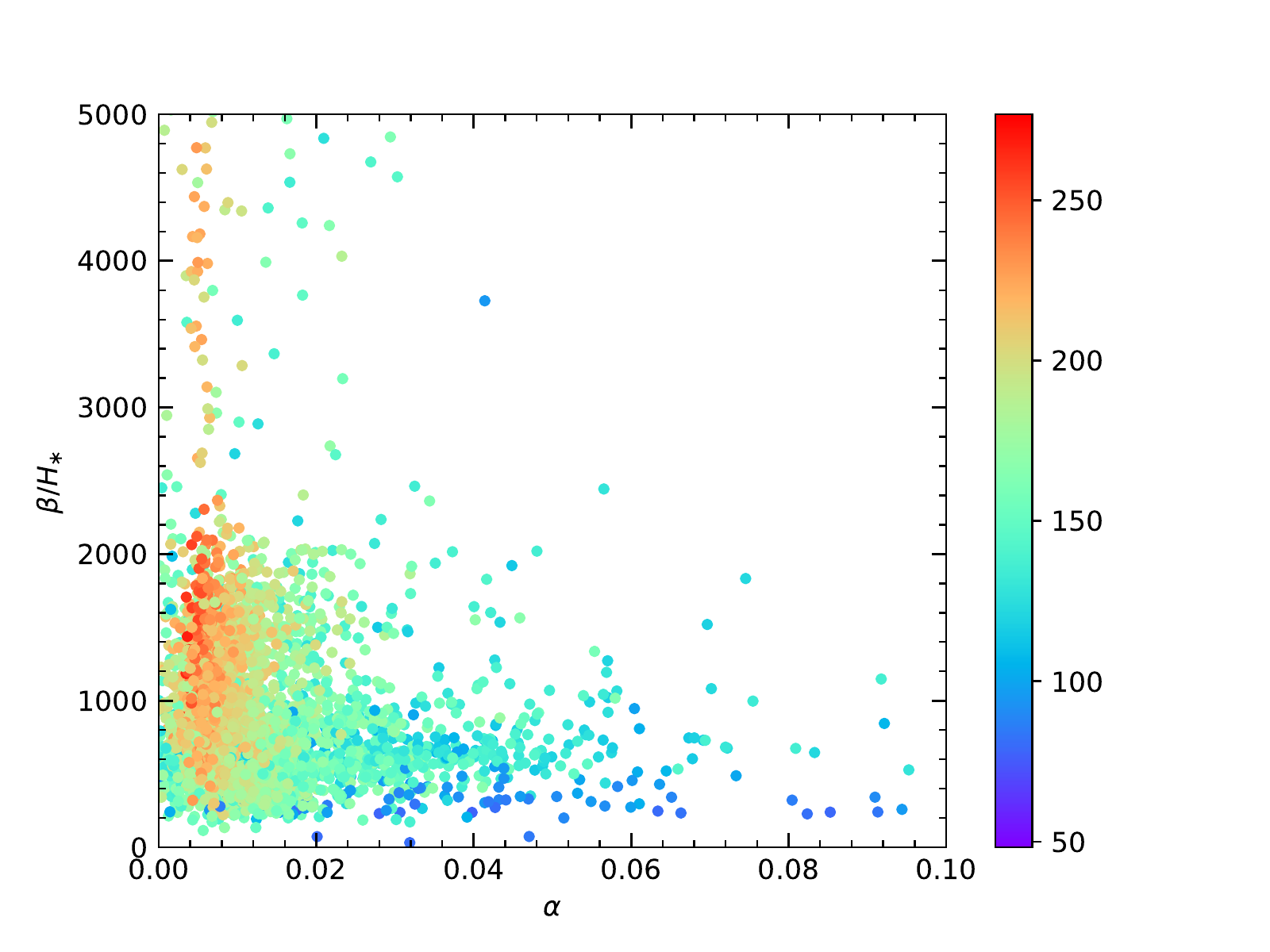}
  \includegraphics[width=75mm,angle=0]{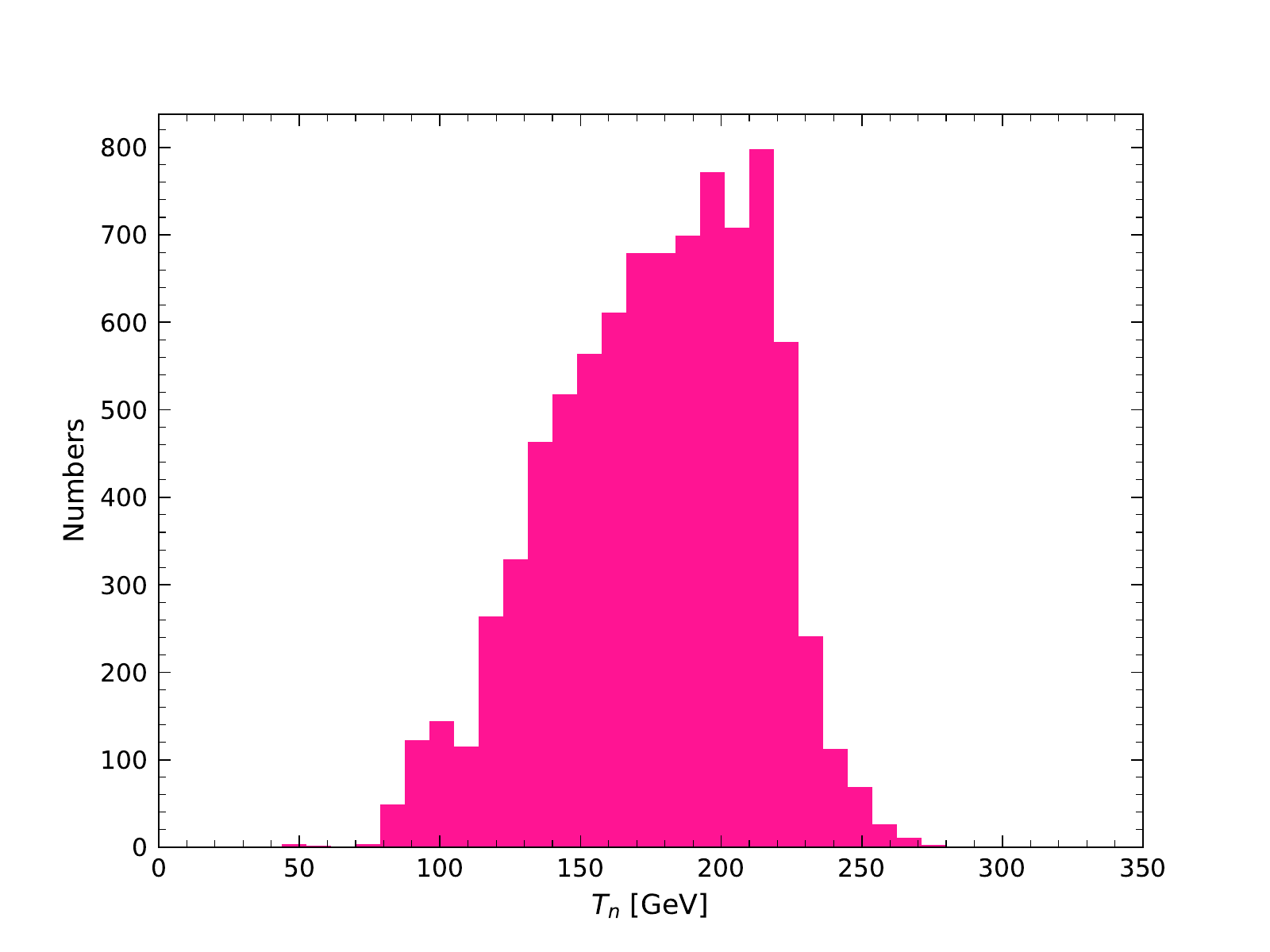}
  \caption{Left plot: Distributions of the GW parameters $\alpha$ and $\beta/H_{\ast}$.  The colored bar indicates
  the nucleation temperature $T_n$.
  Right plot: Distribution of nucleation temperature $T_n$.
  }
  \label{fig:ab}
\end{figure}

We show the calculation results of parameters $\alpha$ and $\beta/H_{\ast}$ in the left plot of Fig.~\ref{fig:ab}, the colored bar indicates the nucleation temperature $T_n$. The distribution of $T_n$ are given in the right plot of Fig.~\ref{fig:ab}.
We find that 8564 out of 18047 sample points satisfy the bubble nucleation condition in Eq.~\eqref{eq:bnc}.
Note that in the calculations of $\beta/H_{\ast}$ with Eq.~\eqref{eq:ab}, we have assumed that reheating 
is not significant for typical transitions.  In this case, the temperature for GW generation $T_{\ast}$ is approximately equivalent to the nucleation temperature $T_n$, i.e., $T_{\ast}\simeq T_{n}$.
A strong supercooling could not only enhance the strength of the phase transition, but also change the evolution of the Universe since the expansion of the Universe would be dominated by vacuum energy in the supercooled phase, instead of radiation.
In this case, there is a lower bound on the temperature of the transition to ensure the successful bubble percolation and completion of the EWPT~\cite{Ellis2019JCAP}.

\subsection{Gravitational wave spectrum}

In what follows we review the three processes that are involved in the production of GW's during 
a first-order phase transition (see Refs.~\cite{Caprini2016JCAP, Cai2017JCAP} and references therein for details):
\begin{itemize}
\item {\bf Collisions of bubble walls and shocks} in the plasma. GW's produced from this process depends only on the dynamics of the scalar field.
The ``envelope approximation'' is used in the numerical simulations to estimate the GW spectrum, given by~\cite{Huber2008JCAP}
(analytical calculations of the GW spectrum from this process can be found in Ref. \cite{Jinno2017PRD})
\begin{equation}
  \label{eq:gwspt1}
  h^{2} \Omega_{\mathrm{\rm col}}(f)=1.67 \times 10^{-5}\left(\frac{H_{*}}{\beta}\right)^{2}\left(\frac{\kappa_{\rm col}\alpha}
  {1+\alpha}\right)^{2}\left(\frac{100}{g_{*}}\right)^{\frac{1}{3}}\left(\frac{0.11 v_{w}^{3}}{0.42+v_{w}^{2}}\right) S_{\mathrm{\rm col}}(f).
\end{equation}
\item {\bf Sound waves} in the plasma generated subsequently after the bubble collisions. Numerical
simulations indicate that the durations of sound waves and turbulence as active sources of GW's are typically much longer than the collisions of the bubble walls. This process contributes a GW spectrum desribed by~\cite{Hindmarsh2015PRD}
\begin{equation}
  \label{eq:gwspt2}
  h^{2} \Omega_{\mathrm{sw}}(f)=2.65 \times 10^{-6}\left(\frac{H_{*}}{\beta}\right)\left(\frac{\kappa_{\rm sw} \alpha}
  {1+\alpha}\right)^{2}\left(\frac{100}{g_{*}}\right)^{\frac{1}{3}} v_{w} S_{\mathrm{sw}}(f).~~~~~~~~~~~~~~~
\end{equation}
\item {\bf Turbulence} in the plasma formation after the bubble collisions. Simulations show that only a small fraction $\epsilon \sim 5-10\%$ of the bulk motion from the bubble walls is converted into turbulence. However, GW's produced from this process could play a dominant role at high frequencies, as the GW signals from sound waves decay much faster.  The GW spectrum from turbulence can be parameterized as~\cite{Caprini2009JCAP}
\begin{equation}
  \label{eq:gwspt3}
  h^{2} \Omega_{\mathrm{turb}}(f)=3.35 \times 10^{-4}\left(\frac{H_{*}}{\beta}\right)\left(\frac{\kappa_{\mathrm{turb}} \alpha}
  {1+\alpha}\right)^{\frac{3}{2}}\left(\frac{100}{g_{*}}\right)^{1 / 3} v_{w} S_{\mathrm{turb}}(f).~~~~~~~~~
\end{equation}
\end{itemize}
The efficiency factors $\kappa_{\rm col}$, $\kappa_{\rm sw}$, and $\kappa_{\rm turb}$ indicate respectively the fractions of latent heat that are
transformed into the kinetic energy of bubbles, the bulk motion of the plasma, and the turbulence and finally into GW's.  Thus they are functions of $\alpha$. 
The total stochastic GW spectrum is approximately given by adding up these three contributions:
\begin{equation}
  h^{2} \Omega_{\mathrm{GW}} \simeq h^{2} \Omega_{\rm col}+h^{2} \Omega_{\mathrm{sw}}+h^{2} \Omega_{\mathrm{turb}}.
\end{equation}

The spectral shapes in Eqs.~\eqref{eq:gwspt1}-\eqref{eq:gwspt3} are given by
\begin{align}
\begin{split}
  S_{\mathrm{col}}(f)&= \frac{3.8\left(f / f_{\mathrm{col}}\right)^{2.8}}{1+2.8\left(f / f_{\mathrm{col}}\right)^{3.8}},\\
  S_{\mathrm{sw}}(f)&= \left(f / f_{\mathrm{sw}}\right)^{3}\left(\frac{7}{4+3\left(f / f_{\mathrm{sw}}\right)^{2}}\right)^{7 / 2},\\
  S_{\mathrm{turb}}(f)&= \frac{\left(f / f_{\mathrm{turb}}\right)^{3}}{\left[1+
  \left(f / f_{\mathrm{turb}}\right)\right]^{\frac{11}{3}}\left(1+8 \pi f / H_{0}\right)},
\end{split}
\end{align}
where the red-shifted Hubble constant observed today is given by
\begin{equation}
  H_{0}=16.5 \times 10^{-3} \mathrm{mHz}\left(\frac{T_{*}}{100 \mathrm{GeV}}\right)\left(\frac{g_{*}}{100}\right)^{\frac{1}{6}}.
\end{equation}
The frequency $f_{\ast}$ with respect to the Hubble scale at the nucleation temperature $T_{\ast}$ is red-shifted to the 
frequency $f$ today by $f_{\ast}/H_{\ast}=f/H_0$. We write out the red-shifted peak frequency today as follows:
\begin{align}
\begin{split}
  f_{\mathrm{col}}&= 16.5 \times 10^{-3}~\mathrm{mHz}\left(\frac{0.62}{1.8-0.1 v_{w}+v_{w}^{2}}\right)
  \left(\frac{\beta}{H_{*}}\right)\left(\frac{T_{*}}{100 \mathrm{GeV}}\right)\left(\frac{g_{*}}{100}\right)^{\frac{1}{6}},
\\
  f_{\mathrm{sw}}&= 1.9 \times 10^{-2}~\mathrm{mHz} \frac{1}{v_{w}}\left(\frac{\beta}{H_{*}}\right)\left(\frac{T_{*}}{100 \mathrm{GeV}}\right)
  \left(\frac{g_{*}}{100}\right)^{\frac{1}{6}},
\\
  f_{\text {turb }}&= 2.7 \times 10^{-2}~\mathrm{mHz} \frac{1}{v_{w}}\left(\frac{\beta}{H_{*}}\right)
  \left(\frac{T_{*}}{100 \mathrm{GeV}}\right)\left(\frac{g_{*}}{100}\right)^{\frac{1}{6}}.
\end{split}
\end{align}

There is a critical phase transition strength $\alpha_{\infty}$ for the phase transition, which is 
estimated as 
\begin{equation}
  \alpha_{\infty} \simeq \frac{30}{24 \pi^{2}} \frac{\sum_{i} n_{i} \Delta M_{i}^{2}\left(\Phi_{*}\right)}{g_{*} T_{*}^{2}},
\end{equation}
where $n_{i}$ is equal to $N_{i}$ for bosons and $N_{i}/2$ for fermions, with $N_i$ already given in section~\ref{sec:ewpt}.
$\Delta M_{i}^{2}\left(\Phi_{*}\right)$ is the difference of the field-dependent squared masses between the symmetric and broken phases.
According to $\alpha_{\infty}$, the phase transition can be divided into two cases~\cite{Caprini2016JCAP}:
\begin{itemize}
  \item Case 1. Non-runaway bubbles, $\alpha<\alpha_{\infty}$. In this case, the bubble walls will reach a terminal velocity and the latent energy transferred into the scalar field is negligible, i.e., $\kappa_{\rm col}\simeq 0$. 
  The efficiency factor for the sound wave contribution is then given by
  \begin{equation}
    \label{eq:ksw}
    \kappa_{\mathrm{sw}}\simeq \frac{\alpha}{0.73+0.083 \sqrt{\alpha}+\alpha},~~{\rm for}~~v_w\sim 1.
  \end{equation}
  The efficiency factor for turbulence $\kappa_{\rm turb}$ is related to $\kappa_{\rm sw}$ by $\kappa_{\rm turb}=\epsilon\kappa_{\rm sw}$,
  where we take $\epsilon=0.1$ in this work.
  \item Case 2. Runaway bubbles, $\alpha>\alpha_{\infty}$. In this case, the bubble walls can accelerate continuously and finally run away.
  The fraction $\kappa_{\rm col}=1-\alpha_{\infty} / \alpha$ of the total latent energy goes into accelerating the bubble wall, and
  the other fraction $\alpha_{\infty} / \alpha$ of the latent energy is transformed into bulk motion and thermal energy.  The efficiency factor
  $\kappa_{\mathrm{sw}}=\kappa\left(\alpha_{\infty}\right) \alpha_{\infty} / \alpha$, where $\kappa(\alpha_{\infty})$ is calculated using Eq.~\eqref{eq:ksw}.
\end{itemize}

A recent study~\cite{Bodeker2017JCAP} suggests that although bubbles may runaway in certain cases, most of their energy is dissipated into the surrounding plasma and very little energy is deposited in the bubble walls.  This in turn leads to a negligible contribution to the GW spectrum from bubble collisions.  Hence, we only take the GW spectra produced by sound waves and turbulence into account in our calculations.

\subsection{Space-based interferometers}

The frequentist approach is normally used for the experimental investigation of the stochastic GW signals from EWPT, where the detectability of the signals is measured by the corresponding signal-to-noise ratio (SNR)~\cite{Caprini2016JCAP, Breitbach2019JCAP}
\begin{equation}
  \rho=\sqrt{\mathcal{N}\mathcal{T}_{\rm obs} \int_{f_{\mathrm{min}}}^{f_{\mathrm{max}}} 
  df\left[\frac{h^{2} \Omega_{\mathrm{GW}}(f)}{h^{2} \Omega_{\mathrm{exp}}(f)}\right]^{2}},
\end{equation}
where $h^2\Omega_{\mathrm{exp}}$ denotes the sensitivity of a GW experiment, $\mathcal{N}$ is the number of independent observatories 
of the experiment, and $\mathcal{T}_{\rm obs}$ is the duration of the mission in units of year.

The peak frequency of GW spectrum produced from the EWPT is red-shifted to around the milli-Hertz band, which
falls right within the range of future space-based GW interferometers. The planned space-based GW experiments considered in this 
work include the LISA~\cite{LISA2017} interferometer as well as the proposed successors B-DECIGO~\cite{Seto2001PRL}, 
DECIGO~\cite{Sato2017JPCS}, and BBO~\cite{Crowder2005PRD}. For the auto-correlated experiments LISA and B-DECIGO, $\mathcal{N}=1$, while  $\mathcal{N}=2$ for the cross-correlated experiments DECIGO and BBO. 
Following Ref.~\cite{Breitbach2019JCAP}, we assume a mission duration of $\mathcal{T}_{\rm obs}=4$~years 
for all of the experiments. The SNR threshold value $\rho_{\rm thr}$, above which the GW signal is detectable for the experiment, and the other detector parameters are summarized in table~\ref{tab:sbi}. The experimental sensitivities are summarized in appendix~\ref{apd:ssbi}.

\begin{table}[tbp]
  \renewcommand\arraystretch{1.5}
  \centering
  \begin{tabular}{|l|l|c|c|c|c|c|}
  \hline
  Experiment & Frequency range & $\rho_{\rm thr}$ & $\mathcal{N}$ & $\mathcal{T}_{\rm obs}$ [yrs] & Refs. \\
  \hline 
  LISA & $10^{-5}-1~{\rm Hz}$ & 10 & 1 & 4 &~\cite{LISA2017, Cornish2018} \\
  B-DECIGO & $10^{-2}-10^{2}~{\rm Hz}$ & 8 & 1 & 4 &~\cite{Seto2001PRL, Isoyama2018} \\
  DECIGO & $10^{-3}-10^{2}~{\rm Hz}$ & 10 & 2 & 4 &~\cite{Sato2017JPCS, Yagi2011PRD, Yagi2013IJMPD} \\
  BBO & $10^{-3}-10^{2}~{\rm Hz}$ & 10 & 2 & 4 &~\cite{Crowder2005PRD, Yagi2011PRD, Yagi2013IJMPD} \\
  \hline
  \end{tabular}
  \caption{\label{tab:sbi} A summary of parameters and assumptions used for the planned space-based interferometers.}
\end{table}

\begin{figure}
  \centering
  \includegraphics[width=75mm,angle=0]{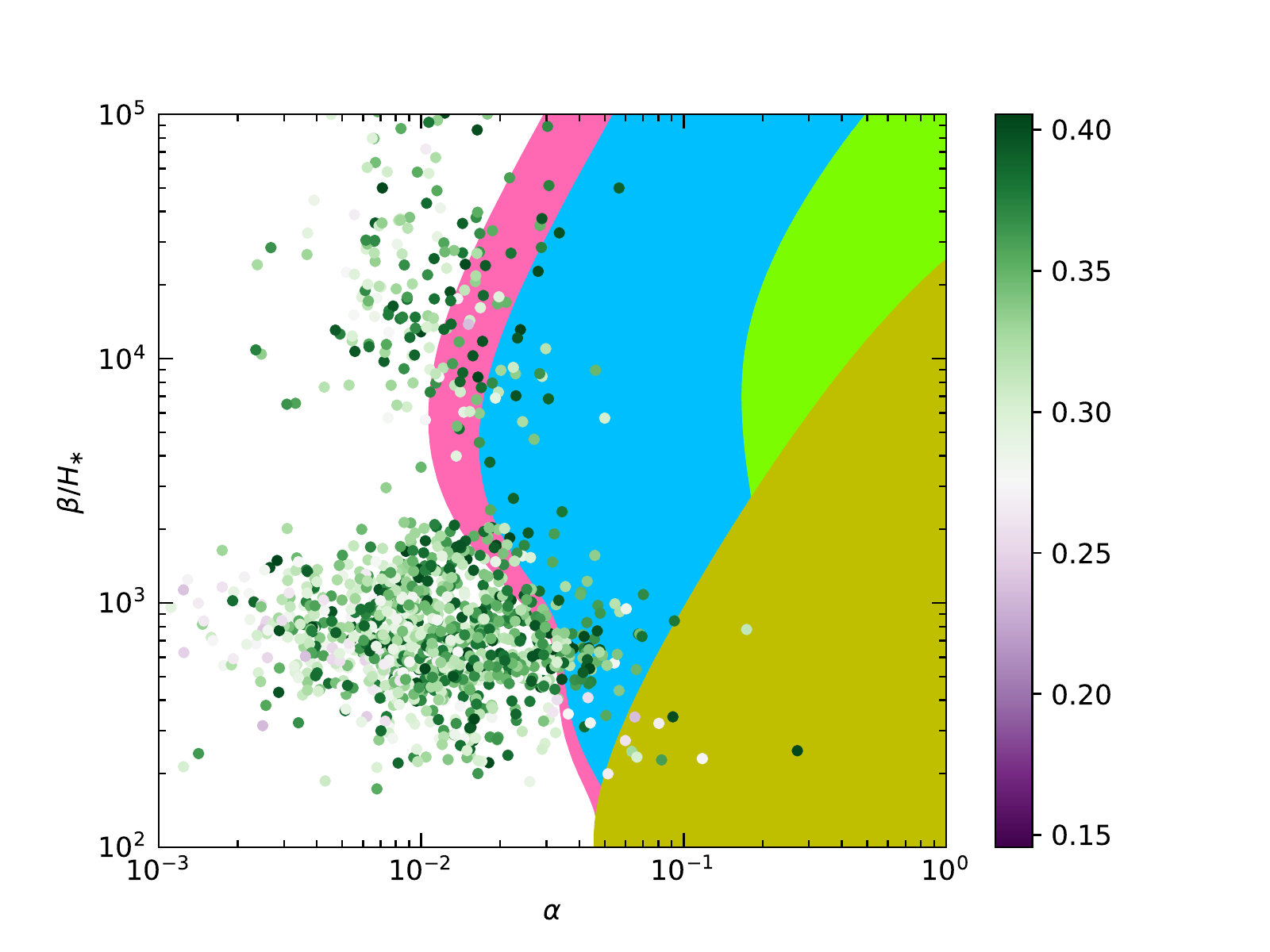}
  \includegraphics[width=75mm,angle=0]{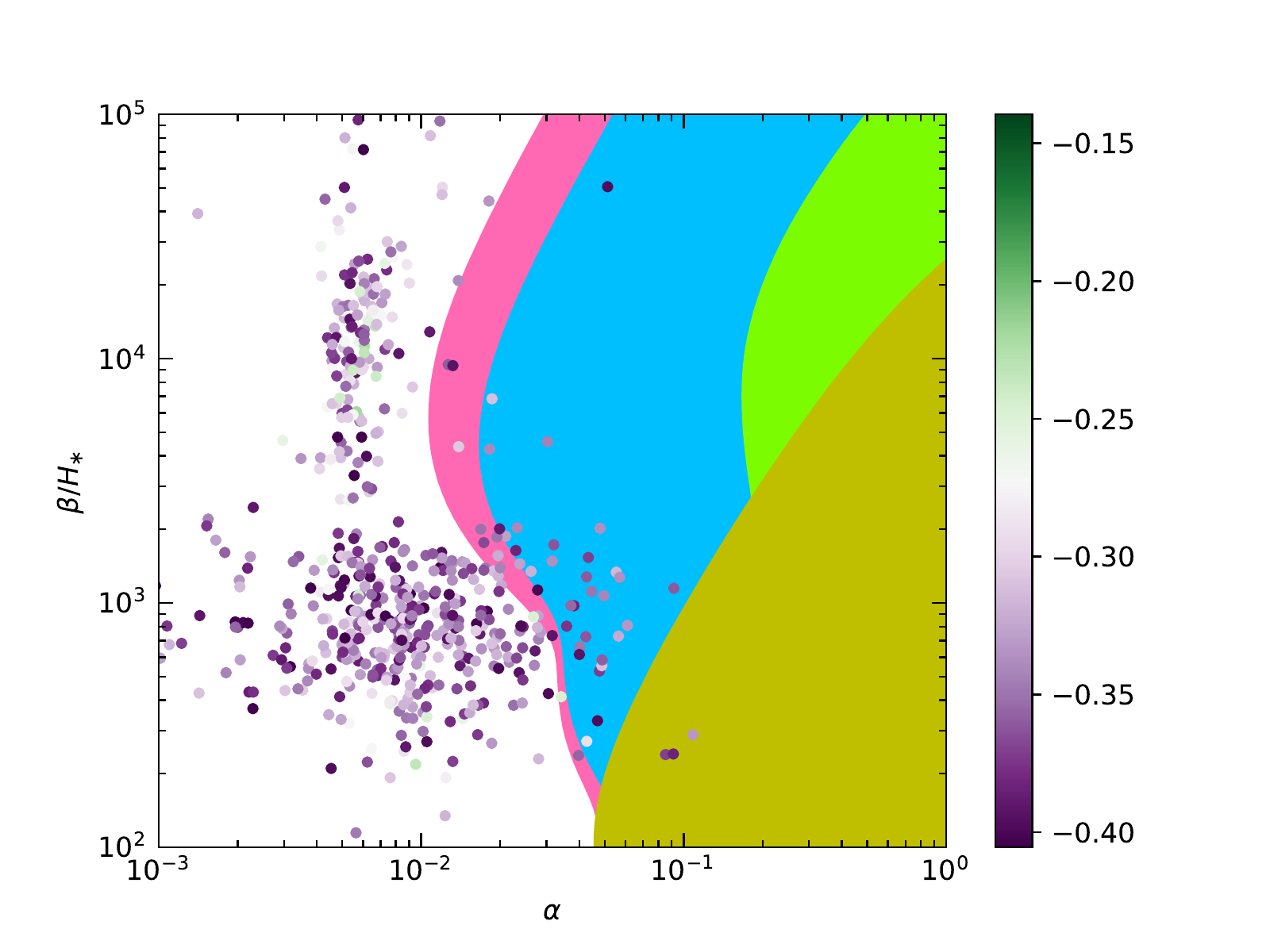}
  \caption{The pink, blue, yellow, and green regions represent respectively the sensitivities of BBO, DECIGO, LISA, and B-DECIGO exceeding 
  the detection threshold, assuming the nucleation temperature $T_{n}=200$~GeV. 
  The scatter points are the samples that both generate a strong first-order EWPT and survive 
  the collider searches and DM experiments.  The colored bar shows the values of $\theta$. 
  The samples with $\theta>0$ are plotted in the left plot, while the samples with $\theta<0$ are plotted in the right plot.}
  \label{fig:gwab}
\end{figure}

In Fig.~\ref{fig:gwab}, the GW experimental sensitivities are shown in the $\alpha$-$\beta/H_{\ast}$ plane by the colored regions,
in which the SNR of a given GW observatory exceeds its detection threshold.
We have assumed a nucleation temperature $T_{n}=200$~GeV for the determination of these regions.
The scatter points are the samples that both generate a strong first-order EWPT and survive the collider searches and DM experiments. 
The colored bar shows values of the mixing angle $\theta$.  The samples with $\theta>0$ ($\theta<0$) are plotted in the left (right) plot of Fig.~\ref{fig:gwab}).
As illustrated by this figure, a considerable portion of sample points with a mixing angle $\left | \theta \right |$ 
in the range $0.25-0.4$ could be detected by BBO and DECIGO. A few samples have extended into the region in which 
the GW signal is large enough to be detected by the LISA experiment.
We thus expect that the EWPT scenario depicted in this work will be tested by the future space-based interferometers.

\section{Gauge dependence}
\label{sec:gaugedepend}

In this section, we scrutinize the issue of gauge dependence in our conclusions drawn above.
It has been pointed out that the SM one-loop effective potential depends on the gauge parameter $\xi$ in the $R_{\xi}$ gauge
due to the thermal corrections to the masses.
Ref.~\cite{Patel2011JHEP} proposed a method to determine the gauge-independent $v_c$ and $v_c/T_c$ in the high-temperature approximation.
With this method, a gauge-invariant, perturbative computation of $v_c$ and $v_c/T_c$ can be made by retaining only the quadratic 
temperature-dependent terms in the effective potential (see Refs.~\cite{Farakos1994NPB,Braaten1995PRD,Kajantie1996NPB,Brauner2017JHEP,Gould2019PRD} for a systematic treatment of the high-temperature effective theory for a number 
of SM extensions).

Here we adopt another gauge-invariant approach introduced in Ref.~\cite{Katz2015PRD}, which is appropriate for the scenarios in which gauge degrees of freedom play a subdominant role in the generation of potential barrier~\cite{Katz2015PRD}, as in our singlet extension of SM. Following this approach, we truncate the one-loop effective potential $V_{\rm eff}(T)$ at the second order in the EW gauge couplings but include terms to all orders in the new couplings. Through this procedure, we obtain the gauge-independent effective
potential by eliminating the gauge-dependent terms that first arise at $\mathcal{O}(g^3)$ \cite{Katz2015PRD}.

\begin{figure}
  \centering
  \includegraphics[width=75mm,angle=0]{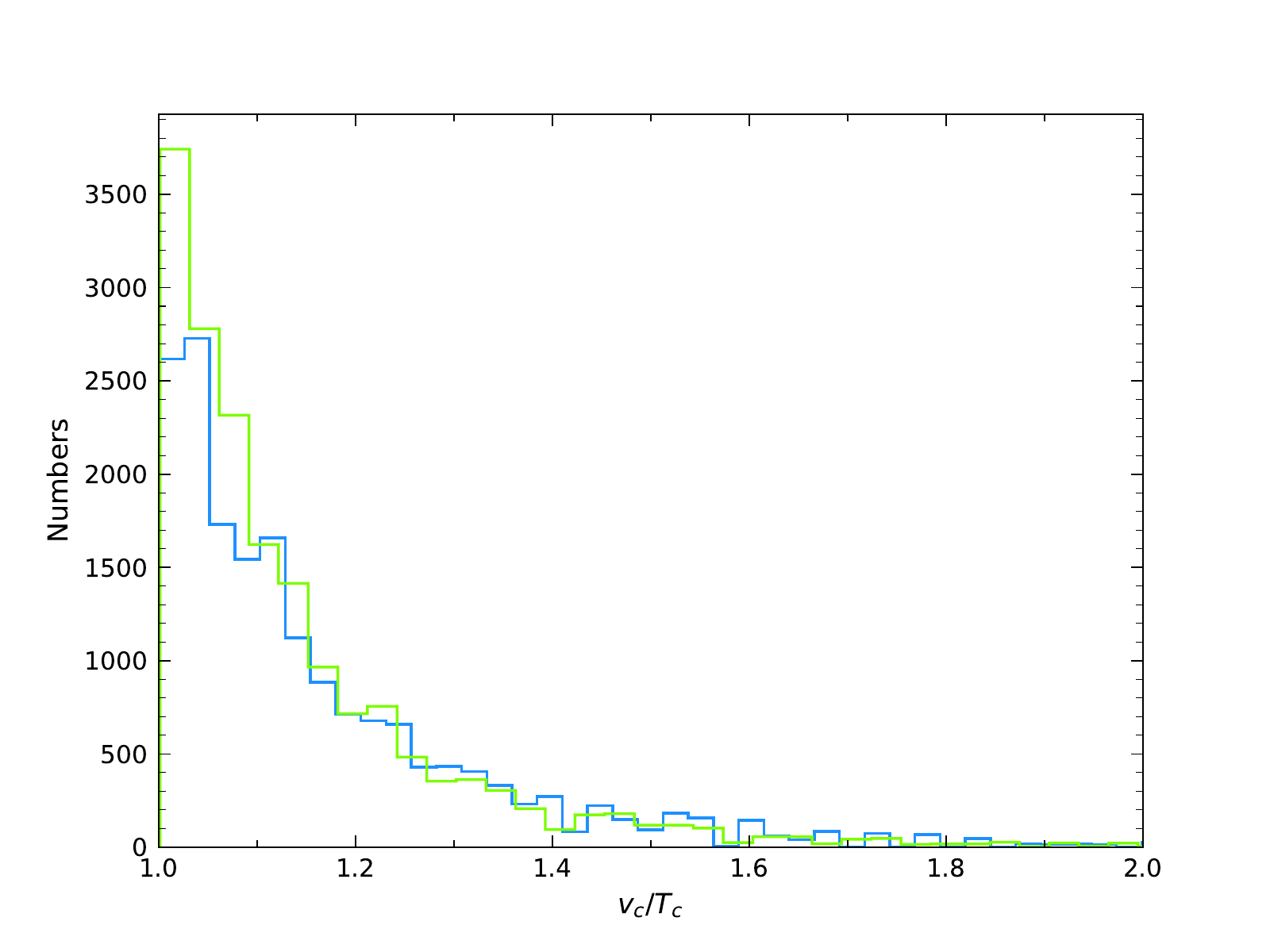}
  \includegraphics[width=75mm,angle=0]{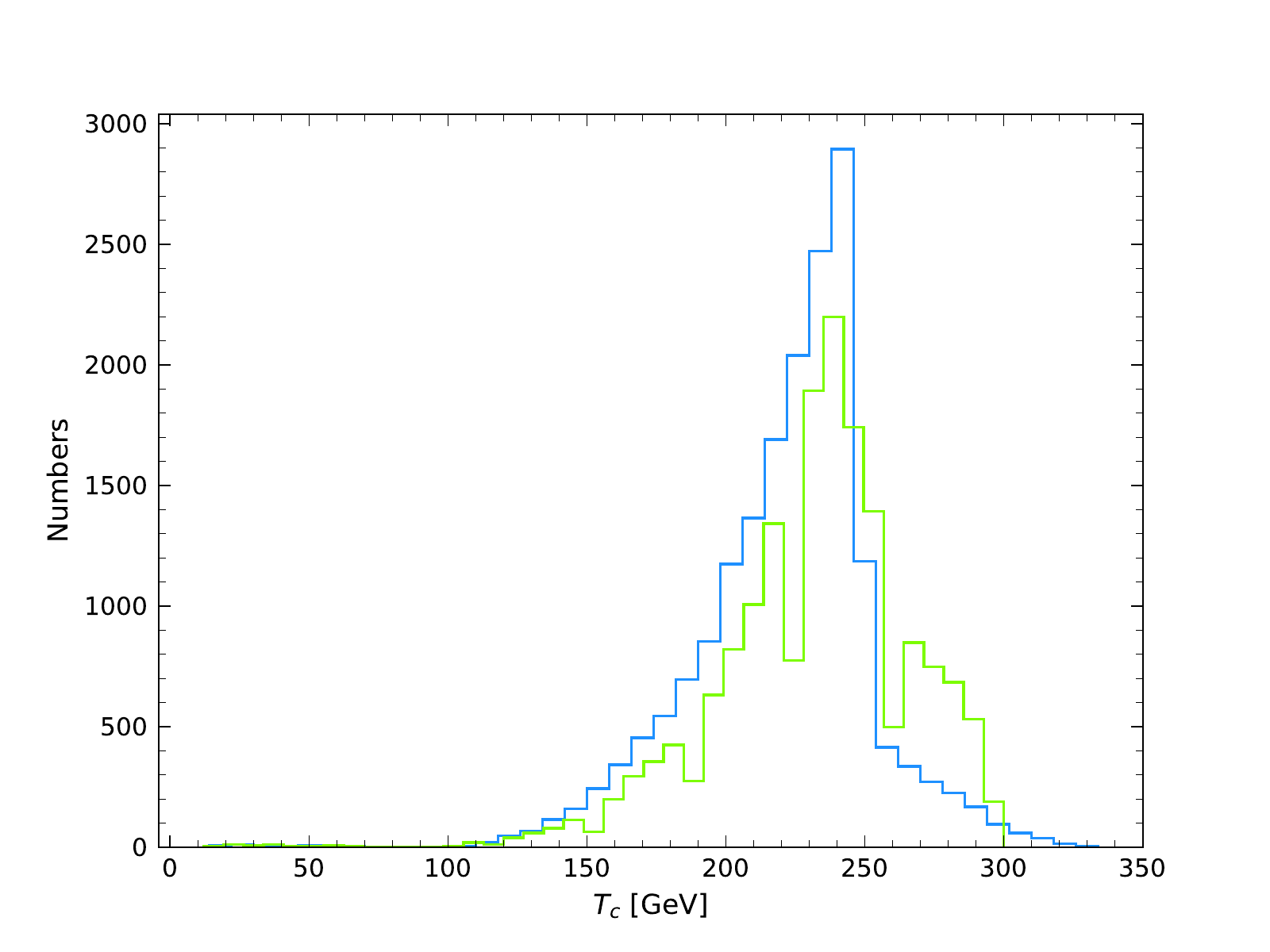}\\
  \includegraphics[width=75mm,angle=0]{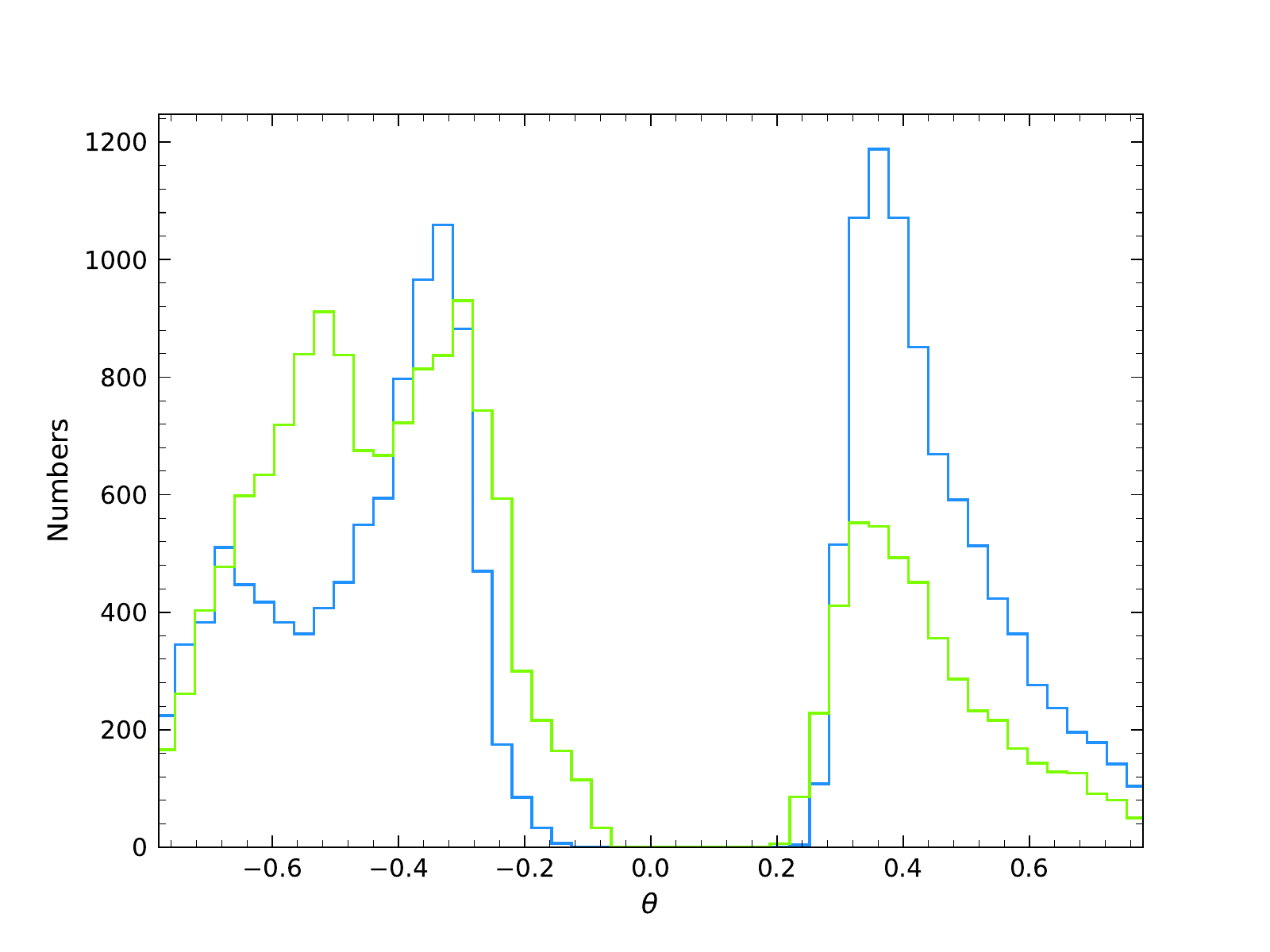}
  \includegraphics[width=75mm,angle=0]{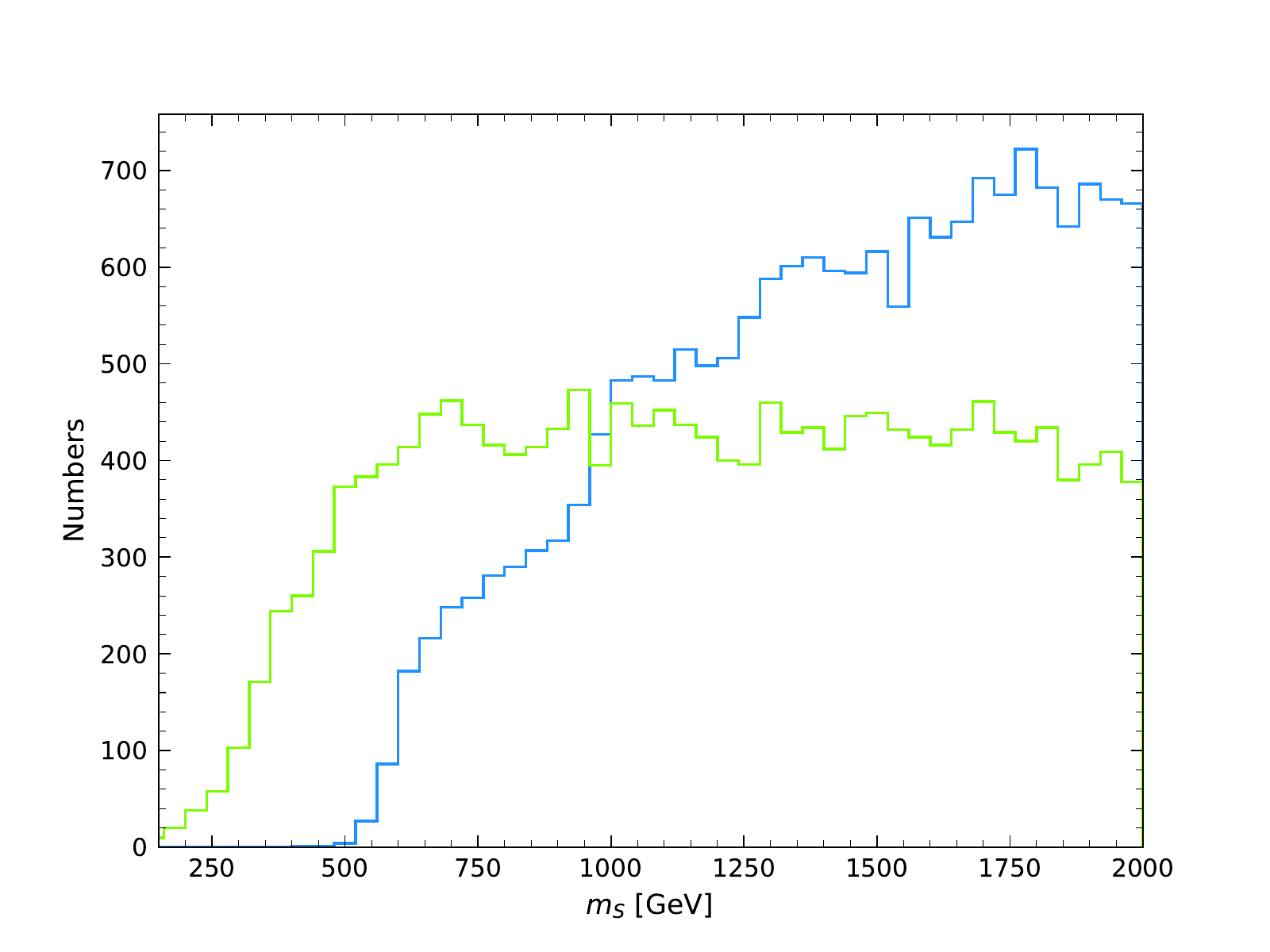}
  \caption{The upper two plots give the distributions in $v_c/T_c$ and $T_c$ and
    the lower two plots give the distributions in $\theta$ and $m_{\mathcal{S}}$ that can generate a sufficiently strong first-order EWPT.
    The blue and green histograms represent the random parameter scan results obtained in the Landau gauge potential and the gauge-independent potential, respectively.
    }
  \label{fig:gaugedepend1}
\end{figure}

\begin{figure}
  \centering
  \includegraphics[width=75mm,angle=0]{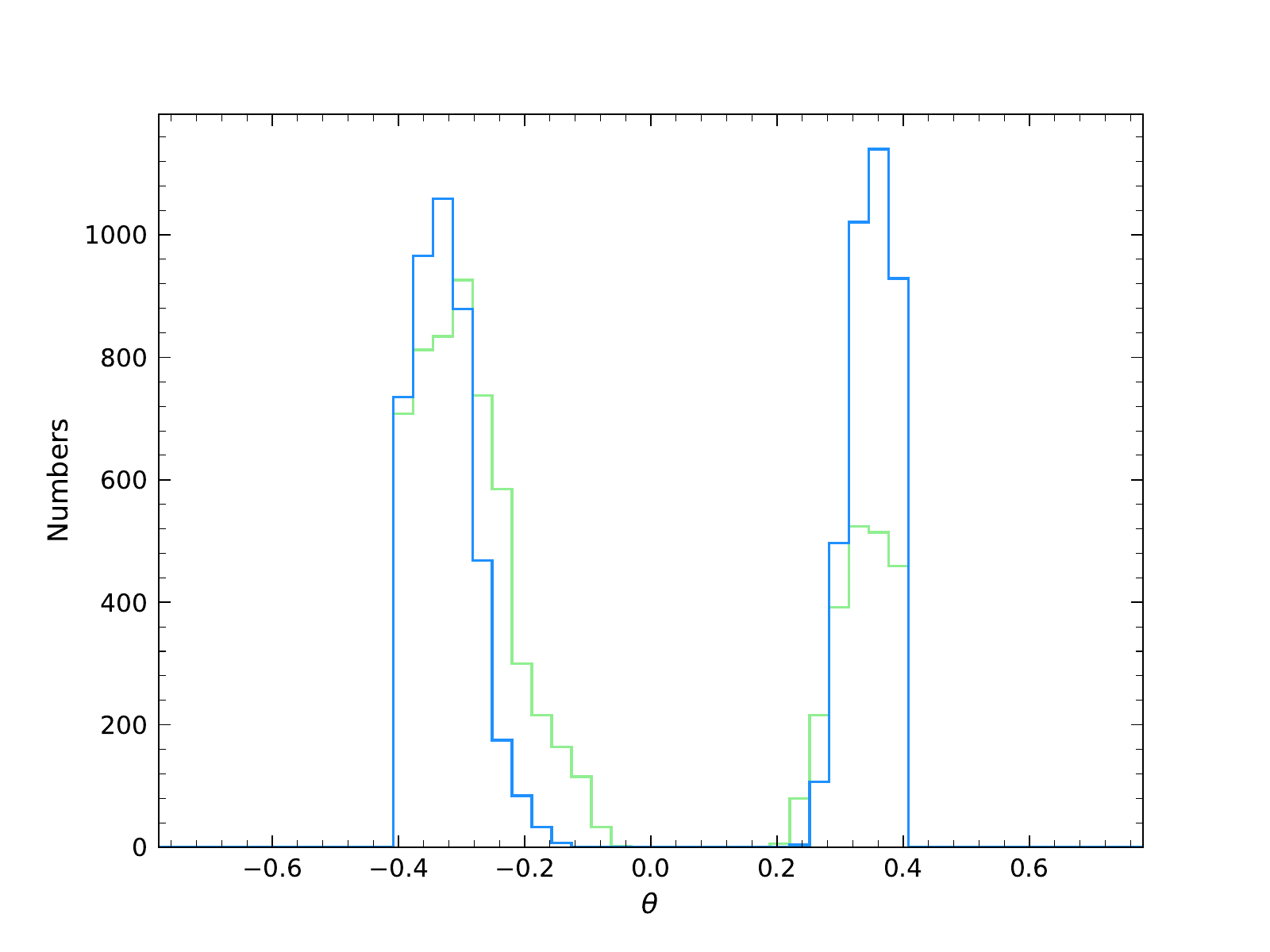}
  \includegraphics[width=75mm,angle=0]{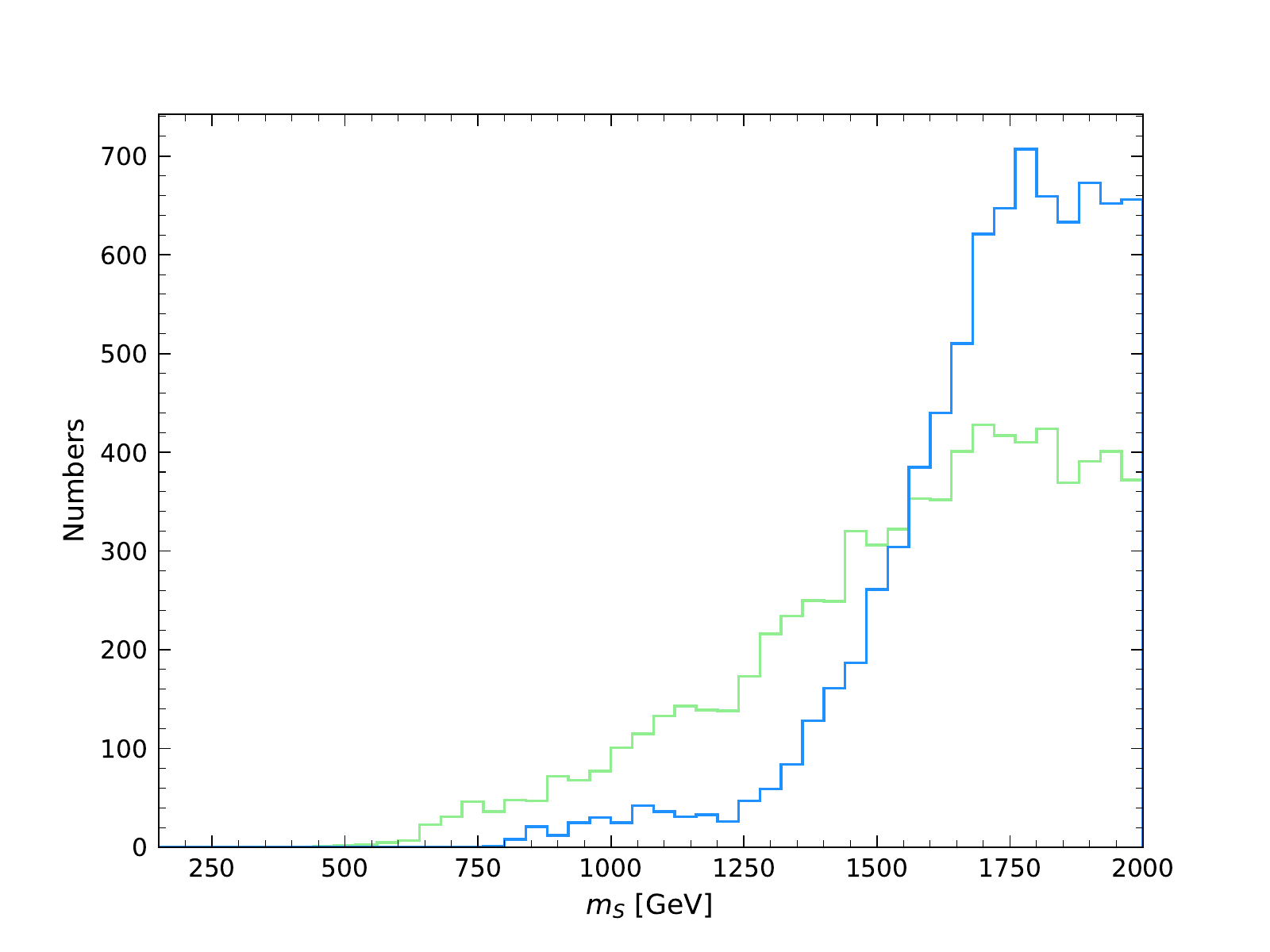}
  \caption{Same as the lower two plots in Fig.~\ref{fig:gaugedepend1} after taking into account the experimental constraints.
  }
  \label{fig:gaugedepend2}
\end{figure}

\begin{figure}
  \centering
  \includegraphics[width=75mm,angle=0]{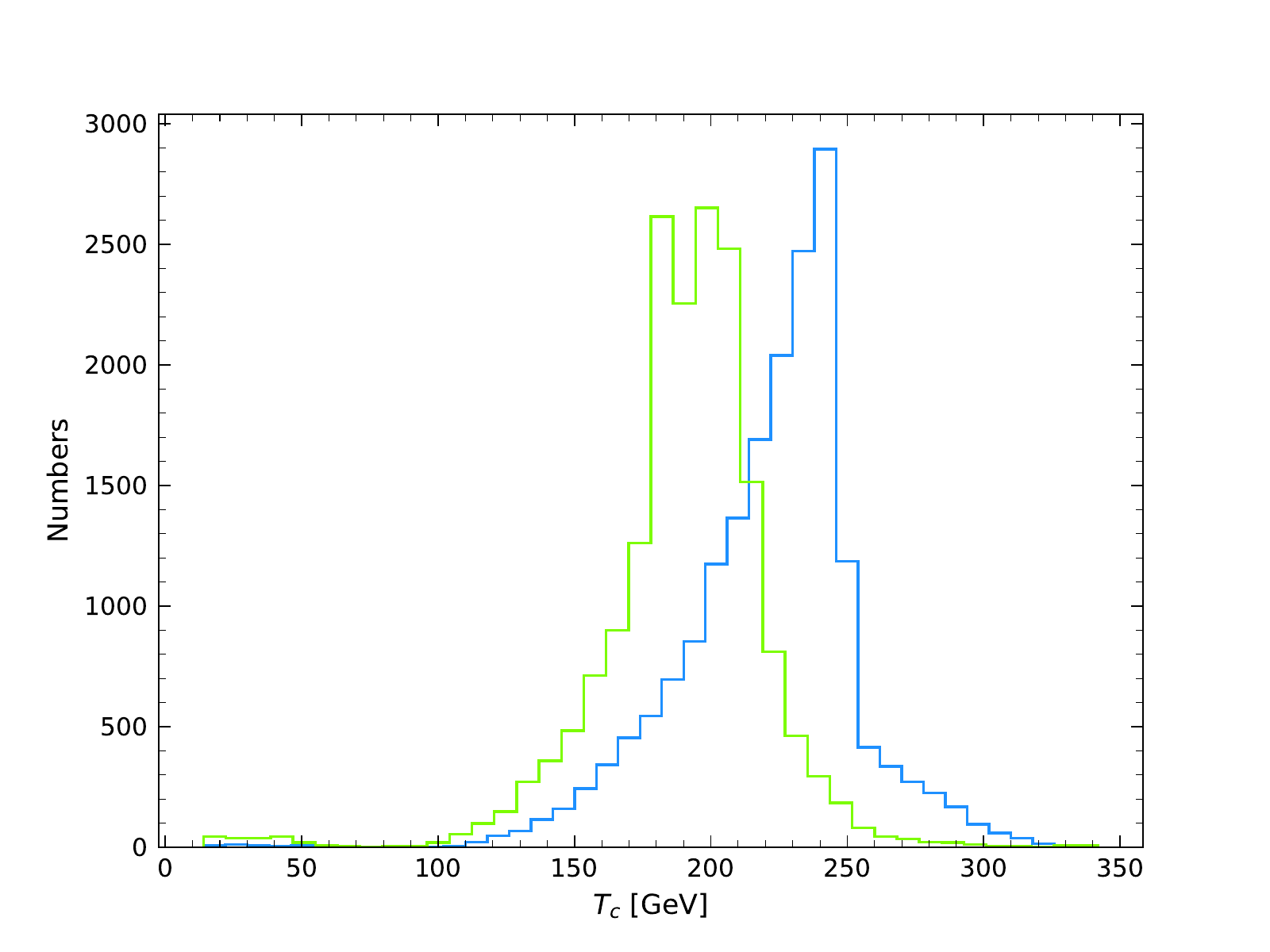}
  \includegraphics[width=75mm,angle=0]{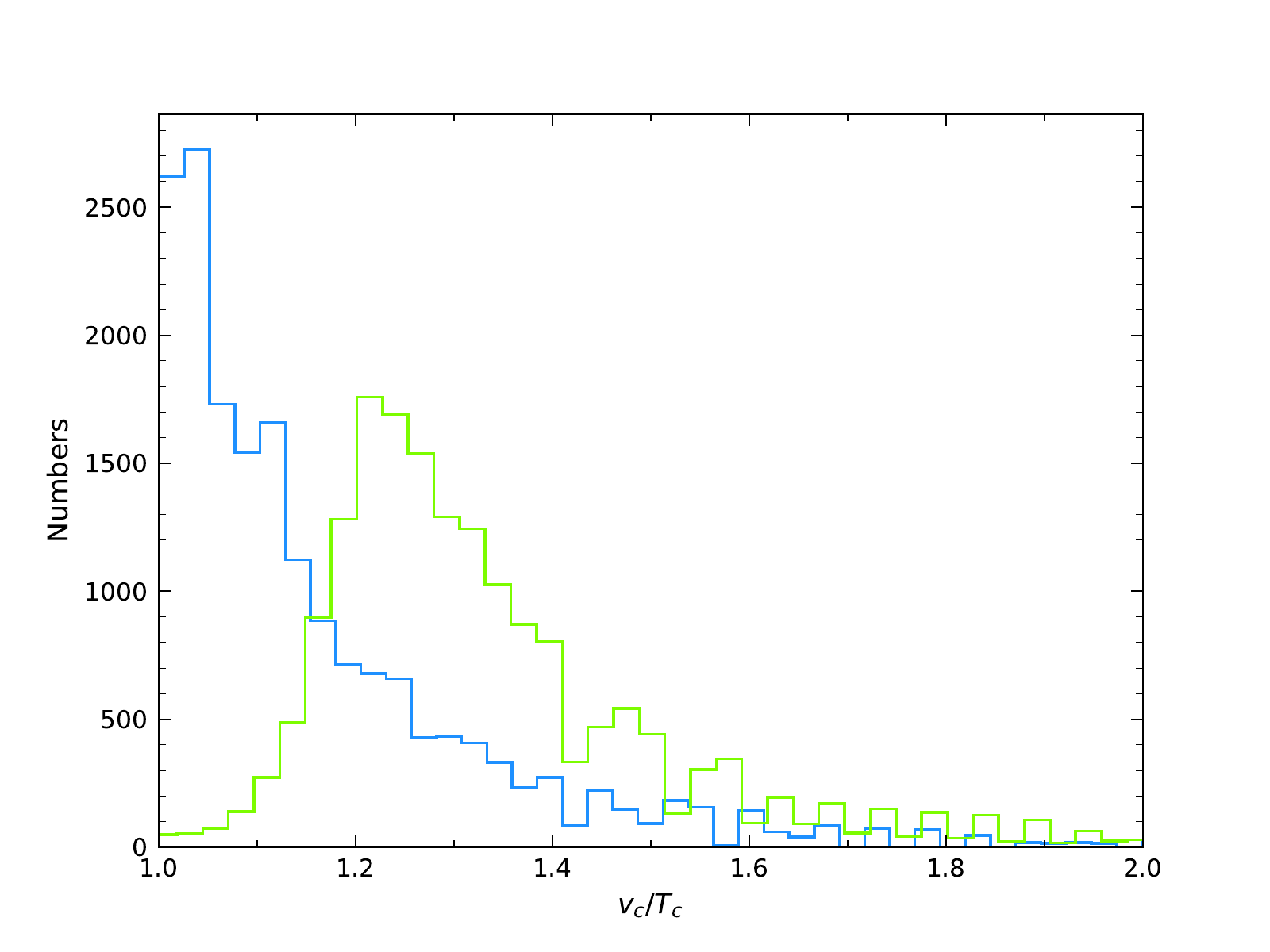}\\
  \includegraphics[width=75mm,angle=0]{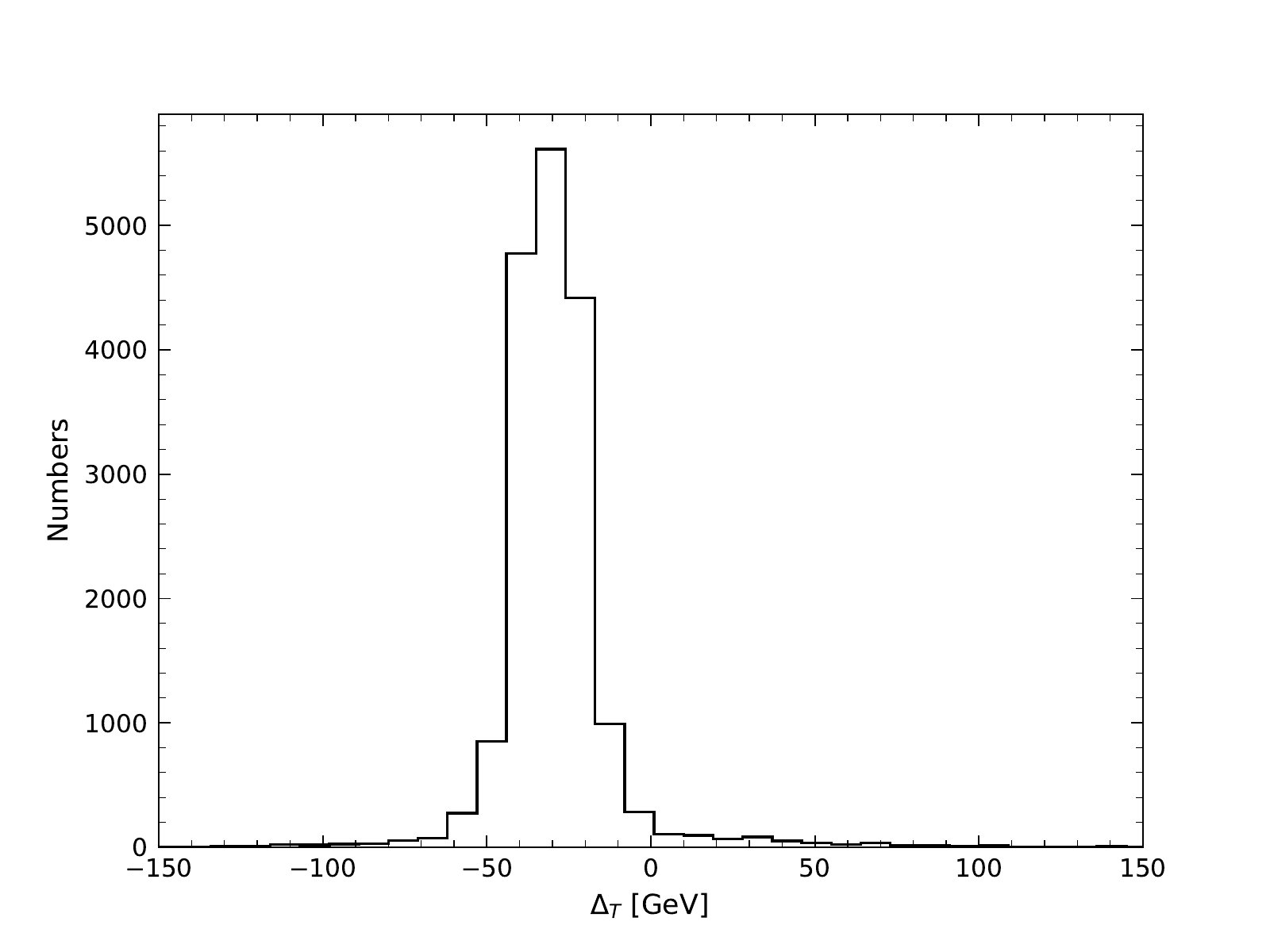}
  \includegraphics[width=75mm,angle=0]{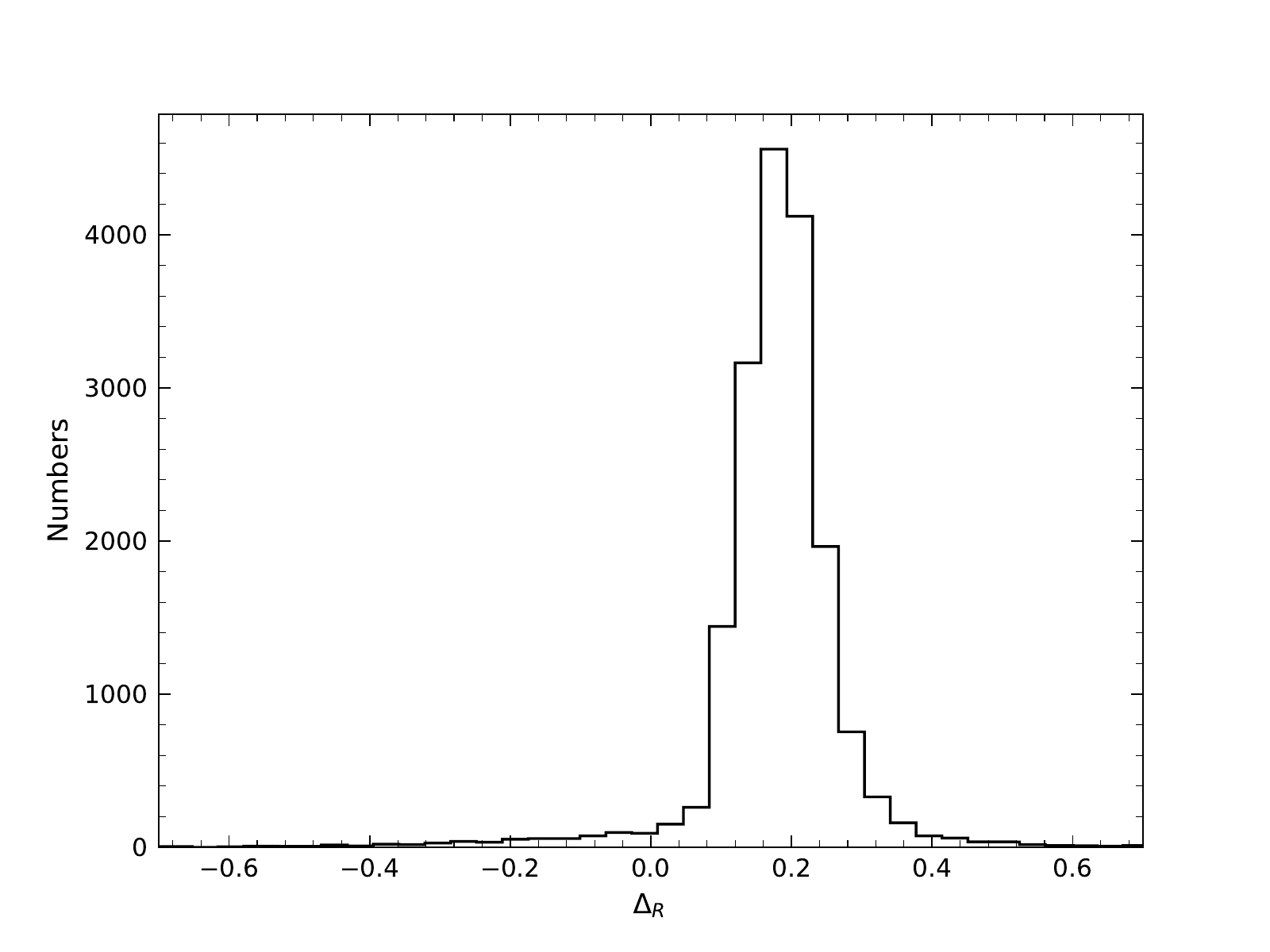}
  \caption{Upper plots: Distributions in $T_c$ and $v_c/T_c$ in the Landau gauge potential (blue histogram) and gauge-independent potential (green histogram). 
  Lower plots: Distribution of difference between the gauge-independent potential and the Landau gauge potential in $T_c$ and $v_c/T_c$.
  The samples of input parameters are the same for both of the potential.
  }
  \label{fig:gaugedepend3}
\end{figure}

\begin{figure}
  \centering
  \includegraphics[width=85mm,angle=0]{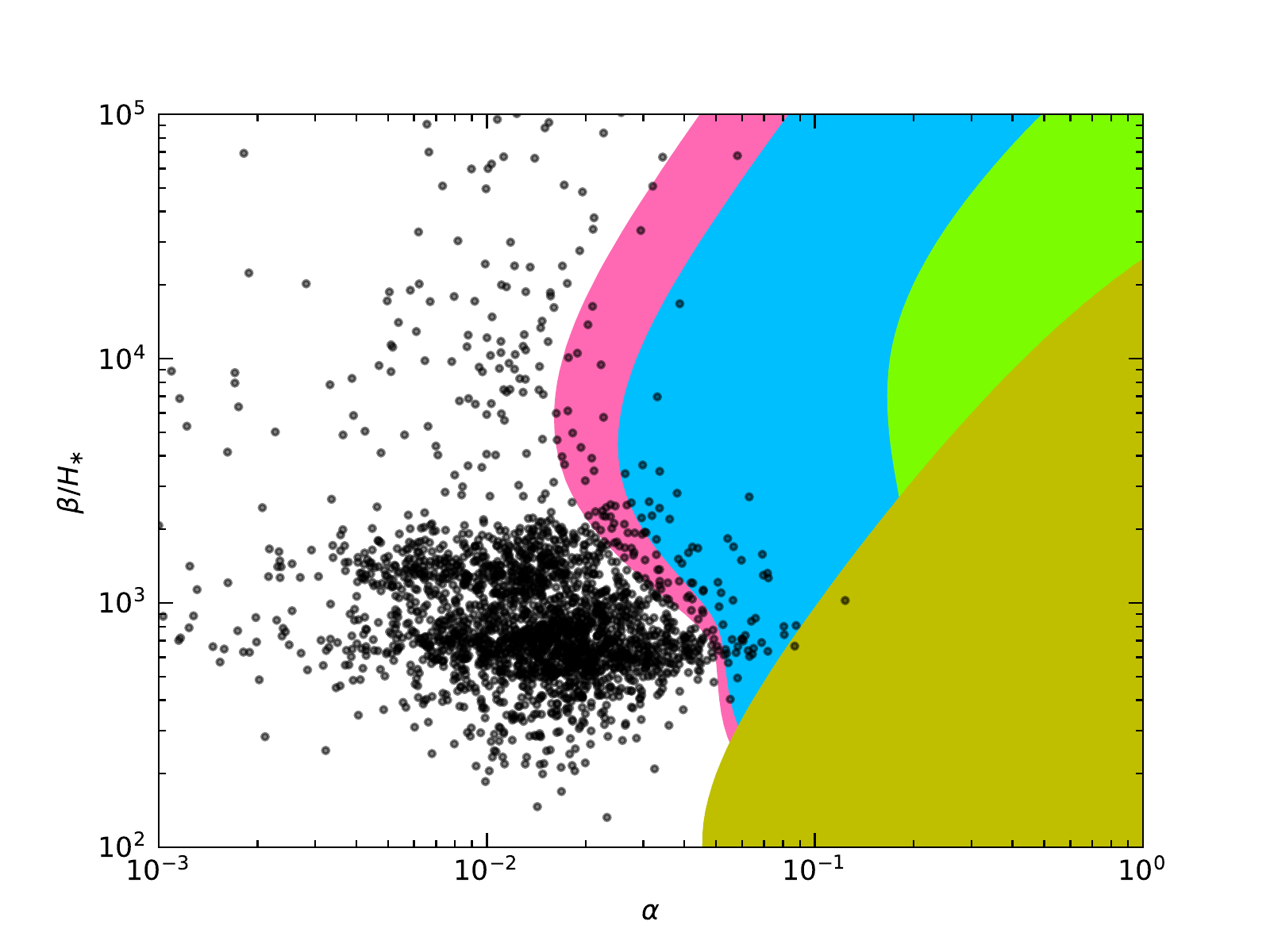}
  \caption{The pink, blue, yellow, and green regions represent respectively the sensitivities of BBO, DECIGO, LISA, and B-DECIGO exceeding 
  the detection threshold, assuming the nucleation temperature $T_{n}$ = 200 GeV. The scatter points are the samples that both generate a strong 
  first- order EWPT and survive the collider searches and DM experiments. The gauge-independent potential is adopted for the calculations}
  \label{fig:gaugedepend4}
\end{figure}

To compare with those results obtained in section~\ref{sec:ss} by using a Landau gauge, we now search for the first-order EWPT-viable parameter space with the gauge-independent effective potential using the same scheme of random parameter scan described in section~\ref{sec:seachsch}. We show the main results in Fig.~\ref{fig:gaugedepend1}. In the figure, the blue histograms are the distributions obtained with a Landau gauge, while the green histograms represent those obtained by using a gauge-independent effective potential, both of them containing about 18000 samples. 
The upper two plots of Fig.~\ref{fig:gaugedepend1} show similar distributions in $v_c/T_c$ and $T_c$. 
From the distribution of $\theta$, we see that the conclusion $|\theta|\gtrsim 0.2$ for a strong first-order EWPT still holds for the gauge-independent potential. However, the lower cutoff in the distribution of $m_{\mathcal{S}}$ is reduced to $\sim 250$ GeV.  The reason for the existence of such a lower bound has been given in section~\ref{sec:paramsdist}.

We now further impose the experimental constraints on the parameter space with the gauge-independent potential, the $\theta$ and $m_{\mathcal{S}}$ distributions are showed in Fig.~\ref{fig:gaugedepend2} (green histograms). As shown in the figure, $0.2\lesssim |\theta|\lesssim 0.4$ and $1.0\lesssim m_{\mathcal{S}}\lesssim 2.0$ TeV are required for a sufficiently strong EWPT while taking into account the constraints from collider experiments when the gauge-independent potential is used. 
Hence, our main conclusions from the effective potential using the Landau gauge remain nearly unchanged.

To estimate the impacts of gauge dependence on criterion \eqref{eq:criterion}, we re-calculate $v_c$ and $T_c$ in the gauge-independent potential 
with those samples that satisfy $v_c/T_c\gtrsim 1$ in the Landau gauge potential, the results are represented by the green histograms in Fig.~\ref{fig:gaugedepend3}.
Compare with the random scan results in Fig.~\ref{fig:gaugedepend1}, all of the samples are the same here.
From the figure we see that for the gauge-independent potential, the distributions in $T_c$ and $v_c/T_c$ mainly fall in the ranges of $\sim 150-250$ GeV and $\sim 1.1-1.4$, respectively. 
For each sample point, we calculate the difference in $T_c$ and $v_c/T_c$ between the gauge-independent potential and the Landau gauge potential, denoted by $\Delta_T$ and $\Delta_R$, respectively.
We show the distributions of $\Delta_T$ and $\Delta_R$ in the lower plots of Fig.~\ref{fig:gaugedepend3}.
We find that when the full one-loop potential with the Landau gauge is replaced by the gauge-independent potential, the critical temperature of 
the phase transition could decrease by $\sim 30$ GeV and $v_c/T_c$ could increase $\sim 0.18$.  Thus, if a sample point triggers a first-order EWPT in the full one-loop potential in the Landau gauge, it also does in the gauge-independent potential.

In Fig.~\ref{fig:gaugedepend4}, we show the results of GW parameters $\alpha$ and $\beta/H_{\ast}$ calculated using the gauge-independent potential.
All the samples generate a strong first-order EWPT and survive the experimental constraints. 
As illustrated in this figure, there is a considerable portion of samples that can be detected by BBO and DECIGO.  However, in contrast with the results obtained from the Landau gauge potential, relatively few sample points fall within the sensitivity of LISA.
This means that the GWs produced from the first-order EWPT in the gauge-independent scalar potential may be too weak to be detected by the LISA interferometer. 
A similar conclusion was also found in Ref.~\cite{Gould2019PRD}, in which a nonperturbative analysis of the GW power spectrum was adopted to show that GWs produced from a first-order EWPT in a beyond-the-SM scenario described by a SM-like effective theory will be so weak that the LISA experiment is unable to probe.

\section{Summary and conclusion}
\label{sec:summary}

In this work, we have considered the extension of SM with a complex singlet scalar field $S$. 
A global $U(1)$ symmetry associated with $S$ in the scalar potential is softly broken by its cubic terms to a $\mathbb{Z}_3$ symmetry.
The real part of the complex scalar field develops a VEV and mixes with the SM Higgs boson after the electroweak symmetry breaking, while the imaginary component, the pseudo-Goldstone boson $\chi$, becomes a DM candidate due to an accidental $\mathbb{Z}_2$ symmetry hidden in the scalar potential.
We focus on the two-step phase transition scenario, in which the location of global minimum of the potential 
moves as $(0,~0)\to (0,~w_0(T))\to (v(T),~w(T))$ with the expansion of the Universe.
We then search for the parameter space for sample points where a sufficiently strong EWPT can occur, taking into account the constraints from collider experiments and DM searches.  The conclusions we have obtained are summarized as follows:
\begin{itemize}
  \item The requirement of a sufficiently strong EWPT demands 
  $\left | \theta \right |\gtrsim 0.2$ ($11.5^{\circ}$) for $m_{\mathcal{S}}\lesssim 2$~TeV.
  The Higgs signal strength measurements, on the other hand, give the constraint $\left | \theta \right |\lesssim 0.4$ ($23^{\circ}$).
  \item To trigger a successful phase transition, the real scalar mass $m_{\mathcal{S}}$ should be larger than about 500~GeV (250~GeV in the gauge-independent potential).
  The constraints of Higgs signal strength measurements further pushes up the lower bound of the scalar mass: $m_{\mathcal{S}}\gtrsim 1.2$~TeV (1.0~TeV in the gauge-independent potential).
  \item A small fraction of DM could be made of the pseudo-Goldstone boson $\chi$, while the constraints from
  current DM direct searches are satisfied.
  \item GW signals from the first-order phase transition in the model with the mixing angle $\theta$ in the range of $\sim0.25-0.4$ can be 
  detectable using future space-based interferometers, such as DECIGO and BBO.
\end{itemize}

We concentrate in this work on the two-step phase transition scenario as described in section~\ref{sec:tspt}.
Other types of multi-step phase transitions can be found in recent works~\cite{Hashino2018JHEP, Alves2019JHEP, Carena2019}.
The requirement of a large mixing angle to trigger a sufficiently strong EWPT is crucial in the two-step phase transition.
The one-step phase transition, $(0,~0)\to (v(T),~w(T))$, can be stronger, and for models with a mixing angle $\lesssim 0.2$, the phase transition may be detectable in planned space-based interferometers~\cite{Alves2019JHEP}.
We also note that in our analysis, we have restricted the heavy scalar mass $m_{\mathcal{S}}$ to be in the range of $150-2000$~GeV ($m_{\mathcal{S}}>m_{\mathcal{H}}$), 
the strong EWPT with a lower scalar mass, $m_{\mathcal{S}}<m_{\mathcal{H}}$, can be found in Refs.~\cite{Carena2019, Kozaczuk2019}.
Future precision Higgs measurements at collider experiments, such as the high-luminosity LHC , the International Linear Collider, and the Circular Electron-Positron Collider, could further probe the mixing angle regions that trigger a strong first-order EWPT
in our scenario~\cite{Profumo2015PRD}. A more systematic and general analysis of the EWPT and the related collider phenomenology has been given in a recent work~\cite{Michael2019}. Our scenario for EWPT and its phenomena at the colliders could be a particular illustration of those more general considerations.

\acknowledgments

We would like to thank E. Senaha and the anonymous Referee for helpful comments and discussions.
This work was supported in part by the Ministry of Science and Technology (MOST) of Taiwan under Grant Nos. MOST-104-2628-M-002-014-MY4, MOST-108-2112-M-002-005-MY3, MOST-108-2811-M-002-524 and MOST-108-2811-M-002-548.

\appendix

\section{Field-dependent mass matrix at finite temperature}
\label{apd:fdms}

The thermally corrected mass matrix of $h$ and $s$ is
\begin{eqnarray}
  \mathcal{M}^2&=&\begin{pmatrix}
    -\mu_{h}^{2}+3 \lambda_{h} h^{2}+\frac{1}{2}\lambda_{m}s^2 & \lambda_{m}hs\\ 
    \lambda_{m}hs & \mu_{s}^{2}+3 \lambda_{s} s^{2}+2 \mu_{3} s+\frac{1}{2} \lambda_{m} h^{2}
    \end{pmatrix}\nonumber\\
    &&\qquad
    +\frac{T^2}{48}\begin{pmatrix}
    9 g^{2}+3 g^{\prime 2}+2[6 (y_{t}^{2}+y_{b}^{2})+12 \lambda_{h}+\lambda_{m}] & 0\\ 
    0 & 4\left(2 \lambda_{m}+3 \lambda_{s}\right)
    \end{pmatrix},
\end{eqnarray}
where $g$ and $g^{\prime}$ are the SM gauge couplings of $SU(2)_{L}$ and $U(1)_{Y}$, $y_t$ and $y_b$ are the top and bottom quark Yukawa couplings.
The thermally corrected masses of SM Goldstone bosons $G^{0,\pm}$ and pseudo-Goldstone boson $\chi$ are
\begin{align}
\begin{split}
  \mathcal{M}_{G}^2&=-\mu_{h}^{2}+\lambda_{h} h^{2}+\frac{1}{2}\lambda_{m}s^2
  +\frac{T^2}{48}\{9 g^{2}+3 g^{\prime 2}+2[6 (y_{t}^{2}+y_{b}^{2})+12 \lambda_{h}+\lambda_{m}]\},\\
  \mathcal{M}_{\chi}^2&=\mu_s^2+\lambda_s s^2+\frac{1}{2}\lambda_m h^2-2\mu_{3} s+\frac{T^2}{12}\left(2 \lambda_{m}+3 \lambda_{s}\right).
\end{split}
\end{align}
The field-dependent masses for the transverse components of SM massive gauge bosons $W$ and $Z$ are
\begin{eqnarray}
  \mathcal{M}_{W,T}^{2}=\frac{1}{4}g^2h^2,~~\mathcal{M}_{Z,T}^{2}=\frac{1}{4}(g^2+g^{\prime 2})h^2.
\end{eqnarray}
The longitudinal components of $W$ and $Z$ receive thermal corrections from the daisy diagrams, and their masses in terms of 
$\left(W_{\mu}^{+}, W_{\mu}^{-}, W_{\mu}^{3}, B_{\mu}^{0}\right)$ basis can be written as 
\begin{eqnarray}
  \mathcal{M}_{L}^2=\frac{h^2}{4}\begin{pmatrix}
    g^2 &  &  & \\ 
     &  g^2 &  & \\ 
     &  & g^2 & g {g}' \\ 
     &  & g{g}' & g^{\prime 2}
    \end{pmatrix}
    +\frac{11}{6}T^2\begin{pmatrix}
    g^2 &  &  & \\ 
     &  g^2 &  & \\ 
     &  & g^2 &  \\ 
     &  &  & g^{\prime 2}
    \end{pmatrix}.
\end{eqnarray}
The field-dependent masses of top and bottom quarks are 
\begin{eqnarray}
  \mathcal{M}_{t}^2=\frac{1}{2}y_th^2,~~\mathcal{M}_{b}^2=\frac{1}{2}y_bh^2.
\end{eqnarray}

\section{Renormalization group equations}
\label{apd:RGE}

The RGEs at the one-loop level are~\cite{Staub2015AHEP}
\begin{eqnarray}
  \label{eq:rges}
  16\pi^2\beta_{\lambda_{h}}&=&\lambda_{h}\left( 24\lambda_h-\frac{9}{5}g_1^2-9g_2^2+\frac{1}{4}\lambda_m^2+12y_t^2 \right )+
  \frac{27}{200}g_{1}^4+\frac{9}{20}g_1^2g_2^2+\frac{9}{8}g_2^4-6y_t^4,~~~~\\
  16\pi^2\beta_{\lambda_{m}}&=&\lambda_m\left( 12\lambda_h-\frac{9}{10}g_1^2-\frac{9}{2}g_2^2+2\lambda_m+2\lambda_s+6y_t^2 \right ),\\
  16\pi^2\beta_{\lambda_{s}}&=&2\lambda_m^2+5\lambda_s^2,\\
  16\pi^2\beta_{\mu_h^2}&=&\mu_h^2\left( 12\lambda_h-\frac{9}{10}g_1^2-\frac{9}{2}g_2^2+6y_t^2 \right )-\frac{1}{2}\lambda_m\mu_s^2,\\
  16\pi^2\beta_{\mu_s^2}&=&2\mu_3^2-4\lambda_m\mu_h^2+2\lambda_s\mu_s^2,\\
  \label{eq:rgee}
  16\pi^2\beta_{\mu_3}&=&3\lambda_s\mu_3.
\end{eqnarray}

\begin{figure}
  \centering
  \includegraphics[width=90mm,angle=0]{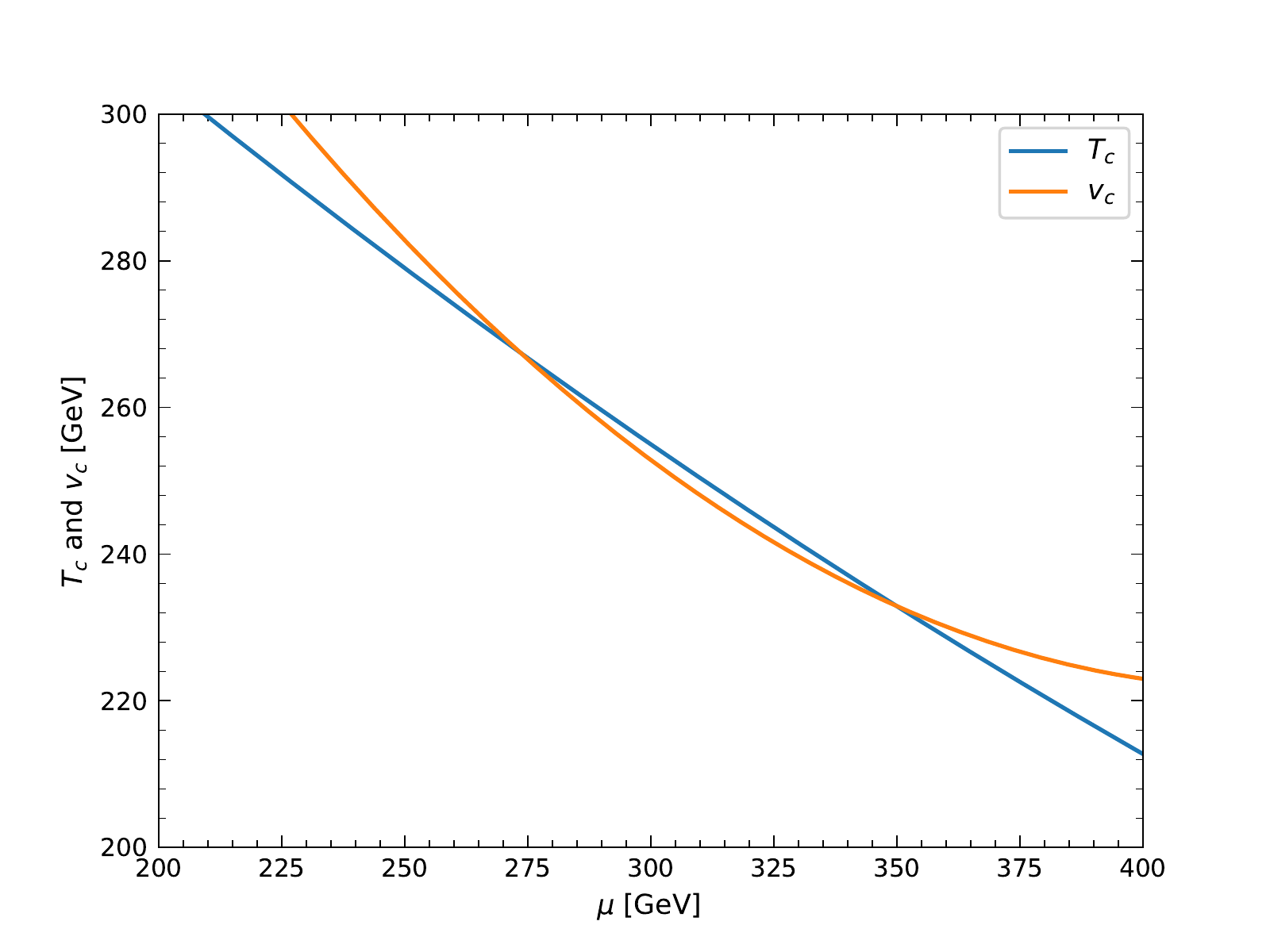}
  \caption{$T_c$ and $v_c$ as functions of $\mu$, with $w(v)=972.7$~GeV, $m_{\mathcal{S}}(v)=655.9$~GeV, $m_{\chi}(v)=460.0$~GeV, and $\theta(v)=0.646$.}
  \label{fig:rgi}
\end{figure}

The parameters in the tree-level potential and $V_{\rm CW}$ have a dependence on the renormalization scale $\mu$, and can lead to a significant $\mu$-dependence in $T_c$ and $v_c$.  We have fixed the value of renormalization scale $\mu=v$ in the results presented in the main text.

In Fig.~\ref{fig:rgi} we estimate the $\mu$-dependence of $T_c$ and $v_c$ by way of example.  To do this, we first take a sample point of the input parameters: $w(v)=972.7$~GeV, $m_{\mathcal{S}}(v)=655.9$~GeV, $m_{\chi}(v)=460.0$~GeV, and $\theta(v)=0.646$ and determine the parameters at another renormalization scale $\mu$ using Eqs.~\eqref{eq:rges}-\eqref{eq:rgee}.  We then use these parameters to calculate an RGI potentials. Following the procedure described in section~\ref{sec:seachsch}, we determine $T_c(\mu)$ and $v_c(\mu)$, as shown in Fig.~\ref{fig:rgi}.
We observe that both $T_c$ and $v_c$ decrease with $\mu$, while the ratio $v_c/T_c$ is seen to have much less dependence on the renormalization scale.

\section{Counter-terms}
\label{apd:ccp}

To maintain the main properties of the tree-level potential derived in the content, 
we add a counter-terms~\eqref{eq:cterm} to renormalize the potential at zero temperature.
The renormalization conditions we used are given by
\begin{align}
\begin{split}
&
  \left(\frac{\partial}{\partial h}, \frac{\partial}{\partial s}, \frac{\partial^{2}}{\partial h^{2}}, 
  \frac{\partial^{2}}{\partial s^{2}}, \frac{\partial^{2}}{\partial h \partial s}\right)
  \left.\begin{matrix}\left(V_{\rm CW}+V_{\rm CT}\right)\end{matrix}\right|_{(h,s)=(v,w)}=0,
\\
&
  \frac{\partial }{\partial s}\left.\begin{matrix}
  (V_{\rm CW}+V_{\rm CT})
  \end{matrix}\right|_{(h,s)=(v,w')}=0,
\end{split}
\end{align}
where $w'=s_-$ is determined by Eq.~\eqref{eq:brokv}. 
With
\begin{align}
\begin{split}
&
F_{10}=\frac{\partial }{\partial h}\left.\begin{matrix}V_{\rm CW}(h,s)\end{matrix}\right|_{(h,s)=(v,w)},~~
F_{01}=\frac{\partial }{\partial s}\left.\begin{matrix}V_{\rm CW}(h,s)\end{matrix}\right|_{(h,s)=(v,w)},
\\
&
F'_{01}=\frac{\partial }{\partial s}\left.\begin{matrix}V_{\rm CW}(h,s)\end{matrix}\right|_{(h,s)=(v,w')},~~
F_{11}=\frac{\partial ^2}{\partial h\partial s}\left.\begin{matrix}V_{\rm CW}(h,s)\end{matrix}\right|_{(h,s)=(v,w)},
\\
&
F_{20}=\frac{\partial ^2}{\partial h^2}\left.\begin{matrix}V_{\rm CW}(h,s)\end{matrix}\right|_{(h,s)=(v,w)},~~
F_{02}=\frac{\partial ^2}{\partial s^2}\left.\begin{matrix}V_{\rm CW}(h,s)\end{matrix}\right|_{(h,s)=(v,w)},
\end{split}
\end{align}
the renormalization conditions fix the counter-terms as
\begin{align}
\begin{split}
\delta u_{h}^{2}
&= \frac{1}{2v}\left ( -3F_{10}+wF_{11}+vF_{20} \right ),
\\
\delta \lambda_{h}
&= -\frac{1}{2 v^{3}}\left ( -F_{10}+vF_{20} \right ),
\\
\delta \lambda_{m}
&= -\frac{1}{vw}F_{11},
\\
\delta\mu_{s}^2
&= \frac{1}{2ww'(w-w')^2}( 4w'^{3}F_{01}-6ww'^2F_{01}-vw'^3F_{11}
\\
&~~ +2vww'^2F_{11}-2ww'^3F_{02}-vw^2w'F_{11}+2w^2w'^2F_{02}+2w^3F'_{01} ),
\\
\delta\mu_3
&= \frac{1}{w^2w'(w-w')^2}\left ( -w^3F_{01}+3w'w^2F_{01}-2w^3F'_{01}+ww'^3F_{02}-w'w^3F_{02} \right ),
\\
\delta\lambda_{s}
&= \frac{1}{w^2w'(w-w')^2}\left ( w'^2F_{01}-2ww'F_{01}-ww'^2F_{02}+w'w^2F_{02}+w^2F'_{01} \right).
\end{split}
\end{align}

\section{Decay width}
\label{apd:dw}

Here we give the partial width formulas for some decays of SM-like Higgs $\mathcal{H}$ and heavy scalar $\mathcal{S}$:
\begin{align}
\begin{split}
  \Gamma _{\mathcal{H}\to \chi\chi}
  &= \frac{g_{\mathcal{H}\chi\chi}^2}{8\pi m_{\mathcal{H}}}\sqrt{1-\frac{4m_{\chi}^2}{m_{\mathcal{H}}^2}},~~
  \Gamma _{\mathcal{H}\to \mathcal{SS}}=\frac{g_{\mathcal{HSS}}^2}{8\pi m_{\mathcal{H}}}\sqrt{1-\frac{4m_{\mathcal{S}}^2}{m_{\mathcal{H}}^2}},
  \\
  \Gamma _{\mathcal{S}\to \chi\chi}
  &= \frac{g_{\mathcal{S}\chi\chi}^2}{8\pi m_{\mathcal{S}}}\sqrt{1-\frac{4m_{\chi}^2}{m_{\mathcal{S}}^2}},~~
  \Gamma _{\mathcal{S}\to \mathcal{HH}}=\frac{g_{\mathcal{SHH}}^2}{8\pi m_{\mathcal{S}}}\sqrt{1-\frac{4m_{\mathcal{H}}^2}{m_{\mathcal{S}}^2}},
\end{split}
\end{align}
where the couplings
\begin{align}
\begin{split}
g_{\mathcal{H}\chi\chi}
&= -2(\mu_3-\lambda_s w)\sin\theta-\lambda_mv\cos\theta,
\\
g_{\mathcal{S}\chi\chi}
&= 2(\mu_3-\lambda_s w)\cos\theta-\lambda_mv\sin\theta,
\\
g_{\mathcal{HSS}}
&= -\lambda_mv\cos^3\theta+2\cos^2\theta\sin\theta(\mu_3-\lambda_mw+3\lambda_s w)
\\
& \qquad +\sin^2\theta[2\cos\theta(-3\lambda_h+\lambda_m)+\lambda_mw\sin\theta],
\\
g_{\mathcal{SHH}}
&= -\lambda_mv\cos^3\theta-2v\cos^2\theta\sin\theta(3\lambda_h-\lambda_m)
\\
& \qquad -\sin^2\theta[2\cos\theta(\mu_3-\lambda_m w+3\lambda_s w)+\lambda_m v\sin\theta].
\end{split}
\end{align}

\section{Sensitivity of space-based interferometers}
\label{apd:ssbi}

Here we summarize the experimental noise power spectral density used in this work.  More details can be found in 
Ref.~\cite{Breitbach2019JCAP} and references therein.

The LISA interferometer is planned to launch in 2034, with the noise power spectral density well approximated 
by~\cite{LISA2017, Cornish2018}
\begin{eqnarray}
  S_{\rm LISA}(f)=\frac{10}{3L^2}\left ( P_{\rm OMS}(f)+2\left [ 1+\cos^2\left ( \frac{f}{f_{\ast}} \right ) \right ]
\frac{P_{\rm acc}(f)}{(2\pi f)^4} \right)\left [ 1+\frac{6}{10}\left ( \frac{f}{f_{\ast}} \right )^2 \right ]+S_c(f),~~~~
\end{eqnarray}
where $L=2.5\times 10^9$~m is the arm length of the LISA detectors and the transfer frequency $f_{\ast}=c/(2\pi L)$,
where $c$ is the speed of light. The instrument noise consists of the optical metrology noise
\begin{eqnarray}
  P_{\rm OMS}(f)=(1.5\times 10^{-11}{\rm m})^2\left [ 1+\left ( \frac{2{\rm~mHz}}{f} \right )^4 \right ]~{\rm Hz}^{-1},
\end{eqnarray}
and the test mass acceleration noise
\begin{eqnarray}
  P_{\rm acc}(f)=(3\times10^{-15}{~\rm m~sec^{-2}})^2\left [ 1+\left ( \frac{0.4~{\rm mHz}}{f} \right )^2 \right ]
  \left [ 1+\left ( \frac{f}{8~\rm mHz} \right )^4 \right ]~{\rm Hz}^{-1}.
\end{eqnarray}
The last term is the confusion noise from unresolved galactic binaries (4 years)
\begin{align}
  S_{c}(f)
  &=9\times 10^{-45}f^{-7/3} \left \{ 1+\tanh\left [ 2.184-\left ( \frac{1680f}{{\rm Hz}}  \right ) \right ] \right \}
  \nonumber \\
  & \qquad\qquad
  \times \exp \left [ -\left ( \frac{f}{{\rm Hz}} \right )^{0.138}-\left ( \frac{221f}{{\rm Hz}} \right )
  \sin\left ( \frac{521f}{{\rm Hz}} \right ) \right ] ~{\rm Hz}^{-1}.
\end{align}

The B-DECIGO as a scaled-down predecessor of DECIGO is planned to launch in 2020~\cite{Sato2017JPCS}. 
Its noise power spectral density can be approximated by~\cite{Isoyama2018}
\begin{eqnarray}
  S_{\rm B-DECIGO}(f)=2.02\times 10^{-48}\left [ 10^3+15.84\left ( \frac{f}{{\rm Hz}}  \right )^{-4}+
  1.584\left ( \frac{f}{{\rm Hz}} \right )^2 \right ]{~\rm Hz}^{-1}.
\end{eqnarray}

The DECIGO and BBO noise power spectral densities can be parameterized as~\cite{Yagi2011PRD, Yagi2013IJMPD}
\begin{eqnarray}
  S_{\rm DECIGO,BBO}(f)=\min\left [ \frac{S_{n}^{\rm inst}(f)}{\exp(-\kappa \mathcal{T}_{\rm obs}^{-1}dN/df)},
  ~S_{n}^{\rm inst}(f)+S_{n}^{\rm gal}(f)\mathcal{F}(f) \right ]+S_{n}^{\rm ex-gal}\mathcal{F}(f),~~~~~
\end{eqnarray}
where $\mathcal{T}_{\rm obs}$ is the duration of the mission,
$\kappa=5$, $dN/df=2\times (f/{\rm Hz})^{-11/3}~{\rm Hz}^{-1}$ is the spectral number density of galactic
white dwarf binaries, and the factor $\mathcal{F}(f)\equiv \exp{\left [ -2(f/0.05{\rm Hz})^2 \right ]}$
corresponds to the high frequency cutoff for the white dwarf confusion noises. 
The non sky-averaged instrumental noise spectral density for DECIGO and BBO are respectively
\begin{align}
\begin{split}
  S_{n,\rm DECIGO}^{\rm inst}(f)
  &= 5.3\times10^{-48}\Bigg [ 1+\left ( \frac{f}{f_p} \right )^2
  +\frac{2.3\times10^{-7}}{1+(f/f_p)^2}\left ( \frac{f}{f_p} \right )^{-4}
  \\
  & \qquad\qquad\qquad\qquad
  +2.6\times10^{-8}\left ( \frac{f}{f_p} \right )^{-4}\Bigg ]~{\rm Hz}^{-1},
  \\
  S_{n,\rm BBO}^{\rm inst}(f)
  &= 10^{-49}\left [ 1.8\left ( \frac{f}{\rm Hz} \right )^2+2.9
  +9.2\times10^{-3}\left ( \frac{f}{\rm Hz} \right )^{-4}\right ]~{\rm Hz}^{-1},
\end{split}
\end{align}
where $f_p=7.36$~Hz. The confusion noises from galactic and extra-galactic white dwarf binaries are given respectively by
\begin{align}
\begin{split}
S_{n}^{\rm gal}
&= 2.1\times10^{-45}\left ( \frac{f}{\rm Hz} \right )^{-7/3}~{\rm Hz}^{-1},
\\
S_{n}^{\rm ex-gal}
&= 4.2\times10^{-47}\left ( \frac{f}{\rm Hz} \right )^{-7/3}~{\rm Hz}^{-1}.
\end{split}
\end{align}
The experimental frequency range, duration of the mission and other parameters for the 
future GW interferometers are summarized in table~\ref{tab:sbi}.


\end{document}